\documentclass[11pt,letterpaper]{article}
\usepackage[T1]{fontenc}
\usepackage[utf8]{inputenc}
\usepackage[margin=1in]{geometry}
\usepackage{mathptmx,microtype}
\usepackage{graphicx,amsmath,amssymb,booktabs,multirow,tabularx}
\usepackage{caption,authblk}
\usepackage[numbers,sort&compress]{natbib}
\usepackage[hidelinks]{hyperref}
\usepackage{placeins}
\graphicspath{{figures/}}
\newcommand{\ManuscriptTitle}{From random patterns to nanophotonic structures: Electromagnetic field prediction without retraining}

\title{\ManuscriptTitle}

\author[1]{Jiwoo Kim}
\author[2]{Mingyu Park}
\author[1]{Chanik Kang}
\author[3]{Hyewon Suk}
\author[1,2,3,*]{Haejun Chung}
\affil[1]{Department of Artificial Intelligence, Hanyang University, Seoul, 04763, Republic of Korea}
\affil[2]{Department of Electronic Engineering, Hanyang University, Seoul, 04763, Republic of Korea}
\affil[3]{Department of Artificial Intelligence Semiconductor Engineering, Hanyang University, Seoul, 04763, Republic of Korea}
\date{*Corresponding author: Haejun Chung, email: haejun@hanyang.ac.kr}
\makeatletter
\let\MergeMaketitle\maketitle \let\MergeAtMaketitle\@maketitle
\let\MergeTitle\title \let\MergeDate\date \let\MergeAnd\and \let\MergeThanks\thanks
\let\MergeAtAuthor\@author
\makeatother
\begin{document}
\maketitle

\begin{abstract}
Electromagnetic surrogate solvers can accelerate nanophotonic design, but their broader use requires accurate predictions for unseen geometries and material arrangements. Training on random dielectric patterns enables direct prediction of complex electromagnetic fields in diverse two-dimensional nanophotonic structures without retraining. The network takes spatial permittivity distributions as inputs, with training samples drawn from a dataset comprising 700,140 combinations of 5,001 random freeform dielectric patterns and 140 relative permittivities from 2.1 to 16.0. The trained model predicts fields for aperiodic bar arrays, manually defined patterns, inverse-designed metalenses, inverse-designed binary structures, and two-material freeform patterns excluded from training. Across 35,441 geometry--permittivity samples, a dilated residual network achieves average relative field errors of 1.4\%--7.6\%. For 6,950 inverse-designed metalens samples, the three most accurate architectures predict transmitted power with median relative errors of 0.7\%--0.9\%. Accuracy depends on refractive index and structural scale, with greater errors for patterns coarser than those used for training. These results demonstrate transfer to independently constructed geometries and to new spatial arrangements of dielectric values not represented in the training. Random-pattern training therefore supports reuse of electromagnetic surrogates across multiple structural families, providing a basis for sharing training costs among material screening, iterative optical design, and fabrication-tolerance analysis.
\end{abstract}

\section{Introduction}
\label{sec:introduction}
Early nanophotonic design relied largely on analytical models, physical intuition, and iterative adjustment of a limited number of geometric parameters~\cite{molesky2018inverse}. For example, photonic crystals were designed using band-structure calculations, whereas many metamaterial designs were guided by effective-medium models and the electromagnetic responses of subwavelength resonators~\cite{joannopoulos2008photonic,smith2004metamaterials}. For periodic structures, Bloch-periodic boundary conditions reduced computational cost by restricting simulations to a representative unit cell~\cite{johnson2001block}. Metasurface design similarly adopted libraries of optical responses calculated for different unit-cell geometries, from which individual elements were selected and arranged to produce a prescribed spatial response~\cite{yu2014flat,khorasaninejad2016metalenses}. This approach commonly relies on the local periodic approximation, which assigns each element the response of a corresponding infinite periodic array. It therefore avoids direct simulation of the entire device, although its accuracy can deteriorate when neighboring elements differ substantially, or their interactions are poorly represented by the periodic model~\cite{pestourie2018inverse,hughes2021perspective}. These discrepancies can introduce stitching errors when locally optimized unit cells are assembled into a larger device~\cite{kim2026suppressing}. In photonic integrated circuits (PICs), waveguide theory, coupled-mode theory, and adiabatic design principles have guided the development of directional couplers, adiabatic transitions, and mode converters, often with further refinement through numerical simulations~\cite{huang1994coupled}. These approaches provide efficient design strategies within their applicable conditions, but extending them to complex, nonperiodic structures requires methods that capture electromagnetic interactions beyond the underlying simplifying assumptions.

\begin{figure}[htbp]
\centering
\includegraphics[width=\textwidth]{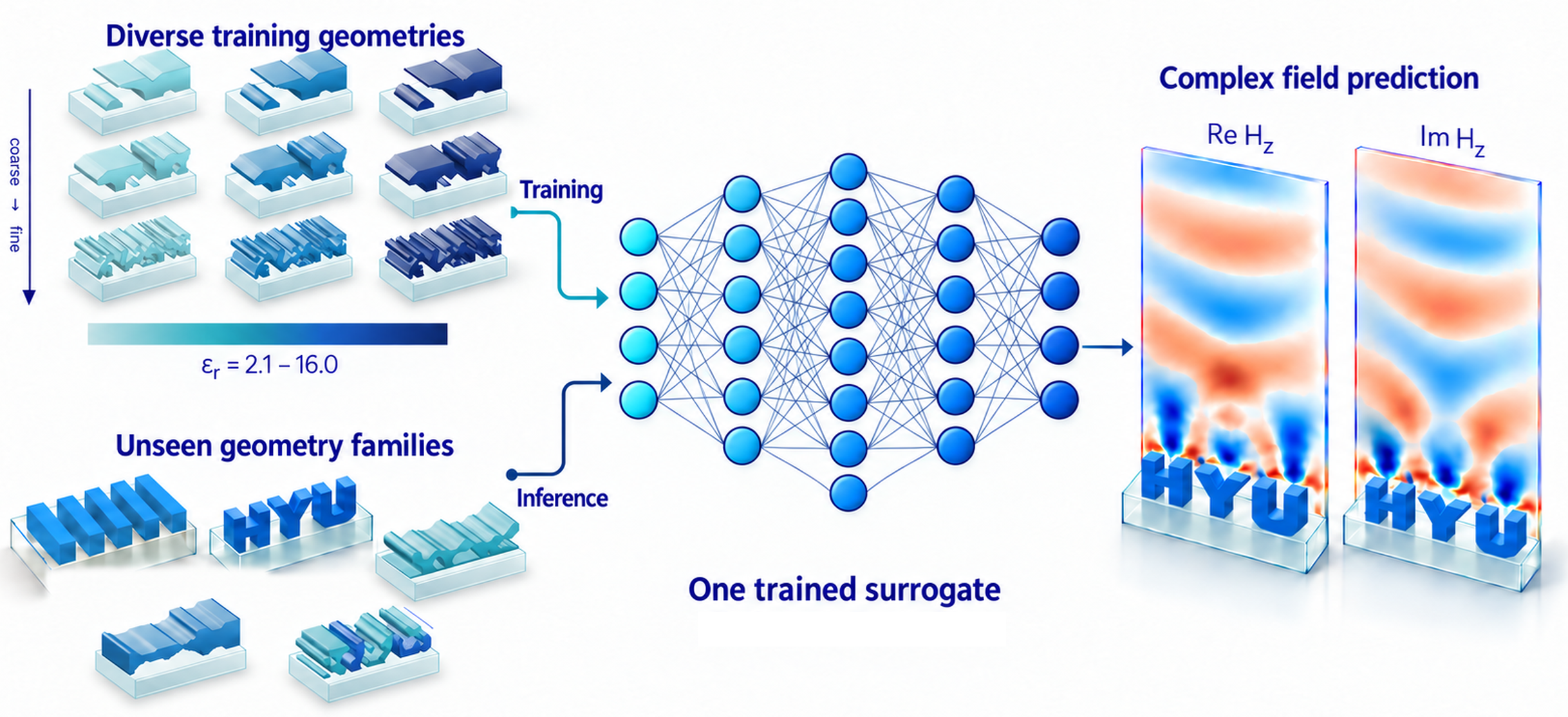}
\caption{\textbf{Electromagnetic field prediction for different materials and nanostructures.} Random freeform permittivity maps spanning $\varepsilon_r=2.1$--16.0 train a mapping to the complex magnetic field. The selected network then predicts $\mathrm{Re}\,H_z$ and $\mathrm{Im}\,H_z$ for aperiodic bar arrays, manually defined patterns, optimized structures, and two-material freeform patterns without retraining. The perspective structures, HYU lettering, network layers, and field textures are conceptual illustrations. Quantitative validation uses lossless, nondispersive two-dimensional transverse-electric (TE) simulations at 500\,nm. An initial draft of this conceptual schematic was produced with OpenAI ChatGPT and was subsequently revised by the authors; it contains no computed field or quantitative result.}
\label{fig:illustration}
\end{figure}

These advances raise the question of whether conventional optical designs approach the best performance attainable under specified material and fabrication constraints~\cite{molesky2018inverse}. In PICs, topology optimization has enabled planar freeform devices with substantially smaller footprints while maintaining high optical performance, demonstrating opportunities beyond conventional layouts~\cite{jensen2011topology,piggott2015inverse}. Freeform metagratings can exploit coupling among spatially overlapping optical modes to achieve efficient large-angle beam deflection~\cite{sell2017large}. Diffusion-based inverse design combines learned distributions of fabrication-compatible binary patterns with adjoint gradients to search for high-performance freeform structures~\cite{seo2026physics}. These approaches expand the accessible design space while accounting for electromagnetic interactions and fabrication constraints. However, full-wave inverse design requires repeated solutions of Maxwell's equations as the material distribution evolves during optimization. Although adjoint methods efficiently evaluate gradients with respect to numerous design variables, repeated forward and adjoint simulations remain computationally demanding, particularly for large devices or optimization over multiple wavelengths and incident angles~\cite{kang2024adjoint,park2026nyquist}. The accumulated cost of these field calculations therefore presents a major obstacle to freeform nanophotonic design. Accelerating electromagnetic field prediction while preserving sufficient accuracy would allow more candidate structures and operating conditions to be evaluated within a practical computational budget.

Deep learning surrogate models address this computational burden by learning mappings from optical structures to electromagnetic responses, allowing new designs to be evaluated without a conventional full-wave simulation for each prediction~\cite{jiang2021deep,so2020deep}. Models that predict spatially resolved, complex fields extend this capability beyond a single performance metric, enabling the calculation of multiple optical quantities from the predicted fields~\cite{wiecha2020deep,chen2022high,kanmaz2023deep}. Neural operators have also been trained to predict dense adjoint-gradient fields from three-dimensional permittivity distributions to accelerate iterative inverse design~\cite{kang2026neuraladjoint}. Recent studies have expanded the materials, structures, and excitation conditions accessible to these solvers. WaveY-Net predicts fields in periodic dielectric metagratings with refractive indices ranging from 2.48 to 3.57~\cite{chen2022high}. Augenstein et al. demonstrated field prediction and inverse design for three-dimensional freeform scatterers with a fixed refractive index of 1.5, representative of polymers used in three-dimensional laser printing~\cite{augenstein2023neural}. Neighborhood-attention models have also supported freeform meta-optical inverse design through pretraining followed by task-specific fine-tuning~\cite{kuang2026rapid}. NeurOLight incorporates device geometry and excitation conditions into neural field prediction for integrated photonics~\cite{gu2022neurolight}. SNAP-DDM combines specialized neural subdomain solvers with iterative domain decomposition to accommodate different geometric layouts and computational domain sizes~\cite{mao2024snapddm}. Multilevel solvers combine neural subdomain models with a global coarse correction to accelerate the iterative solution of Maxwell's equations~\cite{mao2025multilevel}. Our recent wavelength-conditioned solver predicts fields at wavelengths not present in the training data, using reference fields sampled at 20\,nm intervals~\cite{seo2026broadband}.

Despite these advances, the extent to which a trained solver can be reused across different optical design problems requires careful assessment. Accuracy within a prescribed set of materials, geometries, and excitation conditions does not by itself establish reliable transfer to unfamiliar structures or new combinations of these parameters~\cite{jiang2021deep}. Evaluation should therefore distinguish predictions for samples drawn from the training distribution from predictions for structures generated through independent design procedures. This distinction is particularly relevant when a model trained on randomly generated patterns is applied to aperiodic bar arrays or devices produced through inverse design. A broadly reusable electromagnetic surrogate should retain predictive accuracy across diverse material compositions and structural configurations without retraining for each design task. Achieving this capability would allow the cost of data generation and training to be shared across multiple applications, reducing the computational expense of material screening, iterative optimization, and fabrication-tolerance analysis.

Here, we demonstrate prediction of complex electromagnetic fields across various dielectric materials and previously unseen nanostructures without retraining. Figure~\ref{fig:illustration} illustrates how training on random freeform dielectric patterns enables field prediction for new structural designs and material configurations. The training patterns span different feature sizes and relative permittivities of 2.1--16.0. Each spatial permittivity distribution encodes both material properties and their arrangement, and the neural surrogate predicts the real and imaginary components of the magnetic field $H_z$. After training, the network parameters remain fixed when evaluating previously unseen structures. We train four neural-network architectures using a common training protocol and evaluate their predictions on independent test sets comprising aperiodic bar arrays, manually defined patterns, inverse-designed metalenses, inverse-designed binary structures, and two-material freeform patterns. Analyses of complex fields, phase distributions, and transmitted power demonstrate that training on random freeform patterns can enable accurate field prediction for structures generated by different design procedures. We further quantify how the refractive index and structural scale affect prediction error, identifying material and geometric regimes where accuracy deteriorates. These results demonstrate that a trained electromagnetic surrogate can be reused across multiple material choices and structural designs within the tested conditions. Extending this capability to additional wavelengths, illumination conditions, and three-dimensional structures could enable the same trained model to support material screening, geometric optimization, and fabrication-tolerance analysis, reducing the need for separate models for individual design tasks.

\section{Methods}
\label{sec:methods}

Early neural surrogates in nanophotonics learned mappings from geometric parameters to scattering spectra~\cite{peurifoy2018nanophotonic}. Subsequent studies addressed spatial electromagnetic field prediction using physics-augmented networks~\cite{chen2022high} and extended neural surrogate solvers through domain decomposition~\cite{mao2024snapddm}. Physics-driven training also uses Maxwell-equation residuals to constrain predicted fields~\cite{lim2022maxwellnet}. Building on these developments, we examine whether training on random dielectric patterns enables field prediction for conventional, manually specified, and inverse-designed nanostructures without retraining. The methodological emphasis is on constructing the training set and evaluating generalization across distinct structural families.

Training-set construction has also been addressed through error-guided active learning~\cite{pestourie2020active} and the dynamic generation of difficult material and source configurations~\cite{dove2026inductively}. In the present study, we generate random freeform permittivity distributions independently of predefined device geometries or optical design objectives. This approach samples diverse dielectric arrangements without requiring examples of the target device families during training. Four established neural network architectures are trained on the same dataset and follow a common training protocol to assess how consistently this training strategy supports generalization. Aperiodic bar arrays, manually defined patterns, and inverse-designed structures test transfer across geometry families, while two-material freeform patterns test transfer to new material arrangements. Prediction accuracy on these test sets, with all trained parameters held fixed, is the primary measure of generalization. Held-out material--geometry combinations provide a reference for performance within the training distribution. The wavelength, polarization, and illumination are held constant across these comparisons to evaluate structural and material transfer under consistent optical conditions.

\begin{figure}[htbp]
\centering
\includegraphics[width=\textwidth]{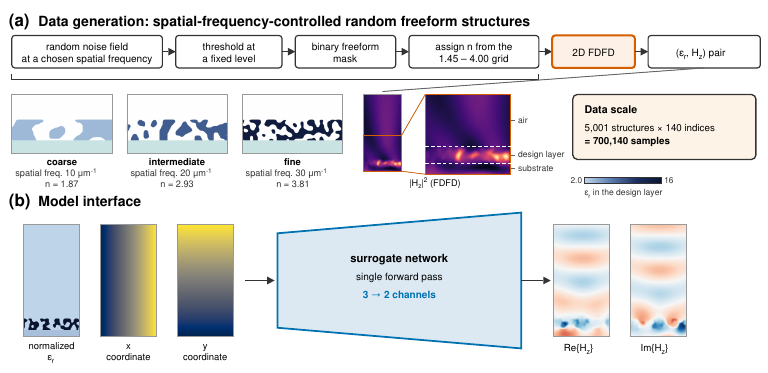}
\caption{\textbf{Training-data construction and shared network interface.} (a) A random noise field at a chosen spatial frequency is thresholded to obtain a binary freeform mask. Each of the 5,001 geometries is assigned all 140 sampled permittivities and simulated using FDFD, yielding 700,140 field samples. The three examples differ in both structural scale and refractive index. Simulations use the 10\,nm grid; for display, structure outlines are smoothed, and field maps are interpolated from that grid. The reference intensity is annotated with the substrate, design-layer, and air regions. (b) A normalized permittivity map and two coordinate channels form the network input. Real and imaginary $H_z$ components form the output on the $64\times128$ retained grid; the output panels show the U-Net prediction for one held-out combination.}
\label{fig:data_interface}
\end{figure}

\subsection{Surrogate formulation and field representation}
\label{sec:formulation}

Neural field solvers have used spatial material maps to represent freeform structures, avoiding a fixed list of geometric parameters~\cite{chen2022high,augenstein2023neural,zhang2023fast}.
This representation can describe both changes in shape and in refractive index.
Its value for model reuse depends on whether a single learned mapping can accommodate both changes over a sufficiently broad training set.
We therefore encode the full relative-permittivity distribution and use a common input--output representation for all evaluated structures.

Figure~\ref{fig:data_interface} shows the prediction process.
Under fixed illumination and simulation conditions, the network approximates
\begin{equation}
\widehat{H}_z(x,y)
=
\mathcal{G}_{\theta}[\varepsilon_r](x,y),
\qquad
\widehat{H}_z(x,y)\in\mathbb{C},
\label{eq:map}
\end{equation}
where $\widehat{H}_z(x,y)$ is the predicted complex amplitude of the out-of-plane magnetic field at the in-plane coordinates $(x,y)$.
The mapping $\mathcal{G}_{\theta}$ represents the neural network with trained parameters $\theta$, and $\varepsilon_r$ denotes the spatial relative-permittivity distribution.
The symbol $\mathbb{C}$ denotes the set of complex numbers.
A normalized permittivity map and two coordinate channels form the network input, while the real and imaginary components of $H_z$ form the output.

\subsection{Electromagnetic simulations for data generation}
\label{sec:reference}

Training and evaluation fields were computed using the two-dimensional finite-difference frequency-domain (FDFD) solver \texttt{fdfd\_hz} in Ceviche 0.1.3~\cite{hughes2019forward}.
FDFD discretizes the frequency-domain Maxwell equations on a staggered grid and expresses them as a sparse linear system for the complex field amplitudes~\cite{rumpf2012fdfd}.
The same solver provided reference fields for all datasets, including structures originally optimized using time-domain or frequency-domain methods.

Each simulation comprised a dielectric design layer on a substrate and an air region on the transmission side.
We used two-dimensional transverse-electric (TE) polarization with $\mathbf{E}=(E_x,E_y,0)$ and $\mathbf{H}=(0,0,H_z)$, where $H_z$ is normal to the simulation plane.
The time dependence was $e^{+j\omega t}$, where $j^2=-1$, $\omega$ is the angular frequency, and $t$ is time.
Unless otherwise stated, the vacuum wavelength was $\lambda_0=500$\,nm, and all physical media were lossless and nondispersive.
These conditions permit comparisons across geometries and assigned refractive indices at a fixed wavelength.

The computational domain contained $124\times213$ cells in the transverse and propagation directions, respectively, with a uniform spacing of $\Delta l=10$\,nm.
Thirty-cell stretched-coordinate perfectly matched layers (PMLs) terminated all four sides to approximate open boundaries~\cite{shin2012pml}.
A unit-amplitude magnetic-current line, uniform in the transverse direction and located in the substrate, excited normally incident illumination.

A centered $64\times128$-cell region, corresponding to $640\times1280$\,nm, was extracted to form paired permittivity and complex-field arrays.
This region retained the structure and transmitted-field region while excluding the source and PMLs.
For the random-pattern training geometries, it contained 8 substrate rows with $\varepsilon_r=2.1$, 13 design-layer rows, and 107 air rows.
For evaluation structures with different thicknesses, the optical readout line was positioned relative to the upper boundary of each structure.

Supporting Note~7 analyzes homogeneous-medium numerical dispersion for the second-order finite-difference stencil.
At $\lambda_0=500$\,nm and $\Delta l=10$\,nm, the axial relative numerical-wavenumber error increases from 0.1387\% at $\varepsilon_r=2.1$ to 1.0839\% at $\varepsilon_r=16$.
These estimates exclude material-interface, PML, finite-domain, source, and linear-solver errors and therefore do not establish the total error relative to the continuum solution.
The reported surrogate errors measure discrepancies from the discrete FDFD reference fields.

\subsection{Random-pattern generation and dataset construction}
\label{sec:random_data}

Random freeform patterns were used to sample dielectric arrangements with varied feature widths, gaps, and connectivity.
These geometric variations affect scattering, near-field coupling, and interference, which contribute to the collective optical response~\cite{albella2013low,sell2017large}.
The aim was to expose the networks to diverse interactions between subwavelength features without restricting the training geometries to a particular device family.

Spatially correlated patterns were generated using gradient noise, which constructs a continuous scalar field from pseudorandom gradients associated with lattice vertices~\cite{perlin2002improving}. We evaluate local gradient contributions on a skewed lattice and smoothly combine them using decaying distance-dependent weights. Each contribution depends on the dot product of the gradient vector and the displacement from the lattice vertex to the evaluation point. The resulting field varies continuously in space, and scaling its input coordinates changes the characteristic length over which the noise varies. Thresholding this field produces binary material regions and gaps with adjustable structural scales.

Figure~\ref{fig:data_interface}(a) illustrates this procedure.
The noise field was sampled on the $64\times13$-cell design layer and thresholded at zero.
For each geometry, the dimensionless coordinate-scaling parameter $s$ was drawn uniformly from $[0.1,0.3]$.
Larger values of $s$ produce more rapid spatial variation and favor finer patterns.
At the prescribed grid spacing, this interval corresponds to a nominal generator spatial-frequency range of 10--30\,$\mu\mathrm{m}^{-1}$.
Varying $s$ across samples therefore introduces a range of structural scales without specifying individual feature shapes.

Each of the 5,001 masks was paired with 140 relative permittivities, $\varepsilon_r=2.1,2.2,\ldots,16.0$, corresponding to refractive indices of approximately 1.45--4.00.
The material pixels were assigned the selected permittivity, while the remaining design-layer pixels were assigned the air permittivity.
This pairing varies the dielectric contrast and the wavelength within the material independently of geometry.
Each complete structure was simulated using the FDFD settings in Section~\ref{sec:reference}, yielding 700,140 geometry--permittivity samples.
The resulting complex fields include multiple scattering and near-field interactions for each configuration under the prescribed illumination.
Independent test sets assess whether training on these responses supports prediction for structures generated by other design procedures.

Structural scale was characterized using horizontal material chord lengths~\cite{torquato1993chord}.
Each chord was defined as a contiguous run of material pixels along a horizontal grid row.
For each mask, the feature size was defined as the median chord length in grid cells multiplied by $\Delta l$.
This quantity was measured from the discretized binary mask used in the FDFD calculation and was not inferred from the inverse generator frequency.

\subsection{Data partition and generalization tests}
\label{sec:datasets}

The random-pattern data pool was partitioned at the geometry--permittivity-pair level.
At each sampled permittivity, samples were randomly assigned to training, validation, and held-out sets in approximate proportions of 80\%, 10\%, and 10\%, respectively, using a permittivity-specific seed derived from the base seed 42.
The resulting sets contained 560,113 training, 70,015 validation, and 70,012 held-out samples.
Identical geometry--permittivity pairs were excluded from crossing the splits, although the same geometry could occur in different splits at different permittivities.
All sampled permittivity values were represented in the training set.
The held-out set therefore evaluates unseen geometry--permittivity combinations within the sampled distribution.

Generalization was assessed separately using four test sets with geometries constructed independently of the training-pattern generator.
A fifth set used the same generator with two design materials to evaluate transfer to material arrangements absent from training.
These test sets were excluded from network training and checkpoint selection, and all predictions used fixed trained parameters.
The test geometries and sampling conditions are introduced alongside the corresponding results.

\subsection{Network training and model selection}
\label{sec:training}

Network architecture specifies how spatial interactions are represented, while the training data provide examples of how geometry and permittivity determine the electromagnetic response. Cross-axis factorized operators, for example, have been developed to connect local device structures with long-range optical-field patterns~\cite{zhu2024pace}. Architectural improvements alone do not establish accurate prediction beyond the training distribution. We therefore focus on whether training on diverse random patterns across a broad permittivity range enables field prediction for unseen geometries and material arrangements without retraining. Established architectures with different spatial representations are used to assess the extent to which this capability is achieved across different network designs.

We compare a Fourier neural operator (FNO), a factorized Fourier neural operator (F-FNO), U-Net, and a dilated residual network (DilResNet)~\cite{li2021fourier,tran2023factorized,ronneberger2015unet,stachenfeld2022learned}. These models represent spatial interactions through Fourier-domain operations, hierarchical convolutions, or dilated convolutions. Their parameter counts range from 5.33 to 5.98 million. All models use the same input--output representation, data partitions, loss function, and optimization procedure. Table~S2 provides the architectural configurations, parameter counts, and training times.

Training uses AdamW with an initial learning rate of $10^{-3}$ and weight decay of $10^{-4}$~\cite{loshchilov2019decoupled}. After 500 warmup steps, the learning rate follows a cosine decay schedule~\cite{loshchilov2017sgdr}. The batch size is 64, the gradient-clipping threshold is 1.0, and mixed precision is used~\cite{micikevicius2018mixed}. Each configuration is trained with a single random seed on one NVIDIA RTX~6000~Ada GPU for up to 200 epochs, with an early-stopping patience of 20 epochs.

The supervised loss is the relative $L_2$ error between the predicted and reference complex fields, evaluated on the normalized real and imaginary output channels (Supporting Note~4). For each sample, the relative error and normalized mean squared error (NMSE) are defined as
\begin{equation}
e =
\frac{
\left(\sum_{(x,y)\in\Omega}
|\widehat{H}_z(x,y)-H_z(x,y)|^2\right)^{1/2}
}{
\left(\sum_{(x,y)\in\Omega}
|H_z(x,y)|^2\right)^{1/2}
},
\qquad
\mathrm{NMSE}=e^2,
\label{eq:error}
\end{equation}
where $H_z$ and $\widehat{H}_z$ denote the reference and predicted complex magnetic fields, respectively, and $\Omega$ contains all grid points in the retained domain. The relative error is computed using the concatenated real and imaginary field components.

For each architecture, the checkpoint with the lowest validation loss is retained and used without further parameter updates for all subsequent evaluations. Fields are de-normalized before calculating the evaluation metrics. The main text reports the relative complex-field error as $100e$ in percent. Figure~S12 presents the learning histories and selected checkpoints. This comparison assesses the selected configurations under a common training procedure and similar parameter budgets. It does not establish the best achievable performance of each architecture or quantify variability across random seeds.

\section{Results}
\label{sec:results}
We evaluate whether a model trained on random freeform patterns can predict fields for other nanophotonic structures without retraining. The results first address independent structure types, then examine refractive-index and feature-size dependence, transmitted optical quantities, predictions beyond the training-scale interval, and computational cost.

\subsection{Field prediction for independent nanophotonic test structures}

\begin{figure}[p]
\centering
\includegraphics[width=\textwidth,height=.755\textheight,keepaspectratio]{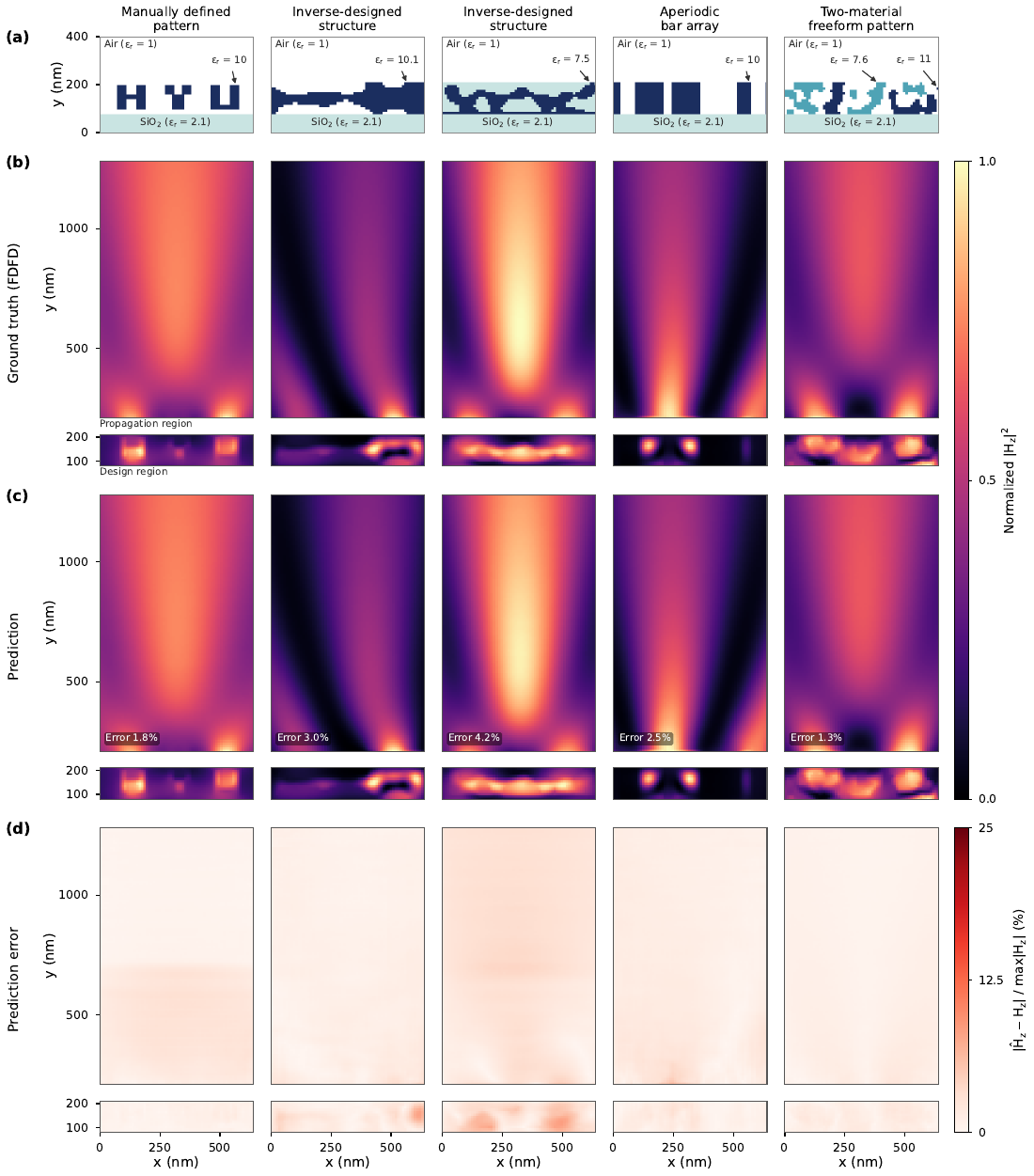}
\caption{\textbf{Field prediction for independent nanophotonic test structures.} (a) Permittivity maps of a manually defined pattern, two inverse-designed structures (a binary structure and a metalens), an aperiodic bar array, and a two-material freeform pattern, one from each test set in Table~\ref{tab:families}. (b) FDFD reference intensities. (c) Intensities predicted by one unchanged DilResNet. Annotations give the full-domain relative complex-field error $e$ for each illustrative example. (d) Magnitude of the complex-field error, $|\widehat H_z-H_z|$, of the same predictions, normalized by the maximum reference $|H_z|$ in each displayed region and shown on a 0--25\% scale (Figure~S4 gives the corresponding maps for all four architectures). Because these maps are computed from the complex field, they include phase differences. Design and propagation regions use separate intensity scales, matched between reference and prediction.}
\label{fig:transfer}
\end{figure}

The aim of random-pattern training is to obtain a field predictor that can be reused across nanophotonic design tasks without preparing a separate training dataset for each device family. We therefore examine whether training on varied dielectric geometries and permittivities supports field prediction for structures generated by human design or numerical optimization. Assessing this capability requires evaluation beyond withheld geometry--permittivity combinations drawn from the training population. The following tests evaluate transfer to independently constructed geometries and material arrangements absent from training, with the network parameters and optical conditions held fixed.

A DilResNet trained on random dielectric patterns predicts complex magnetic fields for independently constructed nanophotonic structures without retraining. Figure~\ref{fig:transfer}(a) shows a manually defined pattern, two inverse-designed structures (a binary structure and a metalens), an aperiodic bar array, and a two-material freeform pattern. The first four geometries were constructed independently of the training-pattern generator. The fifth retains the generator but introduces two dielectric materials within the design layer. All illustrated predictions use one validation-selected DilResNet checkpoint under two-dimensional TE illumination at 500\,nm with lossless, nondispersive media.

The FDFD reference distributions of $|H_z|^2$ in Figure~\ref{fig:transfer}(b) show localized field enhancement within the design layer and distinct intensity patterns in the transmitted region. The predictions in Figure~\ref{fig:transfer}(c) reproduce these features, including the asymmetric distribution of the inverse-designed binary structure, the concentrated intensity above the metalens, and the broadening intensity feature above the bar array. The design and propagation regions use separate display normalizations, with matching reference and prediction scales within each region. From left to right, the relative complex-field errors are 1.8\%, 3.0\%, 4.2\%, 2.5\%, and 1.3\%. These errors are evaluated over the full retained domain using Equation~\eqref{eq:error} and account for both amplitude and phase discrepancies. The complex-field error maps in Figure~\ref{fig:transfer}(d) locate residual differences within the design layer and the propagation region, with more spatially extended residuals visible for the metalens example.

For the five structures illustrated in Figure~\ref{fig:transfer}, the DilResNet predictions are substantially more accurate than those obtained by directly reusing the fields of the nearest training samples selected by the permittivity-map $L_1$ distance. The nearest sample for each structure is identified by an exhaustive search over the 560,113 training samples, based on the mean pixel-wise distance between design-layer permittivity maps. Directly reusing the selected FDFD fields gives relative complex-field errors of 22\%--77\%, compared with 1.3\%--4.2\% for DilResNet. Supporting Note~10 and Figure~S16 describe the selection procedure and comparisons for individual examples.

Table~\ref{tab:families} extends the comparison to 35,441 field samples across the five evaluation sets. Among the tested configurations, DilResNet gives the lowest mean error in four sets, with set-level means ranging from 1.39\% to 7.57\%. U-Net performs best for inverse-designed metalenses, with a mean error of 6.54\%, compared with 7.49\% for F-FNO and 7.57\% for DilResNet. The equally weighted mean across the five sets is lowest for DilResNet at 3.89\%, followed by F-FNO at 4.65\%, U-Net at 5.18\%, and FNO at 13.3\%. This aggregation gives each set equal weight despite differences in sample count.

Prediction accuracy also depends on the structural family. For F-FNO, U-Net, and DilResNet, mean errors are higher for both inverse-designed sets than for aperiodic bar arrays, manually defined patterns, and two-material freeform patterns. The preferred architecture nevertheless differs between the inverse-designed sets. U-Net performs best for metalenses but gives a mean error of 10.1\% for the inverse-designed binary structures, compared with 5.61\% for DilResNet. These differences occur under the same sampled permittivities and optical conditions, indicating sensitivity to the geometric distribution.

The two-material set yields the lowest mean error among the independent sets for every architecture. In the illustrated example, regions with $\varepsilon_r=7.6$ and 11 are predicted with a relative complex-field error of 1.3\%. Both dielectric values occur in training, but their coexistence within the design layer does not. This evaluation therefore demonstrates transfer to new material arrangements. Because the patterns retain the training generator, their lower errors cannot be attributed to material mixing alone.

The architecture ranking changes for the 70,012 held-out geometry--permittivity combinations. U-Net gives the lowest mean error of 0.14\%, followed by DilResNet at 0.19\%, F-FNO at 0.23\%, and FNO at 0.92\%. Although these combinations are excluded from training, their geometries may appear in training at other permittivities. Their lower errors consequently describe prediction within the sampled geometry distribution. U-Net's advantage on held-out combinations does not extend to the average across independent sets, where DilResNet performs best. Reliable assessment of model reuse therefore requires tests on independently constructed structures in addition to held-out combinations.

\begin{table}[htbp]
\centering
\caption{\textbf{Mean relative complex-field errors for independent test sets and held-out combinations.}
Entries report the arithmetic mean of the per-sample relative complex-field error $e$, expressed as a percentage.
$N$ denotes the number of field samples, including repeated geometries evaluated at different permittivities.
The mean across the five test sets assigns equal weight to each set, and its reported $N$ is the total sample count.
Bold values indicate the lowest error in each row.
Each architecture uses one validation-selected checkpoint from a single training run.}
\label{tab:families}
\small
\setlength{\tabcolsep}{3pt}
\begin{tabular*}{\textwidth}{@{\extracolsep{\fill}}lrrrrr@{}}
\toprule
Evaluation set & $N$ & FNO & F-FNO & U-Net & DilResNet \\
\midrule
Aperiodic bar arrays
& 13,900 & $16.0$ & $4.01$ & $4.46$ & $\mathbf{2.91}$ \\

Manually defined patterns
& 2,641 & $11.1$ & $2.71$ & $2.76$ & $\mathbf{1.94}$ \\

Inverse-designed metalenses
& 6,950 & $13.2$ & $7.49$ & $\mathbf{6.54}$ & $7.57$ \\

Inverse-designed binary structures
& 6,950 & $17.0$ & $7.02$ & $10.1$ & $\mathbf{5.61}$ \\

Two-material freeform patterns
& 5,000 & $9.34$ & $2.02$ & $1.98$ & $\mathbf{1.39}$ \\
\midrule
Mean across five test sets
& 35,441 & $13.3$ & $4.65$ & $5.18$ & $\mathbf{3.89}$ \\
\midrule
Held-out combinations
& 70,012 & $0.92$ & $0.23$ & $\mathbf{0.14}$ & $0.19$ \\
\bottomrule
\end{tabular*}
\end{table}

\subsection{Field prediction across refractive indices and structural scales}

\begin{figure}[htbp]
\centering
\includegraphics[width=0.90\textwidth]{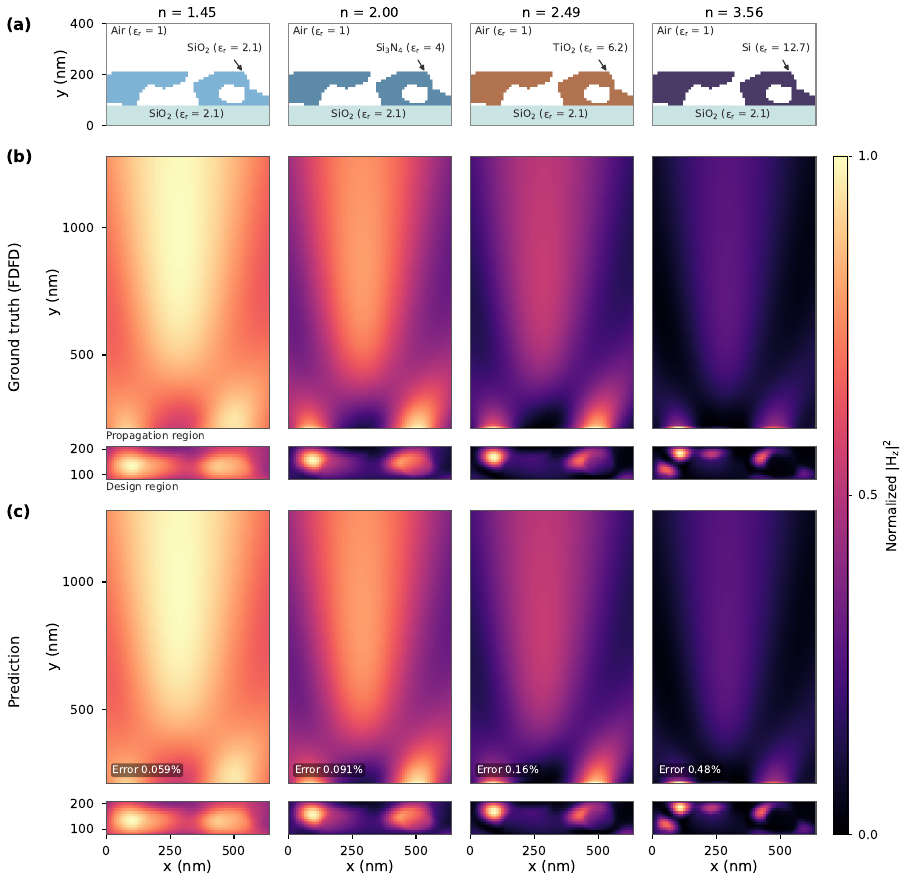}
\caption{\textbf{Prediction for materials with different refractive indices.} (a) A fixed geometry with design-material indices of $n=1.45$, 2.00, 2.49, and 3.56. (b) FDFD reference distributions of $|H_z|^2$. (c) Predictions from one unchanged DilResNet checkpoint. The material labels refer to representative real refractive indices at the reference wavelengths specified in the text. The examples are held-out geometry--permittivity combinations, and all assigned indices are evaluated at 500\,nm under 2D TE illumination using lossless, nondispersive media. Intensity scales are selected separately for the design and propagation regions, with matching reference and prediction scales. Annotations give the relative error of the full complex field.}
\label{fig:refractive_index}
\end{figure}

\begin{figure}[htbp]
\centering
\includegraphics[width=0.90\textwidth]{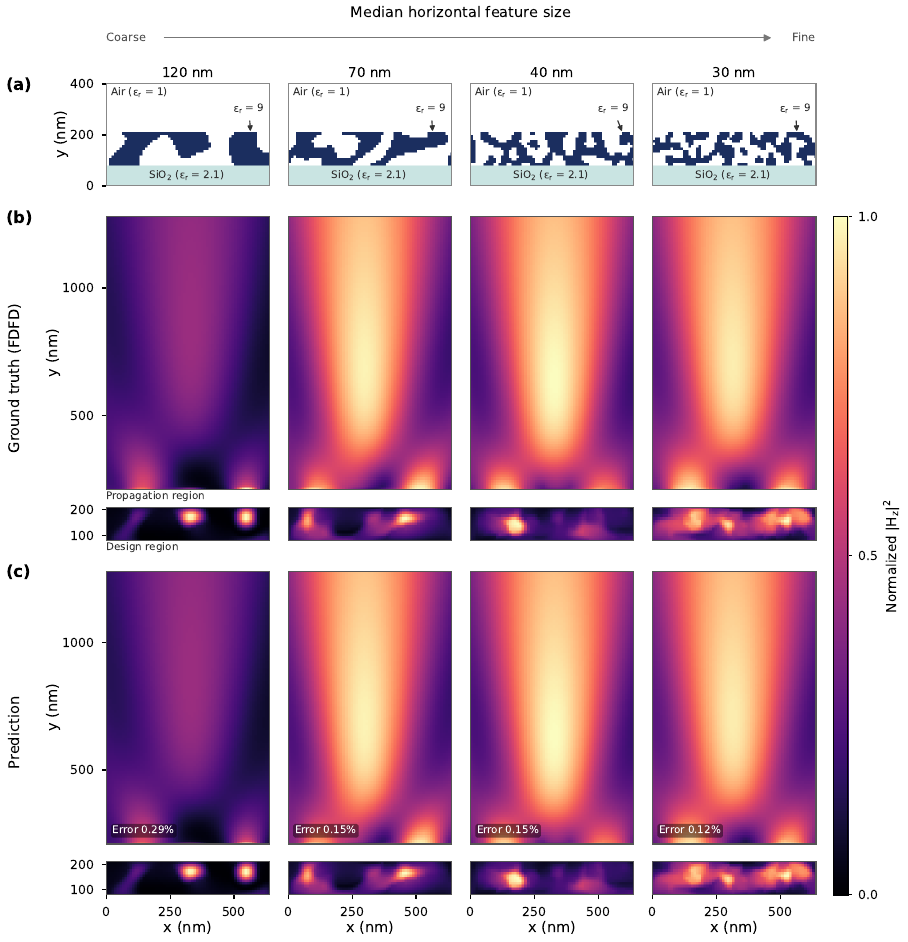}
\caption{\textbf{Prediction across different structural feature sizes.} (a) Four geometries with median horizontal feature sizes of 120, 70, 40, and 30\,nm and a fixed design-material index of $n=3.00$. (b) FDFD reference intensity distributions. (c) DilResNet predictions. These held-out combinations use the simulation and display conventions of Figure~\ref{fig:refractive_index}. Annotations give the full-domain complex-field relative error.}
\label{fig:structural_scale}
\end{figure}

Reusing an electromagnetic surrogate across nanophotonic designs requires accurate prediction of field variations caused by changes in both material properties and geometry. Dielectric metasurfaces employ materials with different refractive indices and structures with different feature sizes to control the spatial distribution of transmitted light~\cite{zhan2016low,devlin2016broadband,yu2014flat}. A predictor intended for reuse across these choices must accommodate their effects through the input permittivity distribution. We examine this capability through two complementary comparisons. The first varies the refractive index of a fixed geometry, while the second compares geometries with different feature sizes at a fixed refractive index. All predictions use one validation-selected DilResNet checkpoint without retraining.

The material comparison evaluates how the predicted field responds to changes in dielectric contrast while the geometry remains unchanged. The four configurations in Figure~\ref{fig:refractive_index}(a) have design-material permittivities of $\varepsilon_r=2.1$, 4.0, 6.2, and 12.7, corresponding to refractive indices of approximately $n=1.45$, 2.00, 2.49, and 3.56. The substrate permittivity remains fixed at $\varepsilon_r=2.1$, and the surrounding medium is air. This arrangement allows the field changes to be attributed to the assigned design-material permittivity under otherwise identical simulation conditions.

The material labels connect these nominal permittivities to familiar optical materials at specified reference wavelengths. The value $n\approx1.45$ is representative of fused silica near a wavelength of $1.0\,\mu\mathrm{m}$~\cite{malitson1965interspecimen}. Silicon nitride provides a representative value of $n\approx2.00$ at 633\,nm~\cite{zhan2016low}. For amorphous titanium dioxide, $n\approx2.49$ corresponds approximately to a wavelength near 470\,nm, as estimated from the reported dispersion curve~\cite{devlin2016broadband}. The value $n\approx3.56$ corresponds to crystalline silicon at approximately $1.03\,\mu\mathrm{m}$~\cite{green1995optical}. These associations provide physical reference points for the assigned real permittivities. In the present comparison, all four values are imposed at the common simulation wavelength of 500\,nm, with absorption and material dispersion excluded. The material labels therefore identify representative dielectric values at their stated reference wavelengths rather than complete material models evaluated at 500\,nm.

The reference fields exhibit substantial spatial changes as the assigned refractive index increases. The FDFD distributions of $|H_z|^2$ in Figure~\ref{fig:refractive_index}(b) show a broad, elongated intensity distribution in the propagation region at $n=1.45$, together with two extended intensity maxima within the design layer. Across the higher-index examples, the transmitted intensity distribution narrows, while the maxima within the patterned layer become more localized. Additional spatial features appear within the design layer at $n=3.56$. The fixed geometry therefore produces distinct field distributions as its dielectric contrast changes, providing a direct test of the material dependence captured by the learned mapping.

The predicted fields reproduce these material-dependent variations throughout the design and propagation regions. Agreement with the reference is evident in Figure~\ref{fig:refractive_index}(c), including the transverse extent of the transmitted intensity distribution and the positions of localized maxima within the structure. The relative complex-field errors are 0.059\%, 0.091\%, 0.16\%, and 0.48\% for $n=1.45$, 2.00, 2.49, and 3.56, respectively. Although the error increases across these four examples, it remains below 0.5\% in every case. These results show that changing the material value in the input is sufficient for the same trained predictor to reproduce the corresponding field variations under the tested conditions. The same trend holds at the dataset level. At these four refractive indices, the median DilResNet error increases from 0.05\% to 0.23\% for the held-out combinations, from 0.09\% to 3.99\% for aperiodic bar arrays, and from 0.16\% to 12.2\% for inverse-designed metalenses (Table~S1), indicating that high refractive indices are the more demanding material regime, particularly for independently constructed structures.

Changes in structural detail provide a complementary assessment at a fixed dielectric contrast. The four random patterns in Figure~\ref{fig:structural_scale}(a) have median horizontal feature sizes of 120, 70, 40, and 30\,nm, with a common design-material refractive index of $n=3.00$. Their geometries range from relatively large dielectric regions to finer arrangements containing smaller material features and air gaps. All four characteristic feature sizes are below the vacuum wavelength of 500\,nm. The substrate, surrounding medium, illumination, and computational grid remain unchanged, so this comparison examines field predictions across geometric variations while keeping the assigned material properties constant.

The geometric differences modify both the localized field distribution and the transmitted spatial profile. Within the design layer, the FDFD reference fields in Figure~\ref{fig:structural_scale}(b) show two compact intensity maxima near the upper boundary of the 120\,nm structure, whereas the 30\,nm structure exhibits several maxima distributed laterally. The intermediate examples have different locations and spatial extents of localized maxima. In the propagation region, the 70, 40, and 30\,nm examples produce elongated intensity distributions, with differences most pronounced near the patterned layer. Their broadly similar profiles farther from the structure coexist with distinct field distributions inside and immediately above the design layer.

The surrogate captures these geometric variations with low complex-field errors. The predictions in Figure~\ref{fig:structural_scale}(c) reproduce the localized maxima within the dielectric patterns and the spatial profiles in the propagation region. The relative errors are 0.29\%, 0.15\%, 0.15\%, and 0.12\% for the 120, 70, 40, and 30\,nm examples, respectively. Thus, accurate prediction is maintained across the illustrated range of structural detail using the same trained parameters. The lower errors in the finer examples do not establish a general dependence on feature size, as the masks also differ in shape, connectivity, and material arrangement.

Both comparisons quantify agreement using the complex fields over the full retained domain, as defined in Equation~\eqref{eq:error}. The reported errors therefore include amplitude and phase discrepancies that cannot be assessed from intensity maps alone. For visualization, the design and propagation regions use separate intensity normalizations, with matching scales for each reference and prediction. These display choices reveal localized features without changing the quantitative error assessment.

Together, these results demonstrate that a single trained field predictor can accommodate changes in the assigned refractive index and subwavelength structural details within the sampled random-pattern population. The illustrated samples are held-out geometry--permittivity combinations. Their geometries may occur in training at other permittivities, and all assigned dielectric values are represented in the training data. These comparisons therefore establish accuracy for withheld combinations and complement the preceding evaluations on independently constructed structures. They also distinguish two requirements for model reuse: accurate prediction across material assignments and reliable transfer across geometric distributions. The following analysis examines the optical significance of the predicted fields through comparisons of transmitted power and phase profiles.

\subsection{Prediction of transmitted amplitude, phase, and power}

\begin{figure}[htbp]
\centering
\includegraphics[width=\textwidth]{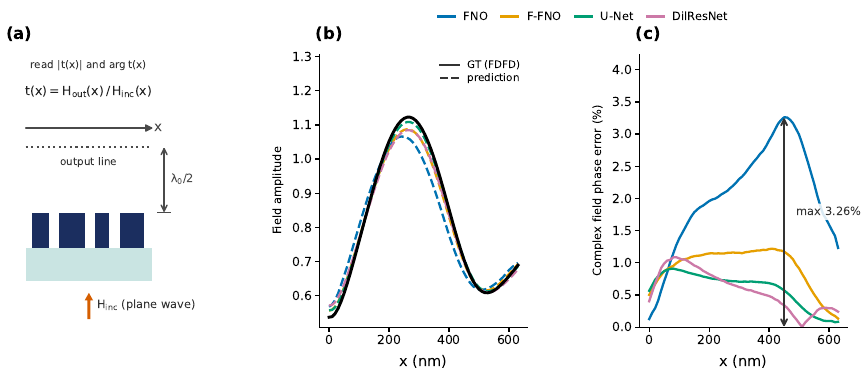}
\caption{\textbf{Transmitted amplitude and phase predicted without retraining.} (a) Optical readout at 0.5$\lambda_0$ above the structure. (b) Normalized magnetic-field amplitude $|t(x)|$ for the selected metalens sample. The FDFD reference is black, and model predictions are colored dashed curves. (c) Absolute wrapped phase difference as a percentage of one phase cycle (Equation~\eqref{eq:phase_error_cycle}). The maximum FNO error is 3.26\%, equivalent to 11.7$^\circ$.
}
\label{fig:readout}
\end{figure}

Accurate prediction of the transmitted amplitude and phase is necessary to evaluate how a nanophotonic structure redistributes an incident optical field. The amplitude describes the spatial variation in field magnitude, while the phase determines the interference that governs subsequent propagation. A full-domain complex-field error summarizes their combined discrepancies but does not show where they occur along the transmitted wavefront. We therefore examine the spatially resolved transmission profiles obtained directly from the predicted complex fields.

The transmitted field is evaluated at a common distance from the structure to compare the optical responses predicted by the four architectures. The readout arrangement in Figure~\ref{fig:readout}(a) places the output line at $\lambda_0/2$ above the upper boundary of the design layer, corresponding to 250\,nm at the simulation wavelength of 500\,nm. The complex transmission profile is $t(x)=H_{\mathrm{out}}(x)/H_{\mathrm{inc}}(x)$, where $x$ is the transverse position, $H_{\mathrm{out}}$ is the complex magnetic field on the output line, and $H_{\mathrm{inc}}$ is the incident-field reference evaluated at the same line. Its magnitude $|t(x)|$ and argument $\arg t(x)$ give the normalized magnetic-field amplitude and transmission phase, respectively. The reference and predicted fields use the same incident-field phase reference.

The four architectures reproduce the main features of the transmitted amplitude for the selected inverse-designed metalens. This example was selected using the median of the per-sample complex-field error averaged across the four architectures. The black reference curve in Figure~\ref{fig:readout}(b) exhibits a broad amplitude maximum near $x=270$\,nm, followed by a minimum near $x=530$\,nm and a subsequent increase toward the right edge. The predicted profiles capture this sequence of spatial features. U-Net closely follows the reference around the amplitude maximum, while F-FNO and DilResNet slightly underestimate its height. FNO shows a larger reduction in peak amplitude and a displacement of the maximum toward smaller $x$. The curves approach one another near the amplitude minimum, showing that the magnitude of the discrepancy varies along the output line.

The phase comparison reveals additional differences that are less apparent from the amplitude profiles. The error plotted in Figure~\ref{fig:readout}(c) is the absolute wrapped phase difference expressed as a percentage of one complete phase cycle,
\begin{equation}
\delta_{\phi}(x)
=
\frac{100}{2\pi}
\left|
\operatorname{Arg}
\left[
\widehat{t}(x)t^{*}(x)
\right]
\right|,
\label{eq:phase_error_cycle}
\end{equation}
where $\delta_{\phi}$ is the phase error in percent, $\widehat{t}$ and $t$ are the predicted and reference complex transmission profiles, respectively, the asterisk denotes complex conjugation, and $\operatorname{Arg}$ returns the principal phase in $(-\pi,\pi]$. This definition accounts for phase periodicity and normalizes the angular difference by $2\pi$.

FNO exhibits the largest phase discrepancy in this example, reaching 3.26\% of a cycle near $x=450$\,nm, equivalent to approximately $11.7^\circ$. This maximum occurs away from the principal amplitude maximum, illustrating why amplitude agreement alone does not fully characterize the predicted transmitted field. F-FNO, U-Net, and DilResNet remain below 1.3\% of a cycle, corresponding to phase differences below approximately $4.7^\circ$. Their spatial error distributions nevertheless differ. F-FNO maintains a relatively broad error plateau across the middle portion of the output line, whereas U-Net and DilResNet generally show smaller errors toward its right side. The curves cross at several positions, indicating that the architecture with the smallest local phase error depends on the transverse position.

These amplitude and phase comparisons demonstrate that training on random dielectric patterns supports recovery of the transmitted complex field for an inverse-designed structure without retraining. The predictions preserve the broad amplitude profile while resolving the spatial phase with substantially smaller discrepancies for F-FNO, U-Net, and DilResNet than for FNO in the illustrated case. Because the curves describe one selected sample, their local differences should be interpreted alongside the dataset-level evaluations. Integrated transmitted power provides a complementary assessment of optical response and is calculated separately from the real Poynting flux using the predicted electric and magnetic fields. Across the 6,950 metalens samples, F-FNO, U-Net, and DilResNet give median relative transmitted-power errors of 0.7\%--0.9\% and Pearson correlations of 0.994--0.996 with the FDFD reference power, compared with 3.7\% and 0.968 for FNO (Supporting Note~8 and Table~S5).

The phase comparisons rely on a consistent incident-field reference across datasets. Fits to the fields in the uniform substrate region give agreement within $0.2^\circ$ between the dataset-mean incident phases across the 14 evaluated datasets, with circular standard deviations no greater than $0.65^\circ$, as detailed in Supporting Note~8. These checks support comparison of the predicted and reference phases under a common illumination convention. The unwrapped transmission-phase profiles in Figure~S13 extend this comparison to selected samples from seven evaluation sets, revealing the broad phase curvature and sharper spatial transitions observed across different structures.

Separating a uniform phase offset from spatially varying discrepancies provides a more specific assessment of wavefront agreement. We estimate the constant offset from the argument of the phasor sum $\sum_q\widehat{H}_{z,q}H^{*}_{z,q}$, which weights each local phase difference by the product of the predicted and reference amplitudes. After subtracting this offset and wrapping the remaining differences to $(-\pi,\pi]$, we calculate the reference-intensity-weighted root-mean-square residual $\sigma_\phi$ in radians (Supporting Note~3). This residual describes differences in the spatial phase profile after uniform phase alignment. The original phase reference is retained in the full complex-field error and the pointwise phase comparison in Figure~\ref{fig:readout}(c).

The residual phase agreement is quantified by $S_\phi=\exp(-\sigma_\phi^2)$, where values close to unity indicate small residual phase discrepancies. This exponential dependence is motivated by the phase-variance approximation used in optical aberration analysis~\cite{mahajan1983strehl}. Here, however, the comparison is between the predicted and FDFD phase distributions, excluding amplitude errors. Accordingly, $S_\phi$ is a diagnostic of residual phase agreement and does not quantify the focal Strehl ratio or focusing efficiency of the structure.

The set of evaluated pixels also affects the phase statistics. The amplitude-mask comparisons in Figure~S11 retain the highest 5\%, 20\%, 50\%, or 100\% of reference-amplitude pixels for the metalens and held-out datasets. Restricting the output-line analysis to high-amplitude pixels reduces the residual phase error, particularly for the higher-index metalens samples. The full-domain residuals show a different, nonmonotonic dependence on the retained fraction. These results demonstrate that masking changes which parts of the field are assessed and can substantially alter the reported phase agreement. The primary full-domain and output-line analysis therefore retains all pixels, with amplitude masking used only to examine the sensitivity of the diagnostic.

The optical assessment further distinguishes local-field errors from the integrated transmitted power. For the 70,012 held-out combinations, the readout-position sweep in Figure~S10 moves the output line from $0.2\lambda_0$ to $1.5\lambda_0$ above the structure. Amplitude and phase errors generally decrease with increasing distance, whereas the reported ratio of predicted to reference transmitted power remains close to unity throughout the sweep. The power comparison is therefore relatively stable over the evaluated distances, even though the local field-error measures remain position-dependent.

These checks support the use of predicted complex fields to evaluate transmitted amplitude, spatial phase, and integrated power under the tested conditions. Their scope aligns with the FDFD optical response for each complete simulated structure. Focal intensity and the coherent combination of separately predicted structures would require additional evaluation of amplitude distributions, relative phases, and electromagnetic coupling. The following analysis examines how prediction accuracy varies with structural scale within and beyond the training-generator interval.

\subsection{Generalization across structural scales}

\begin{figure}[htbp]
\centering
\includegraphics[width=\textwidth]{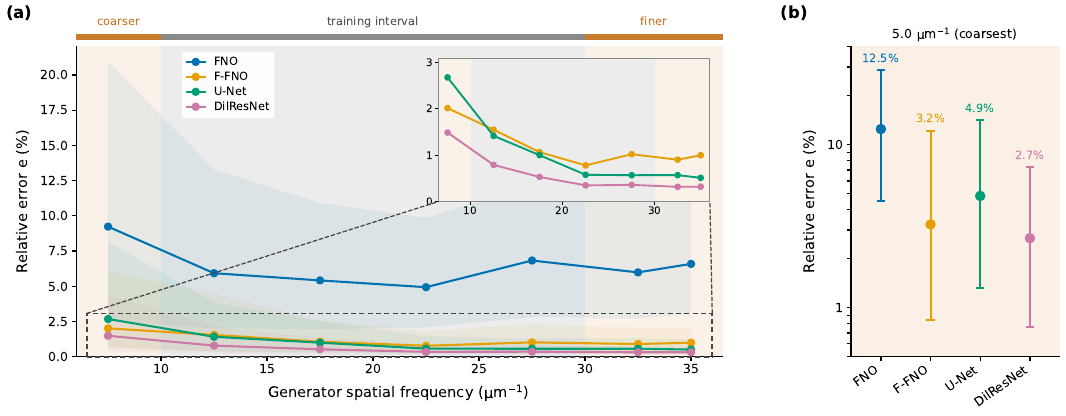}
\caption{\textbf{Generalization across structural scales without retraining.}
(a) Median relative complex-field errors for seven generator spatial frequencies from 7.5 to 35.0\,$\mu\mathrm{m}^{-1}$. Shaded bands indicate interquartile ranges. The gray background marks the training-generator interval of 10--30\,$\mu\mathrm{m}^{-1}$, and the orange backgrounds mark the evaluated extensions toward coarser and finer patterns. The inset resolves the lower errors of F-FNO, U-Net, and DilResNet.
(b) Median errors (markers and values) and interquartile ranges (bars) for the coarsest group at 5.0\,$\mu\mathrm{m}^{-1}$, shown on a logarithmic error axis.
Each group contains 14 independently generated geometries evaluated at 138 relative permittivities, giving 1,932 field samples per group. Errors are expressed as percentages and calculated from the complex fields over the full retained domain. The interquartile ranges describe variation among field samples rather than uncertainty across training runs. Generator spatial frequency does not uniquely specify the physical feature size of a binary mask.}
\label{fig:scale}
\end{figure}

Transfer to new nanophotonic designs requires accurate field prediction when their geometric features differ from those represented in the training data. We therefore examine how prediction accuracy changes as the spatial statistics of random patterns extend beyond the range used for training. The evaluation in Figure~\ref{fig:scale} varies the spatial-frequency parameter of the pattern generator while keeping the trained network parameters, illumination, and computational grid unchanged. This parameter controls the spatial variation of the noise field used to construct the binary masks. It does not uniquely determine a physical feature size, because thresholding produces dielectric regions and air gaps with different dimensions and connectivity.

The scale sweep evaluates independently generated geometries while retaining dielectric values represented in training. The training generator samples a continuous spatial-frequency interval of 10--30\,$\mu\mathrm{m}^{-1}$, whereas the evaluation comprises eight groups spanning 5.0--35.0\,$\mu\mathrm{m}^{-1}$. Each group contains 14 independently generated geometries, evaluated at 138 relative permittivities ranging from 2.2 to 15.9, yielding 1,932 field samples per group and 15,456 samples in total. The four interior groups assess prediction for new patterns within the training-generator interval, while two groups on either side examine extensions toward coarser and finer patterns. Each architecture uses one validation-selected checkpoint throughout the evaluation.

Accurate prediction is maintained across the interior of the training-generator interval for F-FNO, U-Net, and DilResNet. Their median relative complex-field errors in Figure~\ref{fig:scale}(a) remain below 1.6\% at 12.5, 17.5, 22.5, and 27.5\,$\mu\mathrm{m}^{-1}$. DilResNet gives the lowest median error at each frequency, and the inset resolves its performance relative to F-FNO and U-Net. The errors of these three architectures generally decrease toward the finer part of the interval, although the dependence is not strictly monotonic. FNO exhibits higher median errors of approximately 5--7\% and broader interquartile bands over the same groups. Thus, a common training dataset supports substantially different levels of accuracy across the evaluated network configurations.

Low prediction errors persist for the finer patterns outside the training-generator interval. At 32.5 and 35.0\,$\mu\mathrm{m}^{-1}$, the median errors of F-FNO, U-Net, and DilResNet remain approximately 0.32\%--1.0\%, comparable to their errors near the upper end of the training interval. DilResNet remains the most accurate, followed by U-Net and F-FNO. The absence of a pronounced increase in errors in these two groups shows that the learned field mapping remains applicable to the tested extension toward finer patterns. This generalization occurs without changing the trained weights or introducing dielectric values outside the training grid.

Coarser patterns produce a larger increase in error across the sampled extensions. At 7.5\,$\mu\mathrm{m}^{-1}$, the median errors rise to approximately 1.5\% for DilResNet, 2.0\% for F-FNO, and 2.7\% for U-Net, while FNO reaches approximately 9.3\%. The interquartile bands also broaden, indicating greater variation in prediction accuracy among samples. The coarsest group at 5.0\,$\mu\mathrm{m}^{-1}$ produces a further increase, as shown in Figure~\ref{fig:scale}(b). Its median errors are 12.5\% for FNO, 3.2\% for F-FNO, 4.9\% for U-Net, and 2.7\% for DilResNet. DilResNet consequently gives the lowest median error across all eight evaluated frequencies, while F-FNO outperforms U-Net in both groups below the training interval.

The observed dependence reflects changes in the population of random structures rather than the isolated effect of one geometric dimension. Each frequency group contains different masks, so feature dimensions, connectivity, and material arrangement vary together. The sweep therefore differs from uniformly resizing a fixed device, and the lower errors of the finer groups do not establish that smaller features are intrinsically easier to predict. Moreover, each group statistic pools samples across both geometries and permittivities. The resulting distributions describe performance over the sampled conditions rather than repeated realizations of the network training. Supporting Note~9 provides the sampling and aggregation details.

The low errors within the training interval and in the two finer groups demonstrate that training over a finite range of random-pattern scales can support field prediction for independently generated structures beyond the sampled generator interval. The larger errors for coarser patterns identify a more restrictive direction of transfer under the tested conditions. These results connect the intended reuse of a surrogate to the geometric coverage of its training data. Familiar dielectric values alone do not ensure comparable accuracy when the structural statistics change, and performance on held-out combinations does not fully characterize this dependence. Evaluation across independently generated scale groups therefore establishes both the demonstrated range of model reuse and the departures for which prediction becomes less reliable.

\subsection{Accuracy and computational cost of model reuse}

Reusing a trained surrogate across different nanophotonic structures allows the cost of reference data generation and network training to be shared among subsequent field predictions. The preceding evaluations demonstrate this reuse under the tested optical conditions, with each architecture retaining one set of trained parameters across all evaluation sets. Its computational value depends on both the cost of repeated inference and the accuracy obtained for the intended structures. Complementary strategies address the cost of acquiring reference data, including physics-enhanced surrogates that couple trainable models to lower-fidelity physical solvers~\cite{pestourie2023physics}. We therefore compare the measured training and inference times together with the prediction errors.

Repeated field evaluation has a similar measured cost across the four tested configurations. On one NVIDIA RTX 6000 Ada, the amortized processing time is 20.8--22.1\,ms per field sample at a batch size of 32, including data loading. These values describe average time per sample during batched evaluation and should be distinguished from the latency of processing an individual sample. Training costs vary substantially more, requiring 37.7\,h for FNO, 39.1\,h for U-Net, 126.4\,h for DilResNet, and 183.4\,h for F-FNO, as reported in Table~S2. The comparison in Figure~S9 relates these costs to the mean relative complex-field errors in Table~\ref{tab:families}. Under the measured conditions, differences in training cost and prediction accuracy are therefore more pronounced than differences in amortized inference time.

The appropriate balance between training cost and accuracy depends on the structures to be evaluated. U-Net achieves the lowest mean error of 0.14\% on the held-out geometry--permittivity combinations while requiring considerably less training time than DilResNet. For the five independent test sets, however, DilResNet achieves the lowest equally weighted mean error of 3.89\%, compared with 5.18\% for U-Net. Its higher training cost is thus accompanied by better average accuracy across these independently generated evaluation sets. This advantage does not apply uniformly across all structural families, as U-Net remains more accurate for inverse-designed metalenses. A model intended primarily for metalens evaluation may therefore favor a different balance of accuracy and training cost from one intended for use across several structure types. These comparisons describe the selected configurations under the common training procedure, with one training run per architecture.

The benefit of model reuse is avoiding additional parameter fitting for the evaluated structures and material assignments. Once training is complete, the same predictor can evaluate multiple structural families without incurring a separate training cost for each family. As the number of predictions increases, the initial data generation and training costs are spread across more evaluations. Whether this produces a net computational saving relative to repeated FDFD calculations depends on the reference-simulation cost, the total training investment, and the number of subsequent predictions. The present measurements do not establish a solver speedup or the number of predictions required to recover the initial computational cost. Supporting Note~6 and Table~S4 provide conditional estimates for larger datasets, which should be interpreted separately from the measured timings.

These results connect generalization to the computational rationale for surrogate reuse. A low inference cost is most useful when the same trained model retains sufficient accuracy across the structures required by an application. The independent evaluations indicate where this reuse is supported, while the differences among architectures demonstrate that additional training time alone does not guarantee lower prediction error across all structural families. Selecting a reusable surrogate therefore requires considering its accuracy on representative target structures, as well as the initial training cost and the expected volume of subsequent field evaluations.

\section{Conclusion}
\label{sec:conclusion}

We demonstrated that training on random dielectric patterns enables electromagnetic field prediction across diverse nanophotonic structures without retraining. A single learned mapping from spatial permittivity to complex magnetic fields accommodates independently constructed geometries and material arrangements absent from training. Across 35,441 geometry--permittivity samples in five test sets, DilResNet achieves mean relative complex-field errors of 1.4\%--7.6\%, with an equally weighted mean of 3.89\%. Training with one design dielectric per pattern also supports prediction for two-material freeform patterns composed of dielectric values represented in the training data. These results establish that random-pattern training can support transfer to structures produced by different design procedures and to new spatial combinations of familiar materials.

The predicted complex fields provide access to optical quantities beyond spatial intensity distributions. Agreement with FDFD extends to transmitted amplitude and phase profiles, while F-FNO, U-Net, and DilResNet achieve median transmitted-power errors of 0.7\%--0.9\% across 6,950 inverse-designed metalens samples. A common field prediction therefore supports several optical evaluations without separately training a predictor for each quantity. This capability is relevant to comparing candidate structures and material assignments using both their spatial field distributions and integrated optical responses.

Generalization depends on the relationship between the training data and the intended structures. U-Net gives the lowest error on held-out geometry--permittivity combinations, whereas DilResNet gives the lowest average error across the five independent test sets. The structural-scale sweep further shows that low errors persist for the tested extension toward finer random patterns, while coarser patterns produce larger errors. Together, these findings demonstrate why independent geometric and material-arrangement tests are necessary to establish the usable range of a trained surrogate. They also identify the structural conditions under which the present predictors become less reliable.

The demonstrated transfer applies to two-dimensional TE fields at 500\,nm under fixed illumination and computational conditions, with lossless, nondispersive media and scalar permittivities represented in training. Accuracy is measured against discrete FDFD reference solutions. Broader use will require extending and validating the learned mapping for additional optical conditions and three-dimensional fields. Within the evaluated conditions, however, a single trained model already supports multiple nanophotonic design families without additional parameter fitting. This reuse allows the initial investment in reference-data generation and training to be shared across subsequent field evaluations, providing a basis for electromagnetic surrogates that support multiple design tasks through a common trained predictor.

\section*{Conflict of interest}
The authors declare no conflicts of interest.
\section*{Data availability}
The test datasets, trained network weights, and evaluation code supporting the findings of this study are openly available at \url{https://github.com/nanophotonics-lab/random-pattern-em-surrogate} (release v1.0.0)~\cite{kim2026dataset}. The training dataset is available from the corresponding author upon reasonable request.
\section*{Use of generative AI}
OpenAI ChatGPT/Codex was used to assist with language editing and to produce an initial draft of the conceptual schematic in Figure~\ref{fig:illustration}. No scientific data were generated using an image-generation model. The authors reviewed and verified the final manuscript and Figure~\ref{fig:illustration} and take full responsibility for their content.
\section*{Supporting information}
The Supporting Information contains dataset and metric definitions, Tables~S1--S6, Figures~S1--S16, a resource-planning analysis in Supporting Note~6 and Table~S4, and the numerical-error analysis of the FDFD reference in Supporting Note~7. Supporting Notes~8 and 9 provide the optical readout definitions and details of the material- and structural-scale evaluations. Supporting Note~10 compares the test structures with their nearest training samples. These materials cover wavelength and scaling tests, all-model field comparisons, regional and phase diagnostics, error distributions, computational costs, readout-position sensitivity, training histories, physics-residual comparisons, and the simulation domain.

\clearpage
\setcounter{figure}{0}\setcounter{table}{0}\setcounter{equation}{0}\setcounter{footnote}{0}
\renewcommand{\theequation}{S\arabic{equation}}
\renewcommand{\thefigure}{S\arabic{figure}}
\renewcommand{\thetable}{S\arabic{table}}
\captionsetup{font=small,labelfont=bf,singlelinecheck=false,labelsep=period}
\setlength{\parskip}{0.3em}
\makeatletter
\let\maketitle\MergeMaketitle \let\@maketitle\MergeAtMaketitle
\let\title\MergeTitle \let\date\MergeDate \let\and\MergeAnd \let\thanks\MergeThanks
\let\@author\MergeAtAuthor
\let\author\relax\let\affil\relax\newcommand\author[2][]{}\newcommand\affil[2][]{}
\let\MergeCite\cite
\renewcommand\cite[1]{\def\Merge@keys{}\@for\Merge@k:=#1\do{\edef\Merge@keys{\ifx\Merge@keys\@empty\else\Merge@keys,\fi SI:\Merge@k}}\expandafter\MergeCite\expandafter{\Merge@keys}}
\makeatother
\title{Supporting Information for\\[2mm]\ManuscriptTitle}
\author[1]{Jiwoo Kim}
\author[2]{Mingyu Park}
\author[1]{Chanik Kang}
\author[3]{Hyewon Suk}
\author[1,2,3,*]{Haejun Chung}
\affil[1]{Department of Artificial Intelligence, Hanyang University, Seoul, 04763, Republic of Korea}
\affil[2]{Department of Electronic Engineering, Hanyang University, Seoul, 04763, Republic of Korea}
\affil[3]{Department of Artificial Intelligence Semiconductor Engineering, Hanyang University, Seoul, 04763, Republic of Korea}
\date{*Corresponding author: Haejun Chung, email: haejun@hanyang.ac.kr}
\maketitle

\section*{Overview}
This Supporting Information provides additional analyses of the electromagnetic surrogate solver described in the main manuscript. Tables~S1--S3 summarize the refractive-index-dependent errors, network configurations, and dataset composition, and Tables~S5 and S6 report the model-specific transmitted-power statistics and the residual-training comparison. Figures~S1--S14 examine wavelength- and scale-dependence, spatial field fidelity, optical-response metrics, error distributions, and computational cost, and Figure~S15 shows the simulation domain. Figure~S16 and Supporting Note~10 compare the test structures of main-text Figure~3 with their nearest training samples. Supporting Note~6 and Table~S4 give conditional resource estimates for future extensions, and Supporting Note~7 analyzes the numerical error of the finite-difference reference.

Unless specified otherwise, all evaluations use lossless, nondispersive media under two-dimensional TE illumination at a vacuum wavelength of 500\,nm and a grid spacing of 10\,nm, and all reference fields are computed with the finite-difference frequency-domain (FDFD) solver described in the main-text Methods (Figure~S15). The networks predict the complex magnetic field $H_z$. In this two-dimensional setting, the output plane corresponds to a line positioned $0.5\lambda_0$ above the sample-specific structure boundary.

\section*{Supporting Note 1. Dataset construction and generalization tests}
Photonic surrogates have progressed from a prescribed design space to devices parameterized by geometry, material, and excitation~\cite{gu2022neurolight}. This expansion makes the unit of data separation important. The present tests distinguish a new field condition for a previously encountered geometry from an independently generated structural layout.

We distinguish a geometry, which specifies the spatial arrangement of materials, from a field sample, which also includes the assigned permittivity and simulation conditions. Accordingly, evaluating one geometry at several permittivities increases the number of field samples without increasing the number of independent shapes. This distinction is essential when interpreting the dataset sizes and generalization results.

The primary data pool contains 700,140 combinations of 5,001 random freeform geometries and 140 permittivities. Independent splits at each permittivity yield 560,113 training, 70,015 validation, and 70,012 held-out combinations. No identical geometry--permittivity pair occurs in multiple splits. However, a geometry assigned to the held-out set at one permittivity can appear in training at another, and every scalar permittivity is represented in training. The held-out set thus evaluates new combinations within the sampled population.

\subsection*{Random-pattern implementation}

The random freeform masks were generated using two-dimensional OpenSimplex noise. Gradient-noise methods construct a spatially correlated scalar field from local pseudorandom gradients~\cite{perlin2002improving}. In the OpenSimplex construction, scaling the sampling coordinates changes the characteristic spatial variation of the noise field, allowing patterns with different feature widths, gaps, and connectivity to be generated using the same procedure.

The noise field was sampled on a design region containing 64 transverse cells and 13 cells along the propagation direction, with a grid spacing of 10\,nm. For each training geometry, the dimensionless coordinate-scaling parameter $s$ was drawn uniformly from $[0.1,0.3]$, and the sampled field was thresholded at zero to obtain a binary material mask. This interval corresponds to a nominal generator spatial-frequency range of 10--30\,$\mu\mathrm{m}^{-1}$. The generator frequency describes the coordinate scaling of the noise function and does not uniquely specify the feature dimensions of the thresholded mask.

\subsection*{Independent evaluation structures}

Transfer beyond the training population is assessed using five test sets containing 35,441 field samples. The bar arrays, manually defined patterns, and two inverse-designed sets introduce geometric construction rules that differ from those of the training generator. The two-material set retains the generator but introduces material arrangements absent from the single-material training data. None of these sets is used for parameter fitting or checkpoint selection.

\subsubsection*{Aperiodic bar arrays and manually defined patterns}

The aperiodic bar-array set contains 100 geometries, each comprising three to nine dielectric bars. Bar widths and intervening air gaps were selected at random within ranges of 10--330\,nm and 30--240\,nm, respectively. The resulting arrangements have no prescribed lattice period. These structures test prediction for arrangements generated by explicit bar and gap parameters rather than by the random-pattern generator.

The manually defined set contains 19 geometries comprising bars, rectangles, ellipses, rings, stars, chains, oblique bars, and the letters HYU. Their shapes were specified directly rather than obtained by thresholding a noise field. This set therefore introduces geometric motifs constructed independently of the training-pattern generation procedure.

\subsubsection*{Inverse-designed structures}

The inverse-designed metalens set contains 50 binary geometries obtained using a finite-difference time-domain solver. Each geometry was optimized to concentrate the transmitted field at a target point selected at random in the air region above the structure. Varying the target position produces optimized material arrangements with different transmitted-field distributions. In all 50 geometries, the non-material cells of the $64\times13$-cell design layer retain the substrate permittivity of 2.1 rather than air.

The second inverse-designed set contains 50 binary geometries generated for 50 target points using the frequency-domain solver in Meep~\cite{oskooi2010meep}. The time-domain and frequency-domain descriptions identify the solvers used during geometry optimization. They do not identify different reference solvers for surrogate evaluation. The geometries of the second set are binary material--air patterns within the same design layer, and the material of two of them occupies only part of its 13 rows.

After optimization, each binary geometry was retained while its design-material permittivity was varied for evaluation. All reference fields were recomputed using the same Ceviche 0.1.3 FDFD solver and optical conditions as the training data. Consequently, these tests compare surrogate predictions with a common numerical reference, irrespective of the solver originally used to obtain the geometry.

\subsubsection*{Permittivity assignments for single-material test structures}

Each geometry in the four single-material test sets was evaluated at 139 relative permittivities from 2.1 to 15.9 in increments of 0.1. The substrate permittivity remained fixed at 2.1, and the surrounding medium was air. The resulting datasets contain 13,900 bar-array samples, 2,641 manually defined samples, and 6,950 samples in each inverse-designed set.

Repeated permittivity assignments increase the number of field samples without increasing the number of independent geometries. In particular, each inverse-designed set contains 50 optimized geometries evaluated under multiple material assignments, rather than 6,950 separately optimized devices.

\subsubsection*{Two-material freeform patterns}

The two-material set contains 5,000 freeform patterns generated using the same OpenSimplex construction and coordinate-scale interval, $s\in[0.1,0.3]$, as the training patterns. For each sample, two distinct relative permittivities, denoted $\varepsilon_A$ and $\varepsilon_B$, were drawn at random from the discrete set
\begin{equation}
\mathcal{E}_{2}=\{2.2,2.3,\ldots,16.0\}.
\label{eq:si_two_material_values}
\end{equation}
Here, $\mathcal{E}_{2}$ denotes the permitted design-material values for the two-material evaluation.

Each connected component of the material mask, identified with four-neighbor connectivity in the design layer, was assigned one of the two selected permittivities, with a uniform value throughout that component. The two values were assigned to components with approximately equal frequency (49.9\% of all components received $\varepsilon_A$). Every one of the 5,000 samples contains at least two components and both permittivities. The remaining design-layer pixels were assigned air, and the substrate retained its relative permittivity of 2.1. The substrate is common to the training and test configurations and is not counted as a second design material.

This construction introduces spatial combinations of two design materials absent from the single-material training structures. Both assigned dielectric values remain within the sampled training grid, so the evaluation tests transfer to new material arrangements rather than extrapolation to unseen scalar permittivities.

\subsection*{Structural-scale evaluation}

A separate structural-scale sweep contains independent random geometries in eight spatial-frequency groups. Four groups lie within the training-generator interval, and four extend beyond it. This sweep separates the influence of structural scale from transfer to the five independent tests. Together, the held-out combination set, independent structure tests, and scale sweep provide seven primary evaluations comprising 120,909 samples (Table~S3). Figure~S15 shows the simulation domain shared by all of these datasets.

\section*{Supporting Note 2. Wavelength dependence and geometric scaling}
Wavelength-conditioned electromagnetic prediction has made it possible to evaluate spectral conditions absent from a discrete training set~\cite{seo2026broadband}. The current model is trained at one wavelength, so changing illumination must be separated from rescaling the entire nondispersive problem. The following tests determine which form of variation is represented.

Figures~S1 and S2 examine two different forms of wavelength variation. In the fixed-grid sweep, the wavelength changes over ten values between 475 and 525\,nm while the 10\,nm grid and physical geometry remain unchanged. Field error increases as the illumination moves away from the training wavelength. Fitting and removing a global complex factor reduces part of the discrepancy but does not eliminate its wavelength dependence. This correction is used only as a diagnostic. It is not applied to the primary predictions reported in the main manuscript.

In the scaled-grid sweep, eight wavelengths between 300 and 700\,nm are evaluated while the grid spacing is adjusted to keep $\lambda_0/\Delta l=50$. The physical dimensions of a fixed pixel pattern therefore scale with wavelength. For the lossless, nondispersive materials considered here, this transformation preserves the dimensionless geometry, and the reported field and intensity errors remain essentially unchanged. The result illustrates joint wavelength--geometry scaling, rather than broadband prediction for a device with fixed physical dimensions.

\section*{Supporting Note 3. Field-error and phase diagnostics}
Complex-field predictors permit multiple optical quantities to be recovered from one output~\cite{wiecha2020deep}. This advantage requires separate checks of spatial fidelity, transmitted power, and wavefront error. We retain these distinctions so that success in one metric is not taken as evidence for an untested optical quantity.

For a reference field $H$ and prediction $\hat H$ on the retained domain $\Omega$, the per-sample relative error and normalized mean squared error are
\begin{equation}
e=\frac{\left(\sum_{\Omega}|\hat H-H|^2\right)^{1/2}}{\left(\sum_{\Omega}|H|^2\right)^{1/2}},
\qquad \mathrm{NMSE}=e^{2}.
\end{equation}
The main text reports $e$ in percent. Table~S1 reports medians of $e$ at four representative permittivities, and the main-text scale sweep reports the median and interquartile range of $e$ within each group. Several supplementary figures are rendered directly from the evaluation records: axes labeled $\mathrm{rL}_2$ show $e$ as a fraction, and annotations give NMSE, where a value of $10^{-4}$ corresponds to $e=1\%$. Because averaging and taking a square root are different operations, the square root of an average NMSE is not the average or median of $e$. Shaded interquartile bands describe sample-to-sample variation, rather than uncertainty across repeated training runs.

Figures~S3 and S4 compare reference intensities, predicted intensities, and absolute-error maps. Separate display normalizations are applied to the design and propagation regions, with matching reference and prediction scales within each region. These choices improve the visibility of spatial patterns and do not change the complex-field errors reported in the annotations.

Figures~S5 and S7 resolve the error by spatial region. For each region $R$ (substrate, design layer, free space, material interfaces, and high-permittivity pixels), the regional error divides the error energy within $R$ by the reference energy within the same region, $\sum_{R}|\hat H-H|^2/\sum_{R}|H|^2$. A regional value is therefore the NMSE obtained by retaining only that region, and a region/full ratio above one indicates a region that is predicted less accurately than the field as a whole. High-permittivity pixels are those whose permittivity exceeds the 5th-percentile value of the sample by more than 5\% of its permittivity span, and interface pixels are those where the permittivity-gradient magnitude exceeds 5\% of its maximum, dilated by two cells. The two figures display this quantity differently. Figure~S5 plots $\mathrm{rL}_2=\langle\mathrm{NMSE}_R\rangle^{1/2}$, the square root of the per-sample regional NMSE averaged over all samples at the same permittivity, and its panel (c) plots the square root of the averaged per-sample ratio $\mathrm{NMSE}_R/\mathrm{NMSE}$. Figure~S7 reports the averaged per-sample regional NMSE itself, without the square root, within three permittivity intervals bounded by the tertiles of the sampled permittivities. For two-material samples, the permittivity axis of Figures~S5--S7 uses $\varepsilon_A$, the value labeled first when each sample was generated. Because $\varepsilon_A$ and $\varepsilon_B$ are drawn independently, this axis is neither the mean nor the maximum design permittivity of a sample.

\subsection*{Weighted phase diagnostics}

Phase agreement is evaluated separately from amplitude agreement. The transmitted-field profile is referenced to the incident wave reconstructed from the substrate. Electric fields are recovered from $H_z$ in the uniform output region, and transmitted power is calculated from the real Poynting flux as detailed in Supporting Note~8.

At each sampled position, the wrapped phase difference is
\begin{equation}
\Delta\phi_q=\operatorname{Arg}\left(\widehat{H}_{z,q}H_{z,q}^{*}\right),
\label{eq:si_wrapped_phase_difference}
\end{equation}
where $H_{z,q}$ and $\widehat{H}_{z,q}$ are the reference and predicted complex magnetic fields at position $q$, the asterisk denotes complex conjugation, and $\operatorname{Arg}$ returns a phase in $(-\pi,\pi]$.

A constant phase offset is obtained from the weighted circular mean
\begin{equation}
\phi_0=\operatorname{Arg}\left[\sum_{q\in\mathcal{Q}}w_q\exp(j\Delta\phi_q)\right],
\label{eq:si_constant_phase_offset}
\end{equation}
where $\mathcal{Q}$ is the set of evaluated positions, $w_q=|\widehat{H}_{z,q}||H_{z,q}|$, so that Equation~\eqref{eq:si_constant_phase_offset} equals $\operatorname{Arg}\sum_{q}\widehat{H}_{z,q}H_{z,q}^{*}$, and $j^2=-1$.

After removing this offset, the weighted root-mean-square phase residual is
\begin{equation}
\sigma_{\phi}=\left[\frac{\sum_{q\in\mathcal{Q}}v_q\left\{\operatorname{wrap}\left(\Delta\phi_q-\phi_0\right)\right\}^{2}}{\sum_{q\in\mathcal{Q}}v_q}\right]^{1/2},
\label{eq:si_weighted_phase_residual}
\end{equation}
where $v_q=|H_{z,q}|^{2}$ is the reference intensity, $\operatorname{wrap}$ maps the residual phase to $(-\pi,\pi]$, and $\sigma_{\phi}$ is expressed in radians. Positions with vanishing reference amplitude therefore receive zero weight in both the offset and the residual. For the full domain and the output line, the residual $\sigma_\phi$ and $S_\phi$ include all positions. The constant offsets shown in Figure~S6, and the residuals of the design, free-space, and substrate regions, are evaluated on the 20\% of positions with the largest reference amplitude in each region. The amplitude-mask comparisons in Figure~S11 restrict $\mathcal{Q}$ to the selected reference-amplitude subsets.

The phase diagnostic is
\begin{equation}
S_{\phi}=\exp(-\sigma_{\phi}^{2}),
\label{eq:si_phase_diagnostic}
\end{equation}
where $S_{\phi}$ summarizes residual phase agreement after removal of a constant offset. Its exponential form is motivated by an approximation relating the Strehl ratio to phase-aberration variance~\cite{mahajan1983strehl,born1999principles}. Here, the diagnostic compares predicted and reference phase distributions, excludes amplitude errors, and is not a calculated focal Strehl ratio. The constant offset and residual phase error therefore characterize different aspects of agreement with the numerical reference.

The phase errors in main-text Figure~6(c) are expressed as a percentage of one cycle, with 1\% corresponding to $3.6^\circ$. Supplementary plots retain their displayed units. Supporting Note~8 describes the incident-phase consistency checks and optical readout.

\section*{Supporting Note 4. Model implementation, training objectives, and computational cost}
Each architecture is trained once using the common dataset and optimization procedure. The reported costs apply to the configurations in Table S2. Parameter count alone does not determine their training time or prediction accuracy.

\subsection*{Input representation and output normalization}

All architectures use three input channels comprising the normalized relative-permittivity distribution and two spatial coordinates. For each sample, the permittivity channel is calculated as
\begin{equation}
\widetilde{\varepsilon}_r(q,p)=\operatorname{clip}_{[0,1]}\left[\frac{\varepsilon_r(q,p)-\varepsilon_{\mathrm{lo}}}{\varepsilon_{\mathrm{hi}}-\varepsilon_{\mathrm{lo}}+\delta_{\varepsilon}}\right],
\label{eq:si_permittivity_normalization}
\end{equation}
where $\varepsilon_r(q,p)$ is the relative permittivity at row $q$ and column $p$, $\widetilde{\varepsilon}_r$ is its normalized value, and $\operatorname{clip}_{[0,1]}$ restricts the result to $[0,1]$. The numerical stabilizer is $\delta_{\varepsilon}=10^{-8}$, and $\varepsilon_{\mathrm{lo}}$ and $\varepsilon_{\mathrm{hi}}$ are the normalization bounds. These bounds are defined by
\begin{equation}
\varepsilon_{\mathrm{lo}}=\min\left(2.1,\min_{\Omega}\varepsilon_r\right),
\qquad
\varepsilon_{\mathrm{hi}}=\max\left(16.0,\max_{\Omega}\varepsilon_r\right),
\label{eq:si_permittivity_bounds}
\end{equation}
where $\Omega$ denotes the retained computational domain. For the reported samples containing air and design-material permittivities no greater than 16.0, the bounds are 1.0 and 16.0, respectively. The normalization retains the distinction between different assigned material values rather than rescaling each sample to its own maximum permittivity.

The coordinate channels are
\begin{equation}
\xi_p=-1+\frac{2p}{63},
\qquad
\eta_q=-1+\frac{2q}{127},
\label{eq:si_coordinate_channels}
\end{equation}
where $p=0,\ldots,63$ indexes transverse columns and $q=0,\ldots,127$ indexes rows along the propagation direction. Both coordinate channels span $[-1,1]$, including the endpoints. The input array has dimensions $3\times128\times64$, with channels ordered as normalized permittivity, transverse coordinate, and propagation coordinate.

The two output channels represent affinely normalized real and imaginary magnetic-field components. Their normalization is written as
\begin{equation}
u_c=\frac{h_c-\mu_c}{\sigma_c},
\qquad c\in\{\mathrm{R},\mathrm{I}\},
\label{eq:si_output_normalization}
\end{equation}
where $h_{\mathrm{R}}=\operatorname{Re}H_z$ and $h_{\mathrm{I}}=\operatorname{Im}H_z$ are the physical field components, $u_c$ denotes the corresponding normalized component, and $\mu_c$ and $\sigma_c$ are fixed channel-specific offsets and scale factors stored with the trained checkpoint.

The predicted complex field is reconstructed as
\begin{equation}
\widehat{H}_z=\left(\sigma_{\mathrm{R}}\widehat{u}_{\mathrm{R}}+\mu_{\mathrm{R}}\right)
+j\left(\sigma_{\mathrm{I}}\widehat{u}_{\mathrm{I}}+\mu_{\mathrm{I}}\right),
\label{eq:si_output_reconstruction}
\end{equation}
where $\widehat{u}_{\mathrm{R}}$ and $\widehat{u}_{\mathrm{I}}$ are the network outputs and $j^2=-1$. The offsets and scale factors are stored as \texttt{output\_mean} and \texttt{output\_std}, respectively. They remain unchanged across evaluation samples for a given checkpoint. Field errors and optical quantities are evaluated after this inverse transformation. The offsets and scale factors are the channel-wise mean and standard deviation over all pixels of the first 500 shuffled training mini-batches (32,000 training samples); no validation or test sample enters this estimate. All six checkpoints share $\mu_{\mathrm{R}}=-6.67\times10^{-13}$, $\mu_{\mathrm{I}}=-7.88\times10^{-13}$, $\sigma_{\mathrm{R}}=1.114\times10^{-11}$, and $\sigma_{\mathrm{I}}=1.064\times10^{-11}$ in the units of the stored fields. The supervised loss is the per-sample relative $L_2$ error between the normalized predicted and reference channels $(u_{\mathrm{R}},u_{\mathrm{I}})$, averaged over each mini-batch. Because $|\mu_c|/\sigma_c<0.08$ and $\sigma_{\mathrm{R}}/\sigma_{\mathrm{I}}=1.05$, this loss closely follows, but is not identical to, the relative error of the physical field. For each architecture, the retained checkpoint minimizes this loss on the validation set; for the residual-trained variants below, the validation loss includes the residual term.

\subsection*{Computational cost}

Figure S9(a,b) uses the mean held-out relative errors in Table 1. Panel (c) uses the equally weighted mean of the five independent-set means, irrespective of their different sample counts. Panel (d) gives the processing time of each configuration (20.8--22.1\,ms). It includes data loading at batch size 32 on one RTX 6000 Ada and is not a single-sample latency.

For $N$ predictions from one trained model, average total cost can be written as $\overline{t}=t_{\mathrm{eval}}+(T_{\mathrm{ref}}+T_{\mathrm{train}})/N$, using consistent time units and the same accounting of computational resources. Here $T_{\mathrm{ref}}$ is reference-data generation cost and $T_{\mathrm{train}}$ is training cost. Neither a matched FDFD timing nor $T_{\mathrm{ref}}$ is reported, so a numerical acceleration factor or break-even count cannot be calculated.

\subsection*{Training with an additional wave-equation residual}

The comparison in Figure~S14 examines FNO and U-Net variants trained with an additional wave-equation residual. Each variant retains the architecture and parameter count of its corresponding supervised model. The combined objective is written as
\begin{equation}
\mathcal{L}_{\mathrm{total}}=\mathcal{L}_{\mathrm{field}}+\alpha_{\mathrm{res}}\mathcal{L}_{\mathrm{res}},
\label{eq:si_residual_training_loss}
\end{equation}
where $\mathcal{L}_{\mathrm{field}}$ is the supervised relative $L_2$ field loss, $\mathcal{L}_{\mathrm{res}}$ is the wave-equation residual loss, and $\alpha_{\mathrm{res}}$ determines the contribution of the residual term. Both variants use $\alpha_{\mathrm{res}}=0.01$, applied after ten initial epochs trained with $\mathcal{L}_{\mathrm{field}}$ alone.

The residual is that of the source-free form of Equation~\eqref{eq:hz-divergence}, $R=\nabla\cdot(\varepsilon_r^{-1}\nabla u)+k_0^2u$, evaluated separately for the real and imaginary output channels $u$ with $k_0=2\pi/(500\,\mathrm{nm})$ and a cell size of 10\,nm in physical units. Forward differences give the gradient and backward differences the divergence, and the permittivity at each cell face is the arithmetic mean of the two adjacent cells, plus $10^{-5}$, following the Ceviche discretization. The permittivity is recovered from the normalized input channel, whereas $u$ is the normalized network output rather than the de-normalized field. The difference operators are periodic, and a five-cell margin on all four sides is excluded from the loss. The retained domain contains no source. The residual loss is
\begin{equation}
\mathcal{L}_{\mathrm{res}}=\frac{1}{B}\sum_{b=1}^{B}\frac{\sum_{q\in\mathcal{C}}\left(R_{\mathrm{R},bq}^{2}+R_{\mathrm{I},bq}^{2}\right)}{k_0^{4}\sum_{q\in\mathcal{C}}\left(u_{\mathrm{R},bq}^{2}+u_{\mathrm{I},bq}^{2}\right)},
\label{eq:si_residual_loss}
\end{equation}
where $B$ is the mini-batch size and $\mathcal{C}$ is the interior region after the margin is removed.

Across the evaluated datasets, the reported average NMSE changes by less than 15\% in either direction relative to the corresponding supervised models. Table~S6 lists the change for each dataset, defined as the ratio of the mean per-sample NMSE of the residual-trained variant to that of the supervised model, minus one. These results characterize the tested residual definitions and loss weights and do not establish a general advantage or limitation of residual-based training.

\section*{Supporting Note 5. Interpretation of the supplementary comparisons}
Learned optical solvers have been applied to both response prediction and freeform field calculation~\cite{chen2022high}. Broader reuse requires several complementary tests because a selected field image, integrated response, or validation score addresses a different question. The supplementary results are organized according to these questions.

The supplementary analyses connect different aspects of surrogate performance. Representative fields in Figures~S3 and S4 show spatial fidelity, while Figures~S5, S7, and S8 examine its dependence on location, refractive index, and sample population. Figures~S6, S10, S11, and S13 evaluate phase and output-line quantities. Computational cost, convergence, and loss-function comparisons are presented in Figures~S9, S12, and S14. The wavelength sweeps in Figures~S1 and S2 distinguish fixed-device variation from geometric scaling.

Together, these results characterize reuse of one trained model across materials and structures within the evaluated physical setting. Figure~2 of the main text summarizes the procedure used to construct the training data and connect the permittivity representation to the complex-field output, and Figure~S15 shows the corresponding simulation domain.
\section*{Supporting Note 6. Conditional resource estimates}
The following calculations illustrate how training time and field storage scale under fixed assumptions. They are conditional resource estimates, not measurements of generalization or evidence that a particular dataset size guarantees a target accuracy.

\clearpage
\textbf{Conditional scaling of 2D training.} Dataset sizes of $10^6$ and $10^7$ field samples are used as illustrative accounting cases. They are not established data requirements. Actual sampling would depend on the desired material, wavelength, incidence, and geometry ranges and their measured prediction errors. Retaining the current grid, architecture, 80\% training fraction, and 200-epoch schedule gives the linear estimate

\begin{equation}
T_{\mathrm{2D}}(N)=T_0\frac{N}{700{,}140},
\label{eq:cost}
\end{equation}
where $N$ includes training, validation, and held-out samples. The architecture-specific values of $T_0$ are 37.7, 183.4, 39.1, and 126.4 GPU-hours for FNO, F-FNO, U-Net, and DilResNet, respectively. One million samples correspond to approximately 54--262 GPU-hours per run, and ten million to 538--2620 GPU-hours. Eight equivalent GPUs would yield ideal training times of approximately 7--33 and 67--328 hours, respectively, before parallel overhead. These estimates assume unchanged work per sample. They exclude reference simulations, model selection, repeated seeds, and code development. Runs of a different architecture or resolution require fresh timing measurements.

\textbf{Combinatorial growth of simulation conditions.}
For illustration, extending all 700,140 present geometry--permittivity combinations to 20 wavelengths, nine incidence angles, and two polarizations gives
\begin{equation}
N_{\mathrm{dense}}=700{,}140\times20\times9\times2
=252{,}050{,}400.
\label{eq:dense}
\end{equation}
The condition counts are illustrative choices, not proposed minimum resolutions. A wavelength-conditioned time-domain calculation may return several frequency samples from one run, so field-sample count need not equal solver-run count. The dense product nevertheless shows why exhaustive storage and training rapidly become expensive. Error-guided sampling and reuse of learned representations can be tested as alternatives to acquiring every combination.

\textbf{Field-storage cost in two and three dimensions.}
At 32-bit precision, the two real channels for complex $H_z$ on the current $64\times128$ grid require 65,536 bytes per sample. The output-only storage is therefore approximately 65.5\,GB for $10^6$ samples and 655\,GB for $10^7$ samples. For a prospective $128\times64\times64$ 3D grid with all six complex electric and magnetic components, the corresponding storage is
\begin{equation}
B_{\mathrm{3D}}=N\,(128\times64\times64)\times6\times2\times4
\quad\text{bytes}.
\label{eq:storage}
\end{equation}
This amounts to 25.2\,TB for one million field samples. Decimal GB and TB are used throughout. Input maps, coordinates, boundaries, reference-solver intermediate fields, model states, and backups are excluded. Reduced precision, compressed storage, or retaining fewer field components would change this budget and would require validation of the resulting optical accuracy.

\textbf{Reference generation and 3D training.}
The present materials do not provide per-case numerical-solver timings or a 3D network benchmark. Consequently, an absolute 3D GPU-hour requirement cannot be established from the reported 2D runs. A pilot study should measure time per reference case, usable frequency outputs per run, peak solver memory, network memory per sample, and multi-GPU scaling. A possible initial allocation is eight GPUs with the batch size chosen from these measurements. This is an experimental allocation, not a verified hardware minimum. For a pilot reference-solve cost of $t_{\mathrm{ref}}$ GPU-hours per case and $N_{\mathrm{solve}}$ required cases, the simulation budget is $N_{\mathrm{solve}}t_{\mathrm{ref}}$. Adaptive sampling, progressive changes in resolution, and subdomain prediction should then be compared using the same held-out optical targets to determine the necessary data and compute budget.

\section*{Supporting Note 7. Numerical error of the finite-difference reference}
All reference fields were calculated with Ceviche 0.1.3 using its $H_z$-polarized FDFD solver, \texttt{fdfd\_hz}~\cite{hughes2019forward}. In the FDFD formulation, spatial derivatives are represented by finite-difference operators, yielding a sparse linear system for the field amplitudes~\cite{rumpf2012fdfd}. Under the $e^{+j\omega t}$ convention and away from the impressed source, Maxwell's curl equations are
\begin{equation}
\nabla\times\mathbf{E}=-j\omega\mu\mathbf{H},\qquad
\nabla\times\mathbf{H}=j\omega\varepsilon\mathbf{E}.
\label{eq:maxwell-fdfd}
\end{equation}
For the two-dimensional system used here, $\mathbf{E}=(E_x,E_y,0)$ and $\mathbf{H}=(0,0,H_z)$. This is an $H_z$-polarized system, also called two-dimensional TE when $E_z=0$. The electric-field components follow as
\begin{equation}
E_x=\frac{1}{j\omega\varepsilon}\frac{\partial H_z}{\partial y},\qquad
E_y=-\frac{1}{j\omega\varepsilon}\frac{\partial H_z}{\partial x},
\label{eq:e-from-h}
\end{equation}
and substitution gives the scalar divergence-form wave equation
\begin{equation}
\frac{\partial}{\partial x}\left(\frac{1}{\varepsilon}\frac{\partial H_z}{\partial x}\right)
+\frac{\partial}{\partial y}\left(\frac{1}{\varepsilon}\frac{\partial H_z}{\partial y}\right)
+\omega^2\mu H_z=0.
\label{eq:hz-divergence}
\end{equation}
Ceviche implements Eq.~\eqref{eq:hz-divergence} with paired forward and backward finite-difference matrices on a staggered grid and solves the resulting sparse linear system~\cite{hughes2019forward}. At material boundaries, adjacent cell-centered permittivities are arithmetically averaged to the staggered electric-field locations before inversion; the implementation adds $10^{-5}$ as a numerical regularizer. The full grid uses 30-cell stretched-coordinate PMLs on all four sides~\cite{shin2012pml}. A transversely uniform, unit-amplitude magnetic-current line in the substrate excites normal incidence.

In a homogeneous, nonmagnetic material, Eq.~\eqref{eq:hz-divergence} reduces to $\nabla^2H_z+k^2H_z=0$, where
\begin{equation}
k=k_0\sqrt{\varepsilon_r},\qquad
\lambda_m=\frac{\lambda_0}{\sqrt{\varepsilon_r}}.
\label{eq:medium-wave}
\end{equation}
For a square grid with $\Delta x=\Delta y=\Delta$, the standard second-order finite-difference stencil has the numerical dispersion relation
\begin{equation}
\frac{4}{\Delta^2}\sin^2\left(\frac{k_{x,\mathrm{num}}\Delta}{2}\right)
+\frac{4}{\Delta^2}\sin^2\left(\frac{k_{y,\mathrm{num}}\Delta}{2}\right)=k^2.
\label{eq:fdfd-dispersion}
\end{equation}
For propagation along a grid axis, this gives
\begin{equation}
k_{\mathrm{num}}=\frac{2}{\Delta}\sin^{-1}\left(\frac{k\Delta}{2}\right),\qquad
\delta_k=\left(\frac{k_{\mathrm{num}}}{k}-1\right)\times100\%.
\label{eq:axis-error}
\end{equation}
For propagation at $45^\circ$, where $k_x=k_y=k/\sqrt{2}$, the corresponding magnitude is
\begin{equation}
k_{\mathrm{num},45^\circ}=\frac{2\sqrt{2}}{\Delta}\sin^{-1}\left(\frac{k\Delta}{2\sqrt{2}}\right).
\label{eq:diagonal-error}
\end{equation}

The table below evaluates Eq.~\eqref{eq:axis-error} at the simulated $\lambda_0=500$\,nm and $\Delta=10$\,nm. The material wavelength decreases from 345.0\,nm at $\varepsilon_r=2.1$ to 125.0\,nm at $\varepsilon_r=16$, corresponding to 34.50 to 12.50 grid points per wavelength. The axial relative numerical-wavenumber error therefore increases from 0.1387\% to 1.0839\%. At $\varepsilon_r=16$, Eq.~\eqref{eq:diagonal-error} gives 0.534\% for $45^\circ$ propagation. Because the reference problem is solved directly in the frequency domain, there is no time step, Courant factor, or temporal numerical-dispersion term.

\begin{table}[htbp]
\centering
\caption*{\textbf{Homogeneous-medium FDFD numerical dispersion at $\lambda_0=500$\,nm and $\Delta=10$\,nm.} The last column is the axial relative numerical-wavenumber error from Eq.~\eqref{eq:axis-error}.}

\small
\begin{tabular}{rrrr}
\toprule
$\varepsilon_r$ & $\lambda_m$ (nm) & Grid points per $\lambda_m$ & $\delta_k$ (\%)\\
\midrule
2.1 & 345.0 & 34.50 & 0.1387\\
4 & 250.0 & 25.00 & 0.2651\\
6 & 204.1 & 20.41 & 0.3991\\
9 & 166.7 & 16.67 & 0.6018\\
12 & 144.3 & 14.43 & 0.8069\\
16 & 125.0 & 12.50 & 1.0839\\
\bottomrule
\end{tabular}
\end{table}

These values are theoretical bulk-dispersion estimates for a homogeneous medium. They do not quantify additional error from material-interface averaging and staircasing, the PML, finite-domain and source effects, or the sparse linear solve. They also are not an empirical continuum-error measurement for the complete heterogeneous structures. 

\clearpage\subsection*{Separating reference and surrogate errors}
The bulk numerical-wavenumber estimates in Equations~\eqref{eq:axis-error} and~\eqref{eq:diagonal-error} characterize the finite-difference reference. They do not measure the relative complex-field error of the surrogate. Let $H_z^{\mathrm{FDFD}}$ denote the discrete reference and $H_z^{\mathrm{cont}}$ the continuum solution sampled on the same retained grid. The error relative to that continuum solution satisfies

\[\|\widehat H_z-H_z^{\mathrm{cont}}\|_2\leq\|\widehat H_z-H_z^{\mathrm{FDFD}}\|_2+\|H_z^{\mathrm{FDFD}}-H_z^{\mathrm{cont}}\|_2.\]

The reported metric $e$ normalizes the first term on the right by $\|H_z^{\mathrm{FDFD}}\|_2$. It therefore measures the surrogate's discrepancy from the discrete solver. The second term is the reference-solution error, which is not established by the present comparisons. A surrogate can approximate a discrete reference more closely than that reference approximates the continuum solution.

Numerical-wavenumber error and relative field error cannot be used interchangeably or combined as percentages. In particular, the dispersion estimates do not define an error floor for surrogate-to-FDFD agreement. A grid-convergence study of the heterogeneous structures would be needed to assess the reference-field error. The present results establish fidelity to the specified FDFD discretization.

\clearpage

\clearpage
\section*{Supporting Note 8. Transmitted-field and power evaluation}
\textbf{Evaluation set and transmitted power}

The metalens test set contains 50 inverse-designed geometries evaluated at 139 relative permittivities, giving 6,950 field samples. The same trained weights are used for every sample. Material reassignment changes the optical response, so these samples test field prediction for metalens-derived geometries rather than 6,950 independently optimized lenses.

The output line lies 0.5$\lambda_0$ above each structure. The normalized magnetic field is $t(x)=H_\mathrm{out}(x)/H_\mathrm{inc}(x)$, where the incident field is reconstructed from substrate rows and propagated through vacuum to define a common phase reference. Electric fields are recovered using Amp\`ere's law in the uniform output region. The time-averaged real Poynting flux is integrated along the line to obtain transmitted power per unit out-of-plane length.

For reference power $P$ and predicted power $P_\mathrm{pred}$, the relative power error is $\lvert P_\mathrm{pred}-P\rvert/\lvert P\rvert$. The reported statistic is the median of this error over all 6,950 samples. The 0.7\%--0.9\% interval spans the three model-specific medians for F-FNO, U-Net, and DilResNet. Their Pearson correlations span 0.994--0.996. FNO gives a median error of 3.7\% and a correlation of 0.968. Table~S5 lists the values for each architecture.

\textbf{Phase reference and unit conversion}

Phase consistency between independently generated datasets is checked by fitting the $k_x=0$ component of the field in the uniform substrate rows to $c(y)=A\exp(-jky)+B\exp(+jky)$, where $k$ is the substrate wavenumber and the forward-propagating term defines the incident phase. This fit gives an incident phase of 15.35$^\circ$, a circular standard deviation of 0.45$^\circ$, and a relative fit residual of 1.5 $\times$ $10^{-5}$ for the training batches. Across the 14 evaluation datasets included in this check, the mean incident phase agrees within 0.2$^\circ$ and the circular standard deviation is at most 0.65$^\circ$. These checks establish a common incident-phase reference for the compared fields.

The Figure 6 example was selected using the median of the per-sample field error averaged across the four architectures. Panel (c) shows the absolute wrapped phase difference from the FDFD reference as a percentage of one cycle (1\% = 3.6$^\circ$). The FNO maximum of 3.26\% corresponds to 11.7$^\circ$, and errors below 1.3\% for F-FNO, U-Net, and DilResNet correspond to values below 4.7$^\circ$.

\textbf{Interpretation of phase and power diagnostics}

The residual phase diagnostic $\exp(-\sigma_\phi^2)$ uses $\sigma_\phi$ in radians after removal of a constant phase offset. It excludes amplitude error and does not directly evaluate focal intensity. The power statistics quantify flux through the output line. They do not measure focusing efficiency or establish relative phase consistency between separately predicted elements of an assembled optical device.

\clearpage
\section*{Supporting Note 9. Material and structural-scale evaluation}
\textbf{Refractive-index-resolved evaluation}

Table S1 reports medians of the per-sample relative complex-field error $e$ at $\varepsilon_r=2.1$, 4.0, 6.2, and 12.7. Their refractive indices are $n=\sqrt{\varepsilon_r}\simeq1.45$, 2.00, 2.49, and 3.56. These are assigned real dielectric values in a lossless, nondispersive model. Each value is represented in training. The comparison evaluates independently constructed geometries at these permittivities and does not test prediction at withheld scalar material values.

Table S1 covers the five single-material evaluation sets. The held-out set is shown separately from the four independent structure types. The two-material set is excluded because each sample contains two assigned design permittivities. Each table cell is a median at one fixed permittivity. It differs from the mean over all sampled permittivities in Table 1.

\textbf{Structural-scale groups}

The scale sweep contains 14 independently generated geometries at each of the eight generator frequencies below. Each geometry is evaluated at 138 permittivities from 2.2 to 15.9 in increments of 0.1. Every group therefore contains 1,932 field samples and the sweep contains 112 geometries and 15,456 samples. Repeated permittivity assignments increase the number of field samples, not the number of independent geometries.

{
\small\noindent
\begin{tabular*}{\linewidth}{@{\extracolsep{\fill}}lrr@{}}\toprule Position relative to training & Generator frequencies ($\mu$m$^{-1}$) & Groups\\\midrule Coarser & 5.0, 7.5 & 2\\Within interval & 12.5, 17.5, 22.5, 27.5 & 4\\Finer & 32.5, 35.0 & 2\\\bottomrule\end{tabular*}
\par}

Generator frequencies of 10--30\,$\mu$m$^{-1}$ are sampled continuously during training. The four interior test groups sample this interval, while the other groups test coarser or finer mask statistics. Generator frequency is defined by the noise scale and grid spacing. The measured feature size is the median horizontal material chord length multiplied by the 10\,nm grid spacing. Thresholding and connectivity prevent a unique conversion between these quantities.

\textbf{Aggregation and interpretation}

Figure 7(a) shows the group medians and interquartile bands for the seven displayed groups from 7.5 to 35.0\,$\mu$m$^{-1}$, with relative errors expressed as percentages. Figure 7(b) shows the medians and interquartile ranges of all four architectures for the 5.0\,$\mu$m$^{-1}$ group, including medians of 2.7\% for DilResNet and 4.9\% for U-Net. Its masks have a mean per-mask median chord length of approximately 177\,nm. This group is displayed separately from the seven-group curves.

Each group statistic pools geometries and permittivities. Its interquartile range describes variation among field samples and does not estimate uncertainty across training runs. Because the spatial-scale groups contain different geometries, this sweep is not a controlled rescaling of one fixed device. It is also distinct from the simultaneous wavelength and geometry scaling in Figures S1 and S2. All architectures retain the same trained weights throughout these evaluations.

\clearpage
\section*{Supporting Note 10. Nearest training samples for the test structures}

A nearest-sample comparison tests whether the predicted fields are more accurate than directly reusing the field of one training sample selected by similarity in its permittivity map. For each of the five test structures shown in main-text Figure~3, the nearest sample was identified using the mean pixel-wise $L_1$ distance over the design layer,
\begin{equation}
d_1=\frac{1}{N_{\mathrm{d}}}\sum_{(x,y)\in\Omega_{\mathrm{d}}}\left|\varepsilon_r^{\mathrm{test}}(x,y)-\varepsilon_r^{\mathrm{train}}(x,y)\right|,
\label{eq:nearest}
\end{equation}
where $\varepsilon_r^{\mathrm{test}}$ and $\varepsilon_r^{\mathrm{train}}$ are the test and training relative-permittivity maps, $\Omega_{\mathrm{d}}$ is the $13\times64$-pixel design layer, and $N_{\mathrm{d}}=832$ is its number of pixels. The substrate and air regions outside the design layer are excluded. The search is exhaustive over the 560,113 training samples and allows both the training geometry and its assigned permittivity to vary.

The nearest samples identified in Figure~S16 have distances of $d_1=2.63$, 2.92, 2.39, 2.46, and 2.95 for the manually defined pattern, inverse-designed binary structure, inverse-designed metalens, aperiodic bar array, and two-material freeform pattern, respectively. For comparison, the mean design-layer permittivity contrast $|\varepsilon_r-1|$ ranges from 2.6 to 5.3 across these examples. The nearest training structures differ from the test structures in material occupancy over 34\%, 32\%, 45\%, 27\%, and 35\% of the design-layer pixels, respectively.

The adopted distance depends on both material occupancy and dielectric contrast. For example, the nearest training sample for the manually defined HYU pattern at $\varepsilon_r=10$ has an assigned permittivity of 2.1. Selection under this distance therefore does not imply the closest geometric shape at a fixed material value.

Directly reusing the FDFD fields of the selected training samples gives relative complex-field errors of 22\%, 72\%, 36\%, 77\%, and 31\% in the same order. The corresponding DilResNet errors are 1.8\%, 3.0\%, 4.2\%, 2.5\%, and 1.3\%. For the five illustrated structures, the DilResNet predictions are substantially more accurate than directly reusing the fields of the nearest training samples identified by the adopted permittivity-map distance.

This comparison is restricted to five examples and one input-space distance. It does not identify the nearest training field in complex-field space or evaluate predictions formed by combining multiple training samples. It therefore supports an advantage over the specified nearest-sample baseline without establishing that similar fields are absent from the training set or determining how the network combines information learned from training samples.

\clearpage
\begin{table}[p]
\centering
\caption{\textbf{Field-prediction accuracy at representative refractive indices.} Median per-sample relative error $e$ (\%) at four assigned real permittivities, with $n=\sqrt{\varepsilon_r}$ rounded to two decimal places. All values are represented in training and are treated as lossless and nondispersive. The held-out set contains unseen geometry--permittivity combinations from the training pool. The other four sets contain independently constructed geometries. Each architecture uses one unchanged checkpoint. The two-material set is excluded because its samples contain two design permittivities. Bold entries identify all row minima at the displayed precision, including ties.}
\label{tab:materials}
\fontsize{9}{11}\selectfont\setlength{\tabcolsep}{3pt}\renewcommand{\arraystretch}{1.4}
\begin{tabular*}{\textwidth}{@{\extracolsep{\fill}}lrrrrrr@{}}
\toprule
Structure type & $n$ & $\varepsilon_r$ & FNO & F-FNO & U-Net & DilResNet\\
\midrule
Held-out combinations & 1.45 & 2.1 & 0.26 & 0.07 & \textbf{0.05} & \textbf{0.05}\\
 & 2.00 & 4.0 & 0.42 & 0.10 & \textbf{0.06} & \textbf{0.06}\\
 & 2.49 & 6.2 & 0.60 & 0.13 & \textbf{0.08} & 0.09\\
 & 3.56 & 12.7 & 1.17 & 0.28 & \textbf{0.19} & 0.23\\
\midrule
Aperiodic bar arrays & 1.45 & 2.1 & 0.96 & \textbf{0.09} & 0.10 & \textbf{0.09}\\
 & 2.00 & 4.0 & 3.43 & 0.29 & 0.31 & \textbf{0.23}\\
 & 2.49 & 6.2 & 8.02 & 0.79 & 1.06 & \textbf{0.66}\\
 & 3.56 & 12.7 & 23.7 & 5.37 & 5.95 & \textbf{3.99}\\
\midrule
Manually defined & 1.45 & 2.1 & 0.42 & 0.12 & \textbf{0.09} & 0.13\\
 & 2.00 & 4.0 & 1.33 & \textbf{0.23} & \textbf{0.23} & 0.24\\
 & 2.49 & 6.2 & 2.83 & 0.45 & 0.40 & \textbf{0.36}\\
 & 3.56 & 12.7 & 13.7 & 2.45 & 2.43 & \textbf{2.18}\\
\midrule
Inverse-designed metalenses & 1.45 & 2.1 & 0.73 & 0.13 & \textbf{0.09} & 0.16\\
 & 2.00 & 4.0 & 2.82 & 1.12 & \textbf{0.49} & 1.13\\
 & 2.49 & 6.2 & 6.47 & 3.27 & \textbf{1.86} & 3.30\\
 & 3.56 & 12.7 & 20.9 & 12.1 & \textbf{11.2} & 12.2\\
\midrule
Inverse-designed binary & 1.45 & 2.1 & 0.63 & \textbf{0.17} & 0.58 & 0.30\\
 & 2.00 & 4.0 & 2.39 & \textbf{0.49} & 1.53 & 0.59\\
 & 2.49 & 6.2 & 7.14 & \textbf{1.39} & 3.56 & 1.45\\
 & 3.56 & 12.7 & 27.2 & 11.5 & 16.9 & \textbf{7.32}\\
\bottomrule
\end{tabular*}
\end{table}
\clearpage
\begin{table}[p]
\centering
\caption{\textbf{Neural-network configurations and training cost.} Parameter counts and wall-clock training times for 200 epochs on one NVIDIA RTX~6000~Ada. FNO and F-FNO retain $9\times9$ and $20\times20$ Fourier modes, respectively. GN($g$) denotes group normalization with $g$ groups. All models use the common training procedure described in Methods. The comparison concerns these configurations rather than an architecture-wide optimum.}
\label{tab:architectures}
\small
\setlength{\tabcolsep}{4pt}
\renewcommand{\arraystretch}{1.2}
\resizebox{\textwidth}{!}{\begin{tabular}{lllcc}
    \toprule
    Network & Class & Hyperparameters & Parameters & Training \\
    \midrule
    FNO            & spectral            & modes $9\times9$, width 64, depth 4    & 5.33\,M & 37.7\,h \\
    F-FNO     & factorized spectral & modes $20\times20$, width 96, depth 8  & 5.98\,M & 183.4\,h \\
    U-Net    & hierarchical CNN    & base 32, 4 downsamplings, GN(8)        & 5.49\,M & 39.1\,h \\
    DilResNet & dilated CNN     & width 192, depth 8, dilation cycle 4   & 5.35\,M & 126.4\,h \\
    \bottomrule
  \end{tabular}
}
\end{table}
\clearpage

\begin{table}[p]
\centering
\caption{\textbf{Dataset composition and evaluation objectives.} Each sample denotes a geometry--permittivity--condition combination. Repeated simulations of one geometry at different permittivities are therefore distinct field samples, not independent shapes. The primary total combines the held-out set, five independent structure tests, and structural-scale sweep. The four single-material independent test sets use 139 permittivities from 2.1 to 15.9, and the structural-scale sweep uses 138 permittivities from 2.2 to 15.9, both in increments of 0.1.}
\label{tab:inventory}
\small\renewcommand{\arraystretch}{1.25}
\begin{tabularx}{\textwidth}{@{}p{34mm}p{38mm}rX@{}}
\toprule
Population & Geometry/condition coverage & Samples & Interpretation\\
\midrule
Training & From a pool of 5,001 geometries at 140 permittivities & 560,113 & Parameter fitting\\
Validation & Same pool and permittivity range & 70,015 & Checkpoint selection\\
Held-out combinations & Same pool with an independent split per permittivity & 70,012 & Unseen combinations of geometries and permittivities represented in training\\
Aperiodic bar arrays & 100 geometries $\times$ 139 permittivities & 13,900 & Prediction of independent structures\\
Manually defined & 19 geometries $\times$ 139 permittivities & 2,641 & Prediction of independent structures\\
Inverse-designed metalenses & 50 geometries $\times$ 139 permittivities & 6,950 & Transfer to optimized shapes\\
Inverse-designed binary & 50 geometries $\times$ 139 permittivities & 6,950 & Transfer to optimized shapes\\
Two-material freeform patterns & 5,000 geometry/material-pair samples & 5,000 & Unseen two-material arrangements\\
Structural-scale sweep & $8\times14$ geometries $\times$ 138 permittivities & 15,456 & Independent scales inside/outside training interval\\
Primary evaluation total & Held-out plus five structure types plus scale sweep & 120,909 & Seven evaluations\\
Fixed-grid wavelength & 10 wavelengths, 475--525\,nm & 19,600 & Source-wavelength change\\
Scaled-grid wavelength & 8 wavelengths, 300--700\,nm & 15,680 & Joint geometry/wavelength rescaling\\
\bottomrule
\end{tabularx}
\end{table}
\clearpage
\begin{table}[p]
\centering
\caption{\textbf{Conditional resource budgets for expanding the surrogate dataset.} Counts refer to stored field samples. The 2D training ranges span the four reported architectures and assume the same resolution, 80\% training fraction, 200 epochs, and throughput. Storage includes field outputs only in 32-bit precision. All expanded cases are planning estimates, and the 3D case requires a separate timing benchmark.}
\label{tab:resources}
\small
\setlength{\tabcolsep}{4pt}
\renewcommand{\arraystretch}{1.25}
\begin{tabularx}{\textwidth}{@{}p{37mm}rXX@{}}
\toprule
Case & Field samples & Output storage & Training per run\\
\midrule
Current 2D dataset & 700,140 & 45.9\,GB & 37.7--183.4 GPU-hours\\
2D scaling scenario & $10^6$ & 65.5\,GB & 54--262 GPU-hours\\
Expanded 2D scenario & $10^7$ & 655\,GB & 538--2620 GPU-hours\\
Dense 2D condition product & 252,050,400 & 16.5\,TB & 13,572--66,024 GPU-hours\\
3D vector-field example & $10^6$ & 25.2\,TB & Requires pilot measurements\\
\bottomrule
\end{tabularx}
\end{table}

\begin{table}[htbp]
\centering
\caption{\textbf{Transmitted-power prediction for the inverse-designed metalens set.} Median relative transmitted-power error, $|P_\mathrm{pred}-P|/|P|$, and Pearson correlation between predicted and FDFD reference power over all 6,950 samples (Supporting Note~8). Power is integrated from the real Poynting flux on the output line $0.5\lambda_0$ above each structure. Each architecture uses one validation-selected checkpoint.}
\label{tab:power}
\small
\begin{tabular}{lcc}
\toprule
Architecture & Median relative power error (\%) & Pearson correlation\\
\midrule
FNO & 3.67 & 0.968\\
F-FNO & 0.87 & 0.996\\
U-Net & 0.73 & 0.995\\
DilResNet & 0.80 & 0.994\\
\bottomrule
\end{tabular}
\end{table}

\begin{table}[htbp]
\centering
\caption{\textbf{Effect of the additional wave-equation residual on the average NMSE.} Each entry is $\overline{\mathrm{NMSE}}_{\mathrm{res}}/\overline{\mathrm{NMSE}}_{\mathrm{sup}}-1$, in percent, where $\overline{\mathrm{NMSE}}$ is the mean per-sample NMSE of a dataset, and the subscripts denote the residual-trained variant and the corresponding supervised model. Negative values indicate lower error with the residual term (Supporting Note~4 and Figure~S14).}
\label{tab:residual}
\small
\begin{tabular}{lrr}
\toprule
Evaluation set & FNO (\%) & U-Net (\%)\\
\midrule
Held-out combinations & $-3.2$ & $-7.2$\\
Aperiodic bar arrays & $+1.8$ & $-5.2$\\
Manually defined patterns & $+4.0$ & $+7.0$\\
Inverse-designed metalenses & $-4.0$ & $+0.7$\\
Inverse-designed binary structures & $+8.9$ & $-14.0$\\
Two-material freeform patterns & $+8.5$ & $-3.0$\\
Structural-scale sweep & $-5.7$ & $-9.5$\\
\bottomrule
\end{tabular}
\end{table}
\clearpage

\begin{figure}[p]
\centering
\includegraphics[width=\textwidth]{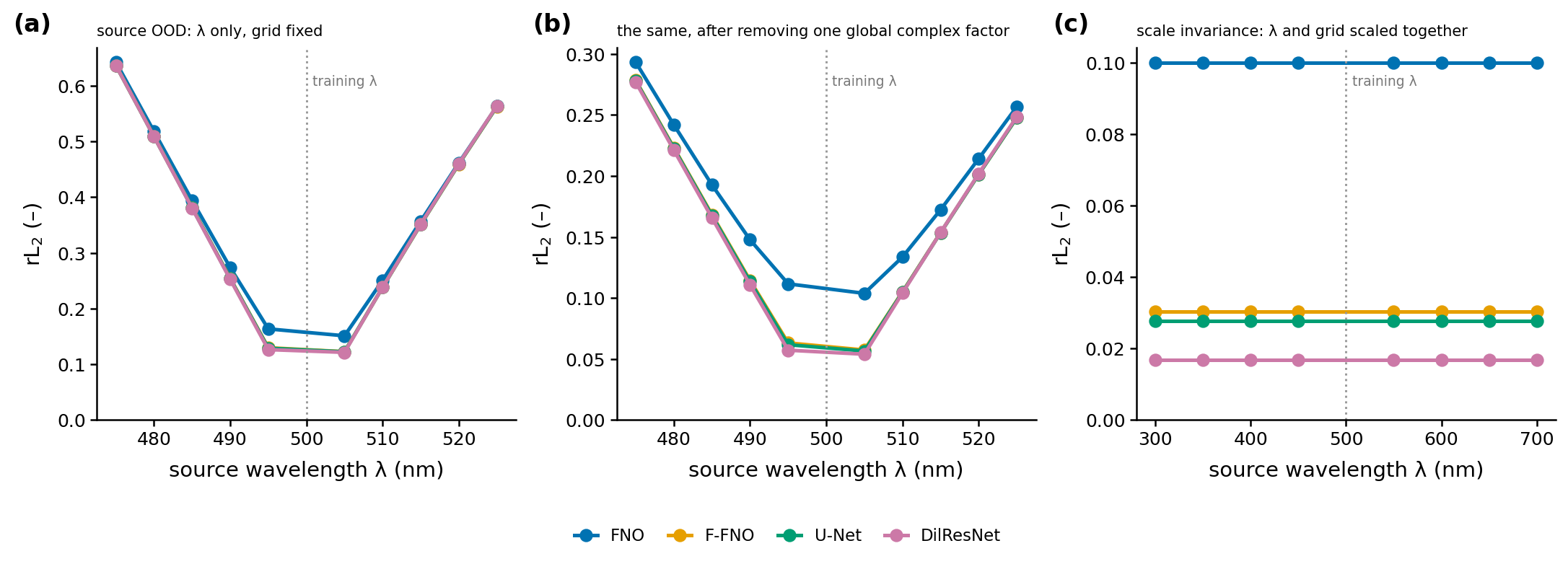}
\caption{\textbf{Wavelength variation at fixed and scaled physical dimensions.} Relative $L_2$ field error for (a) a fixed 10\,nm grid and physical geometry, (b) the same fixed-grid predictions after fitting and removing a global complex factor, and (c) a grid and geometry scaled with wavelength to maintain $\lambda_0/\Delta l=50$. The fixed-grid and scaled-grid sweeps contain 19,600 and 15,680 samples, respectively. Dotted vertical lines mark the training wavelength. Panel (c) preserves the dimensionless geometry and therefore evaluates scaling equivalence rather than broadband prediction of a fixed device.}
\label{fig:s1}
\end{figure}
\clearpage

\begin{figure}[p]
\centering
\includegraphics[width=\textwidth]{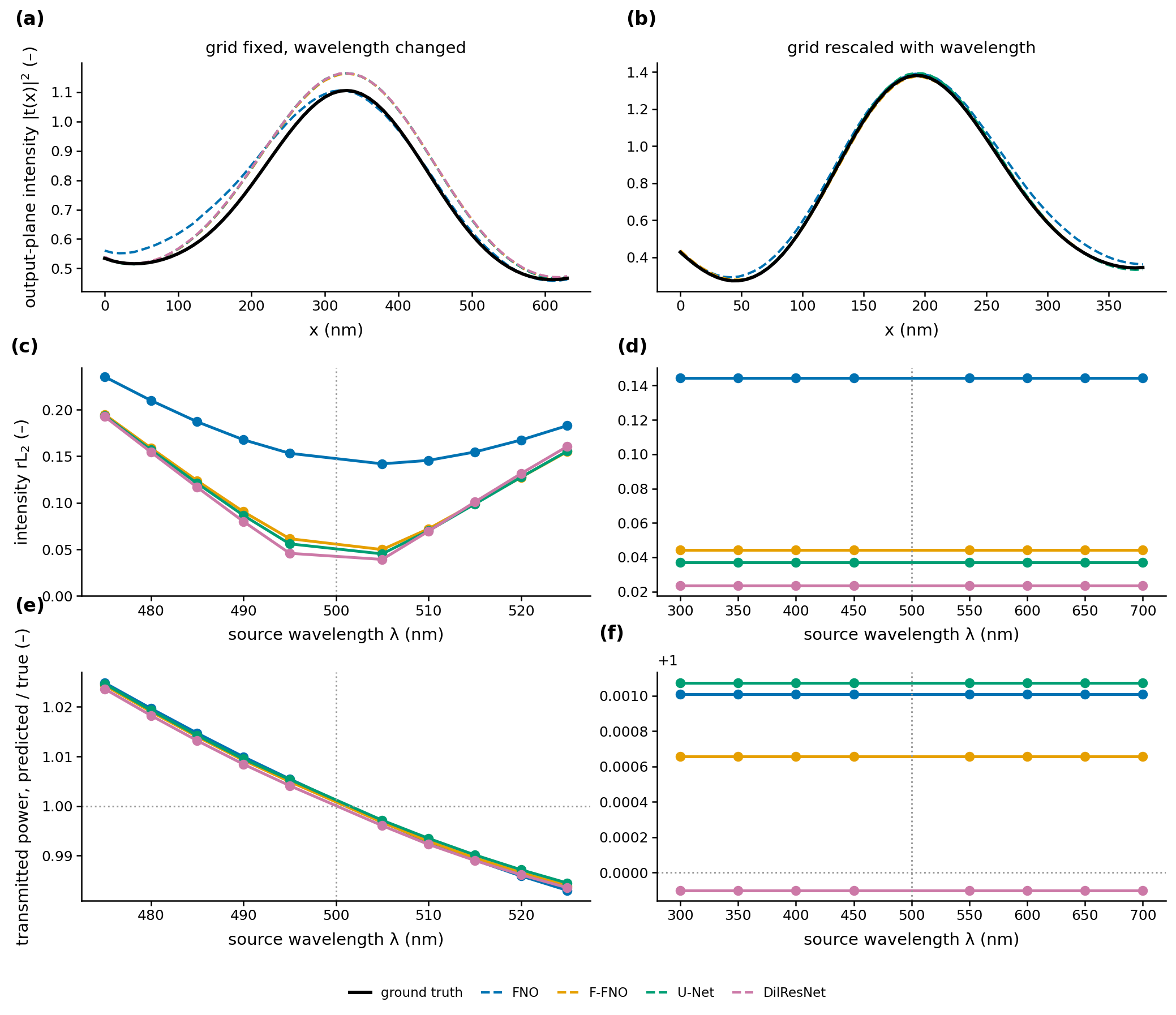}
\caption{\textbf{Optical-response profiles under wavelength variation.} The left column uses a fixed grid and physical geometry. The right column scales both grid spacing and geometry with wavelength. (a,b) Representative magnetic-field intensity profiles $|t(x)|^2$. (c,d) Relative $L_2$ error of the profiles. (e,f) Predicted-to-reference ratios of integrated transmitted power, calculated from the real Poynting flux. Black curves indicate numerical references, and colors distinguish the networks. The 35,280 samples are those evaluated in Figure~S1.}
\label{fig:s2}
\end{figure}
\clearpage

\begin{figure}[p]
\centering
{\small\bfseries Predicted field intensities}\par\vspace{1mm}
\includegraphics[width=\textwidth,height=190mm,keepaspectratio]{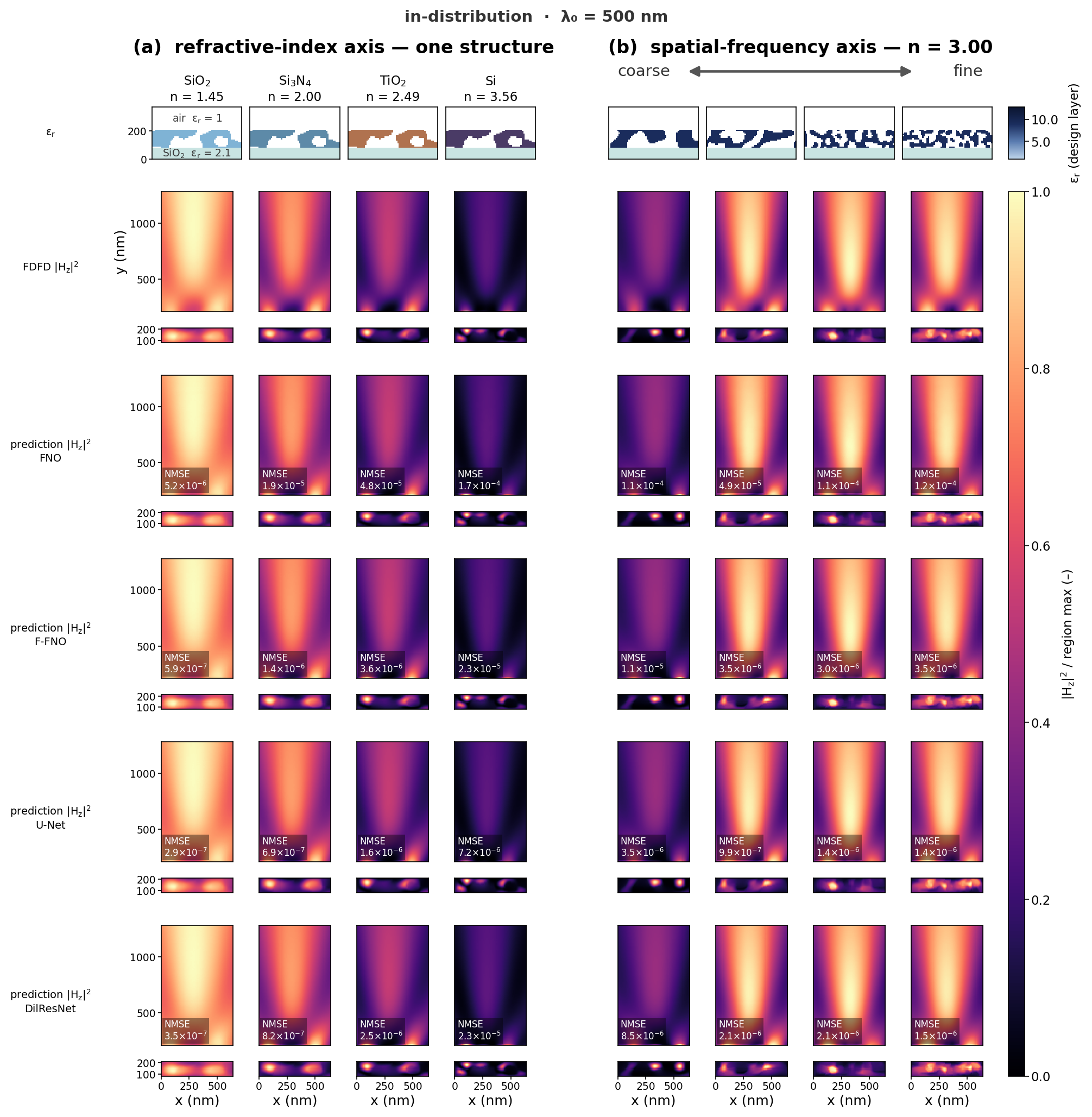}
\caption{\textbf{Refractive-index and structural-detail comparisons for all four architectures.} The examples in main-text Figures~4 and 5 are extended to FNO, F-FNO, U-Net, and DilResNet. The left block varies refractive index at fixed geometry, and the right block varies geometry at $n=3.0$. Rows show permittivity, reference intensity, and predicted intensity. The following page contains the corresponding absolute-error maps. Annotations report per-sample complex-field NMSE ($10^{-4}$ corresponds to a relative error of 1\%). Design and propagation regions use separate, matched reference/prediction display scales. All samples belong to the 70,012 held-out combinations.}
\label{fig:s3}
\end{figure}
\clearpage

\begin{figure}[p]
\ContinuedFloat
\centering
{\small\bfseries Absolute-error maps}\par\vspace{1mm}
\includegraphics[width=\textwidth,height=200mm,keepaspectratio]{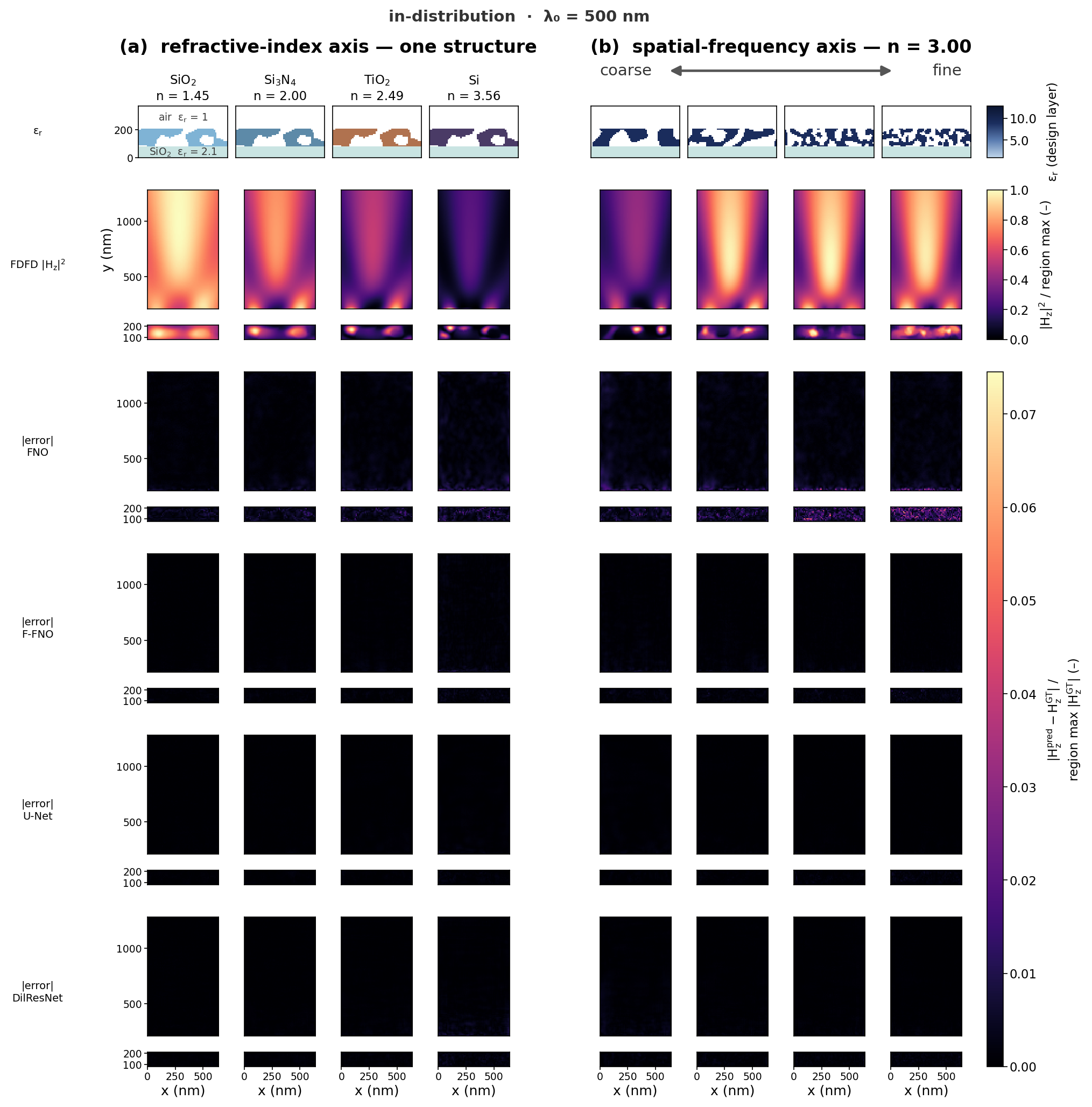}
\caption[]{\textbf{Continued.} Absolute-error maps for the Figure~S3 samples, using the same field and display conventions as the preceding page.}
\end{figure}
\clearpage

\begin{figure}[p]
\centering
{\small\bfseries Predicted field intensities}\par\vspace{1mm}
\includegraphics[width=\textwidth,height=196mm,keepaspectratio]{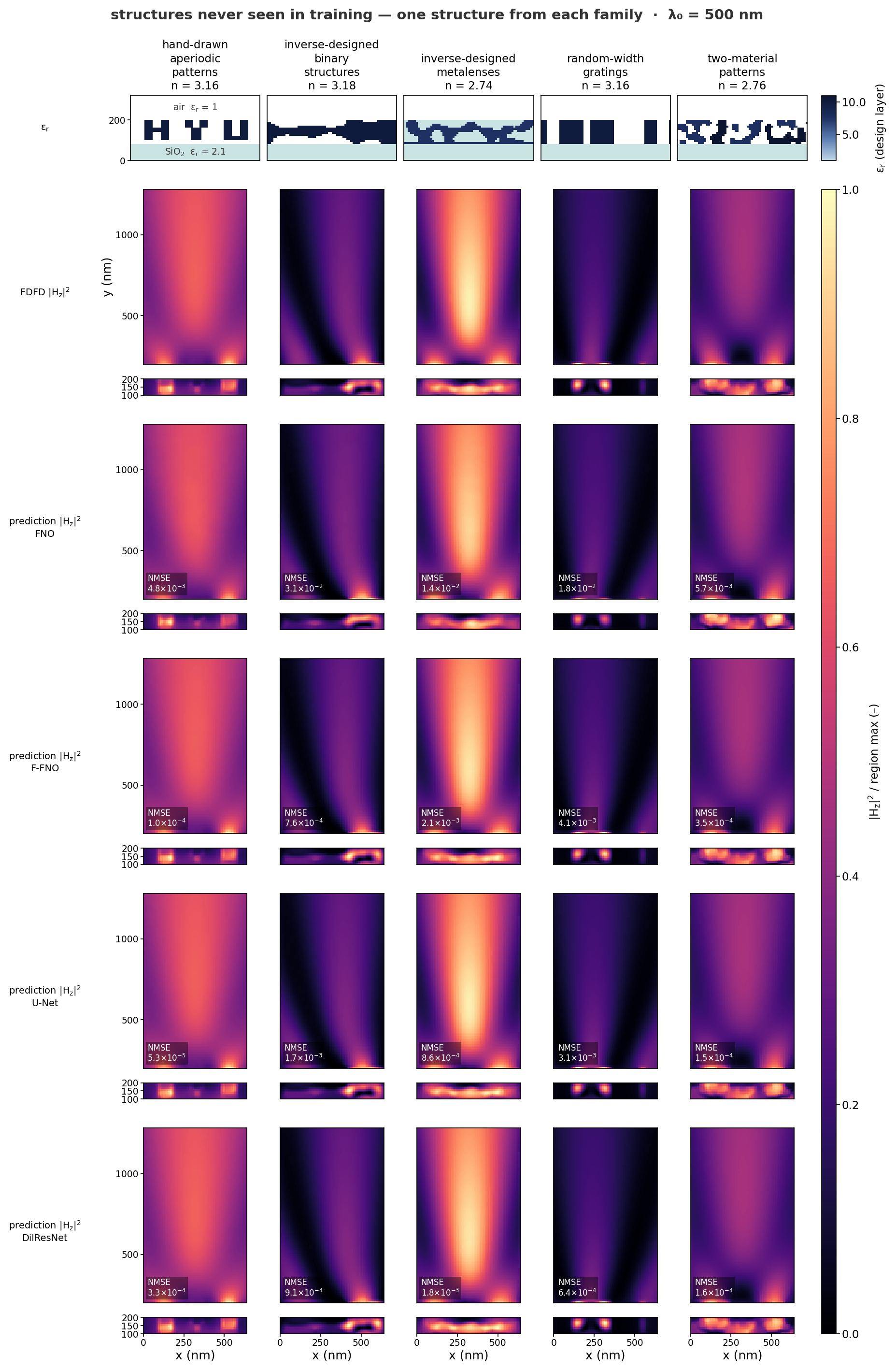}
\caption{\textbf{All-model predictions for independent nanostructures.} The five examples in main-text Figure~3 are shown for each architecture. Columns correspond to a manually defined pattern, an inverse-designed binary structure, an inverse-designed metalens, an aperiodic bar array, and a two-material freeform pattern. The lettering in the manually defined example lies in the simulated 2D plane. None of these structure types contributes to training or checkpoint selection. Predicted intensities are followed by absolute-error maps on the next page. Annotations give sample NMSE, and display normalization follows Figure~S3. Reference fields are FDFD solutions. Aggregate evaluations comprise 35,441 samples.}
\label{fig:s4}
\end{figure}
\clearpage

\begin{figure}[p]
\ContinuedFloat
\centering
{\small\bfseries Absolute-error maps}\par\vspace{1mm}
\includegraphics[width=\textwidth,height=204mm,keepaspectratio]{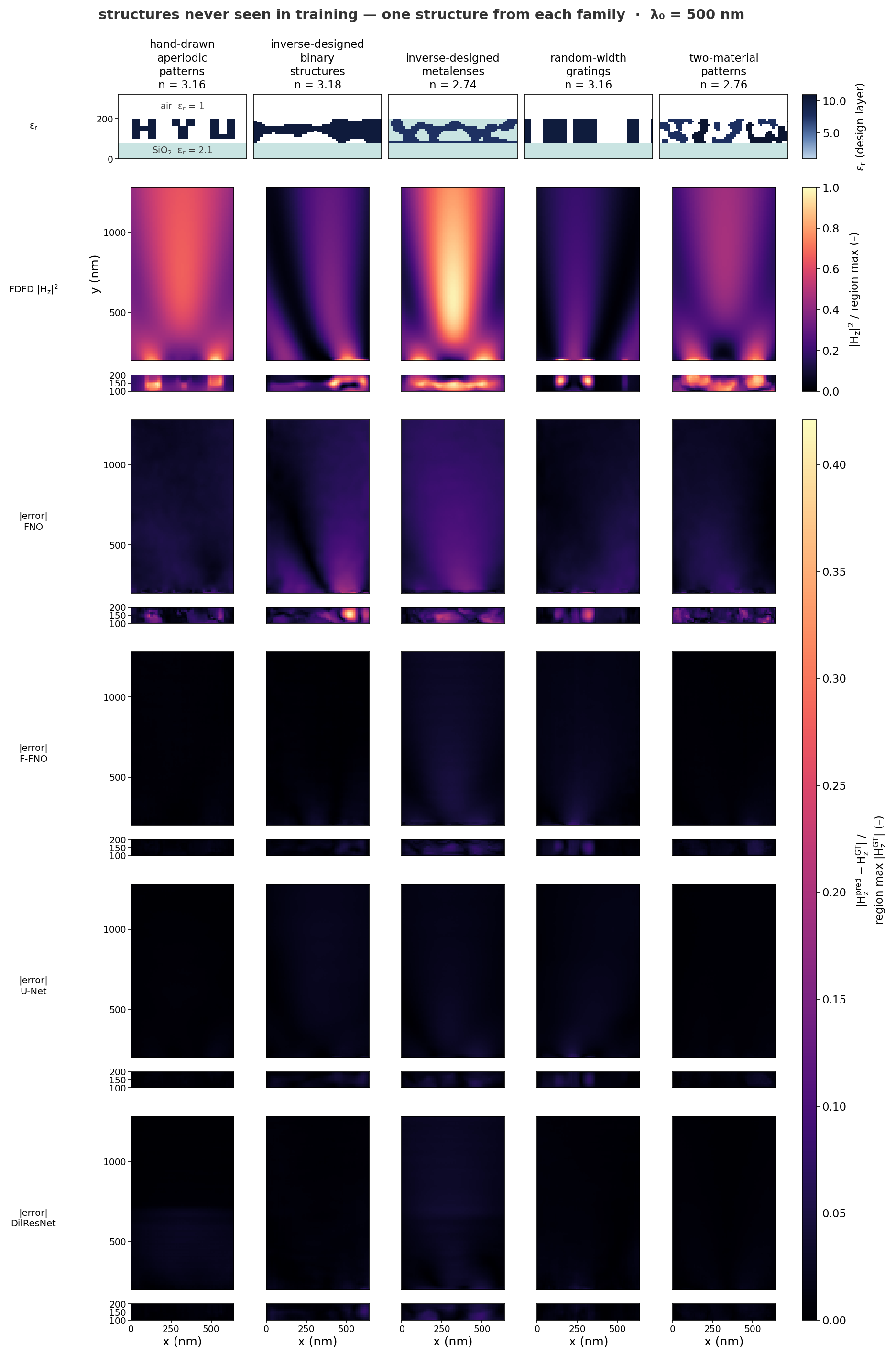}
\caption[]{\textbf{Continued.} Complex-field error maps, $|\widehat H_z-H_z|$, for the five preceding examples, normalized by the maximum reference $|H_z|$ in each displayed region. Because they are computed from the complex field, they include phase differences. The DilResNet maps appear in main-text Figure 3(d) with the same normalization and a 0--25\% color scale.}
\end{figure}
\clearpage

\begin{figure}[p]
\centering
{\small\bfseries Held-out geometry--permittivity combinations}\par\vspace{1mm}
\includegraphics[width=\textwidth]{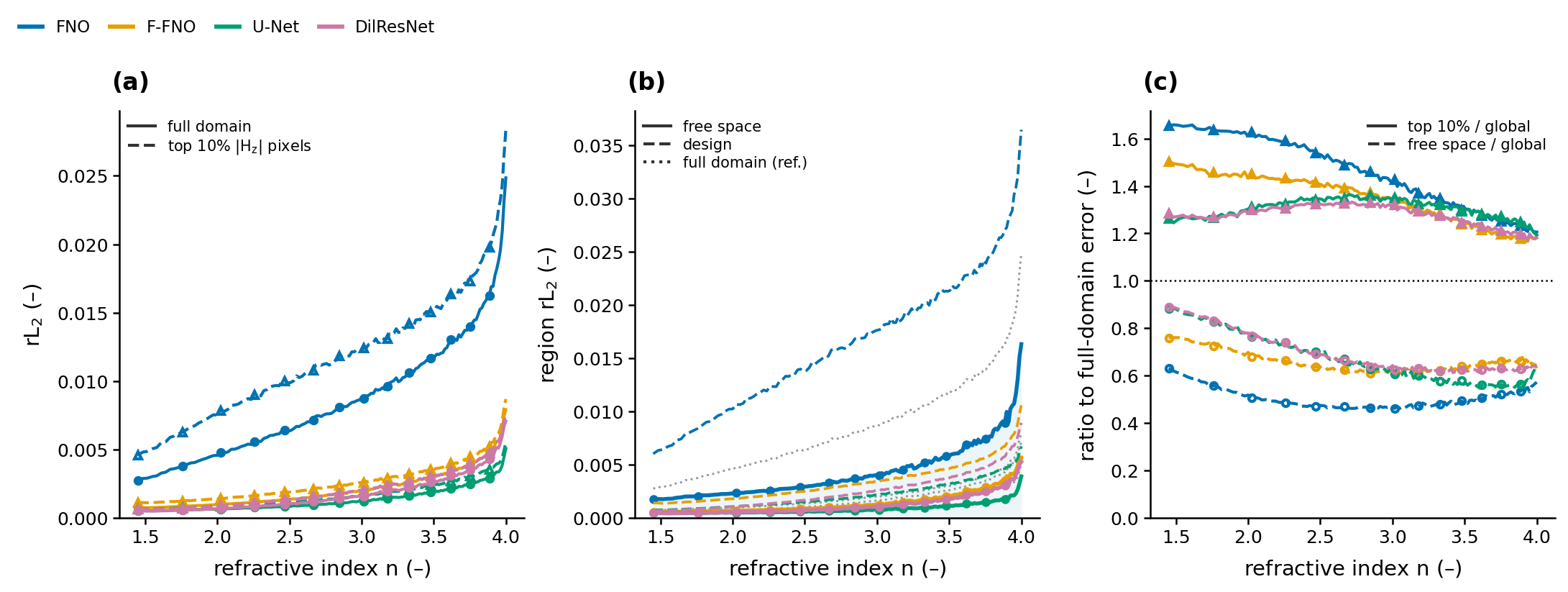}
\caption{\textbf{Spatial distribution of field-prediction errors.} For each evaluation set, (a) full-domain relative $L_2$ error and error on the 10\% of pixels with the highest reference $|H_z|$. (b) free-space and design-region errors, with the full-domain curve repeated for comparison. (c) region/full error ratios. Curves resolve refractive-index dependence across the seven primary sets, totaling 120,909 samples. The regional metrics follow the definition in Supporting Note~3; each curve is the square root of the per-permittivity mean of the per-sample regional NMSE, and two-material samples are placed at $\varepsilon_A$. The ratios are not interpreted as fractions of total error energy.}
\label{fig:s5}
\end{figure}
\clearpage

\begin{figure}[p]
\ContinuedFloat
\centering
{\small\bfseries Aperiodic bar arrays}\par\vspace{1mm}
\includegraphics[width=\textwidth]{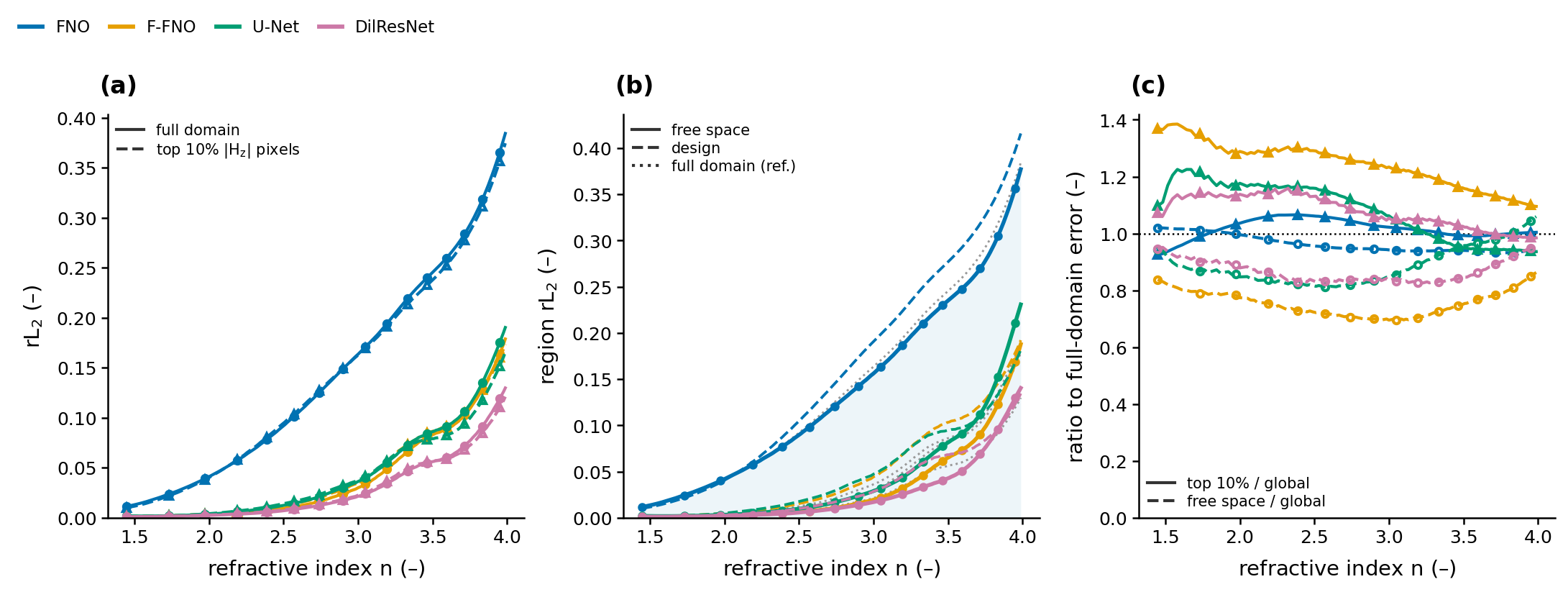}
\par\vspace{4mm}
{\small\bfseries Manually defined patterns}\par\vspace{1mm}
\includegraphics[width=\textwidth]{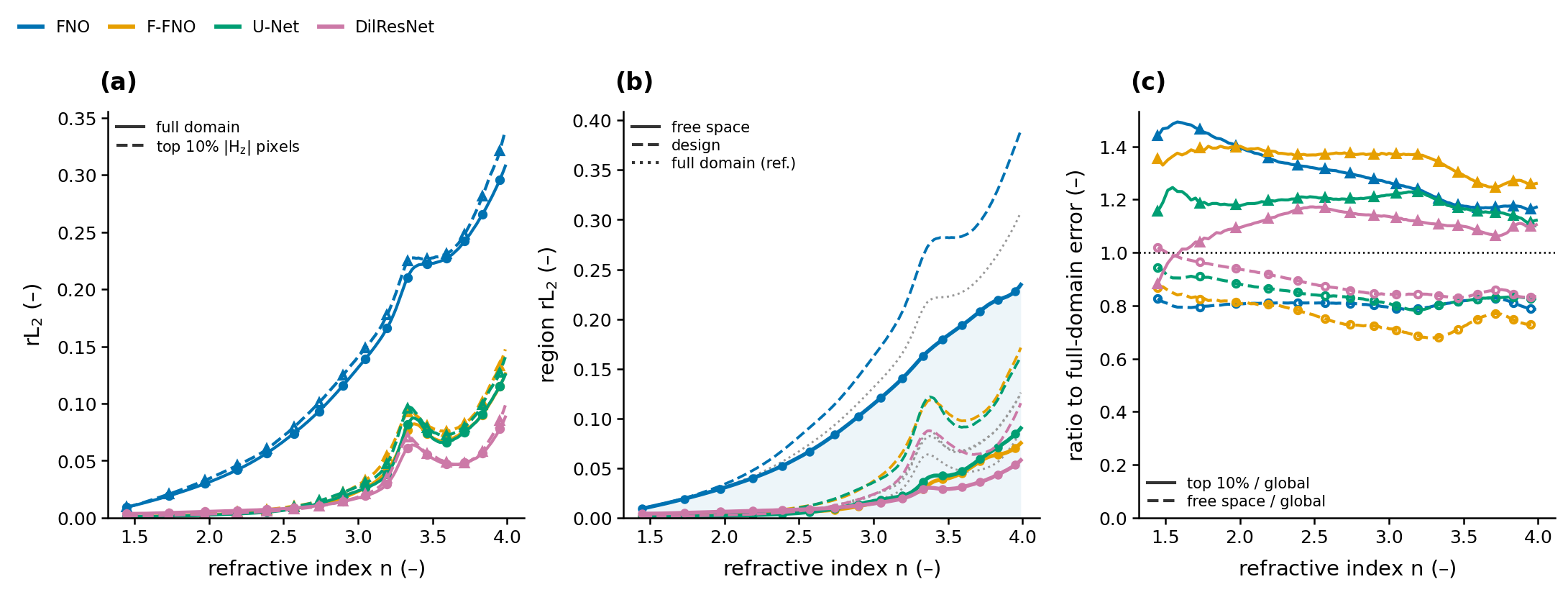}
\caption[]{\textbf{Continued.} Additional test sets using the same quantities and normalization as the preceding panels.}
\end{figure}
\clearpage

\begin{figure}[p]
\ContinuedFloat
\centering
{\small\bfseries Inverse-designed metalenses}\par\vspace{1mm}
\includegraphics[width=\textwidth]{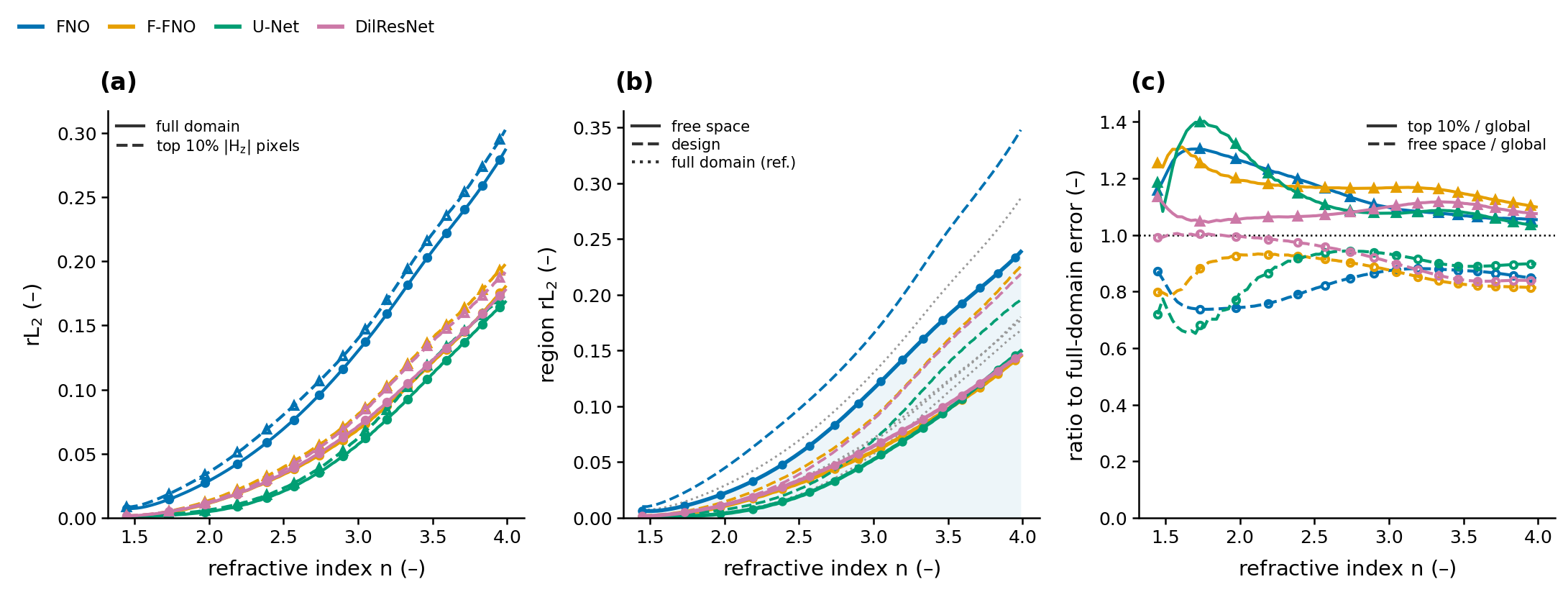}
\par\vspace{4mm}
{\small\bfseries Inverse-designed binary structures}\par\vspace{1mm}
\includegraphics[width=\textwidth]{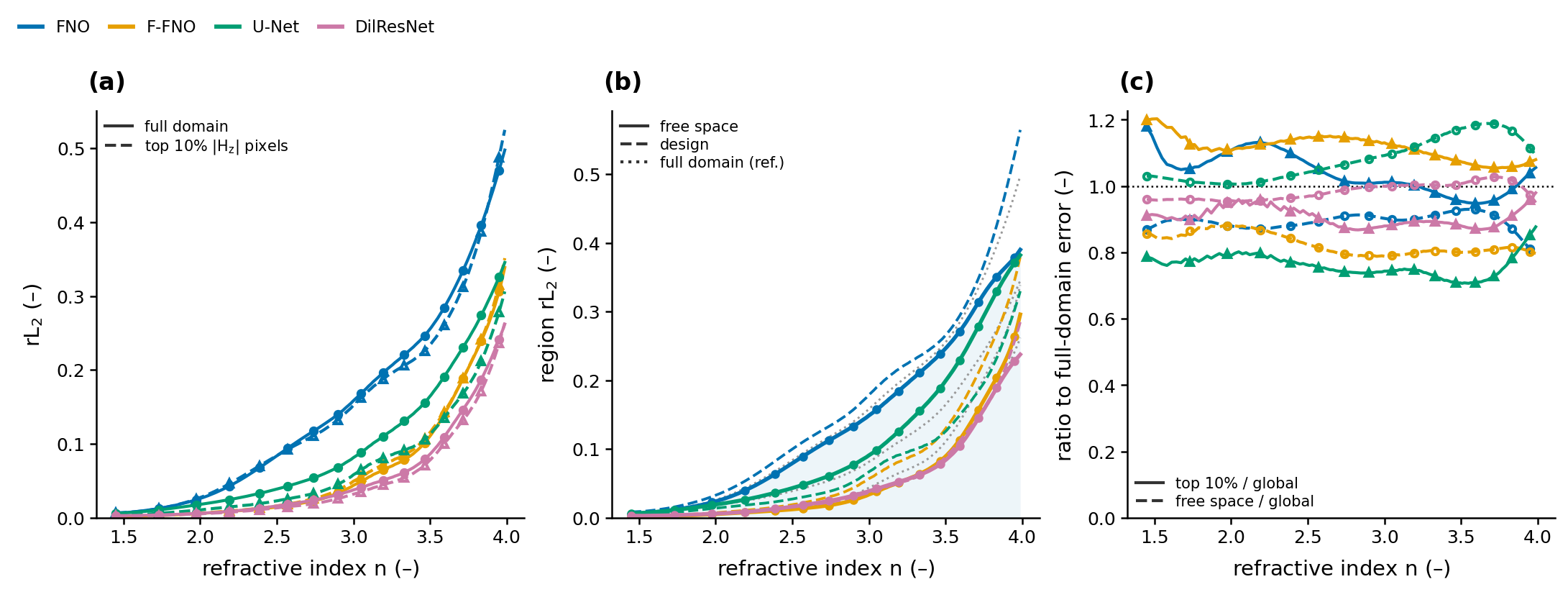}
\caption[]{\textbf{Continued.} Additional test sets using the same quantities and normalization as the preceding panels.}
\end{figure}
\clearpage

\begin{figure}[p]
\ContinuedFloat
\centering
{\small\bfseries Two-material freeform patterns}\par\vspace{1mm}
\includegraphics[width=\textwidth]{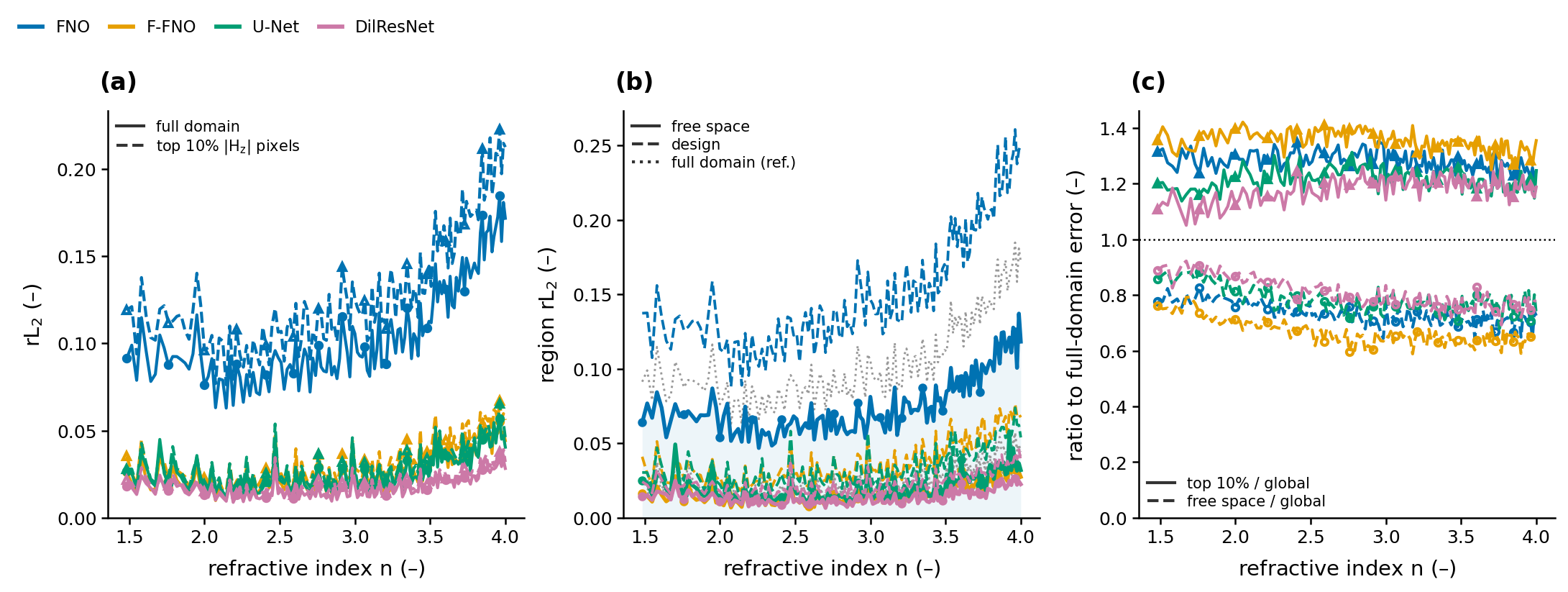}
\par\vspace{4mm}
{\small\bfseries Freeform structures across the structural-scale sweep}\par\vspace{1mm}
\includegraphics[width=\textwidth]{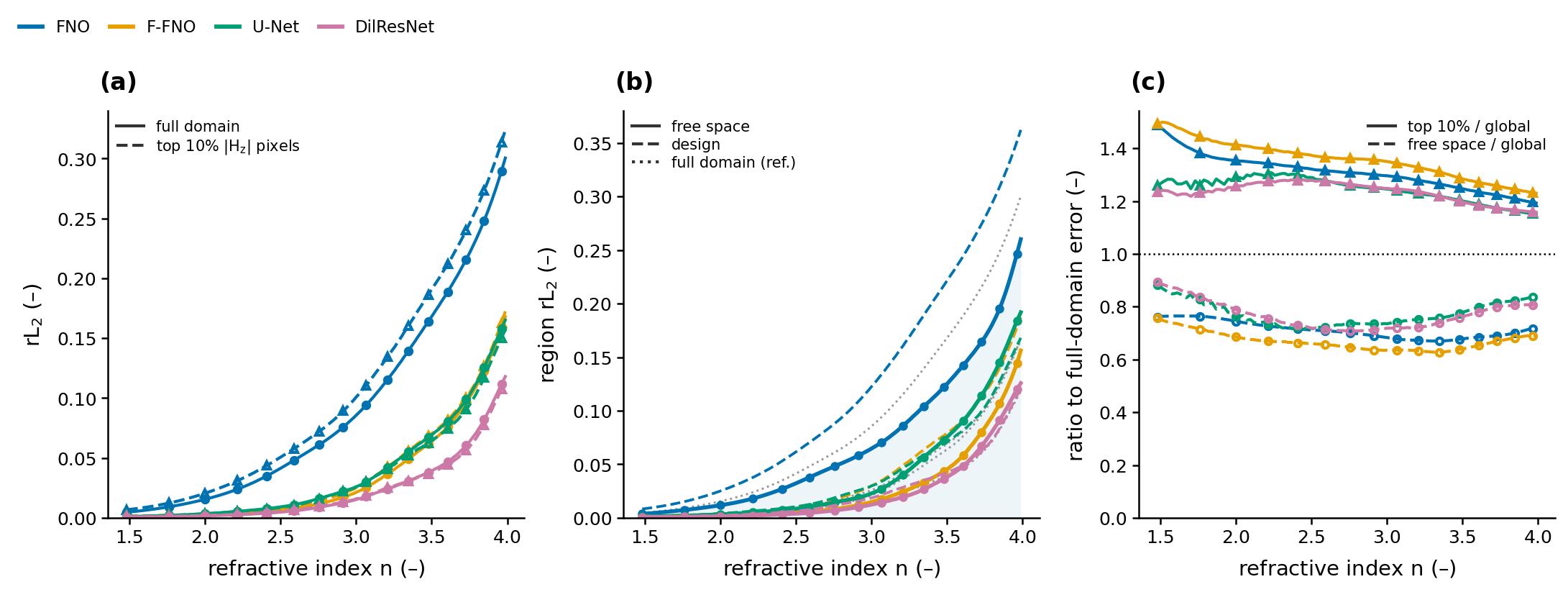}
\caption[]{\textbf{Continued.} Additional test sets using the same quantities and normalization as the preceding panels.}
\end{figure}
\clearpage

\begin{figure}[p]
\centering
{\small\bfseries Held-out geometry--permittivity combinations}\par\vspace{1mm}
\includegraphics[width=\textwidth]{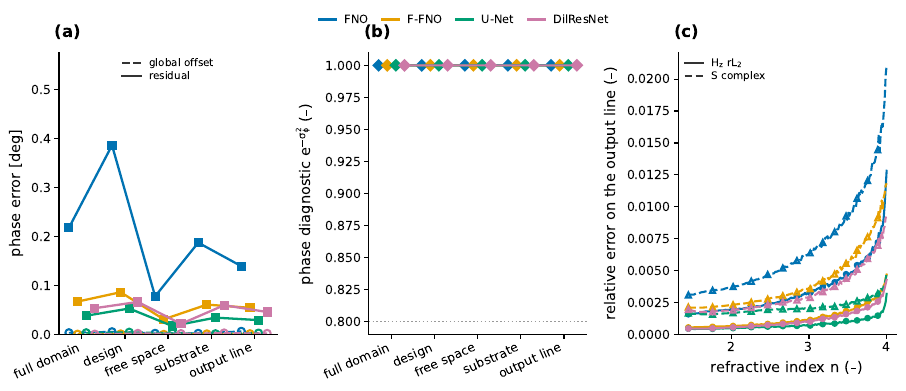}
\caption{\textbf{Phase agreement and output-line field fidelity.} For each evaluation set, (a) amplitude-weighted constant phase offset and residual phase error, (b) the phase-derived diagnostic $\exp(-\sigma_\phi^2)$, and (c) output-line relative $L_2$ errors for $H_z$ and the complex Poynting quantity. Residuals and phase diagnostics include all positions for the full domain and the output line; constant offsets, and the residuals of the other regions, use the 20\% of positions with the largest reference amplitude (Supporting Note~3). The dotted 0.8 level is a visual reference and does not represent measured focal quality. Offset removal within one sample does not validate relative phase between independently assembled elements.}
\label{fig:s6}
\end{figure}
\clearpage

\begin{figure}[p]
\ContinuedFloat
\centering
{\small\bfseries Aperiodic bar arrays}\par\vspace{1mm}
\includegraphics[width=\textwidth]{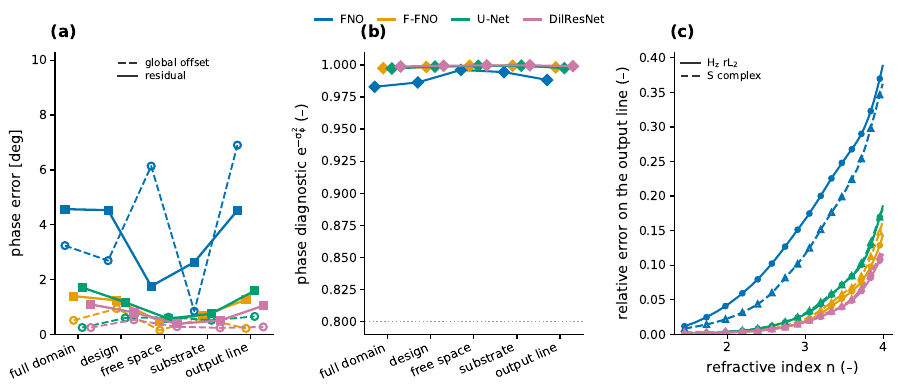}
\par\vspace{4mm}
{\small\bfseries Manually defined patterns}\par\vspace{1mm}
\includegraphics[width=\textwidth]{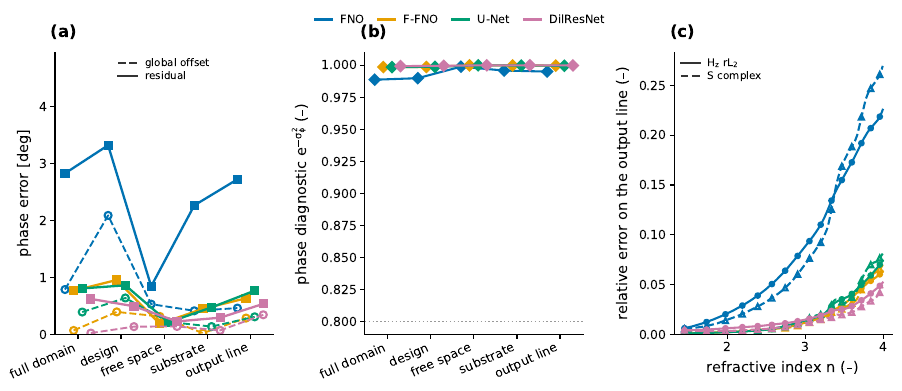}
\caption[]{\textbf{Continued.} Additional test sets using the same quantities and normalization as the preceding panels.}
\end{figure}
\clearpage

\begin{figure}[p]
\ContinuedFloat
\centering
{\small\bfseries Inverse-designed metalenses}\par\vspace{1mm}
\includegraphics[width=\textwidth]{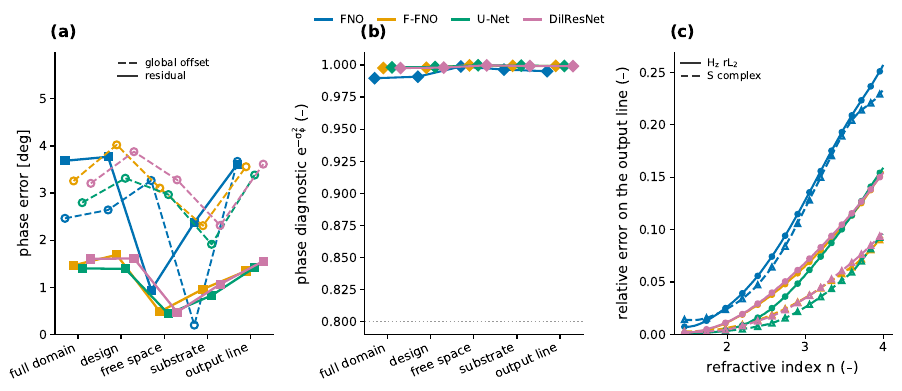}
\par\vspace{4mm}
{\small\bfseries Inverse-designed binary structures}\par\vspace{1mm}
\includegraphics[width=\textwidth]{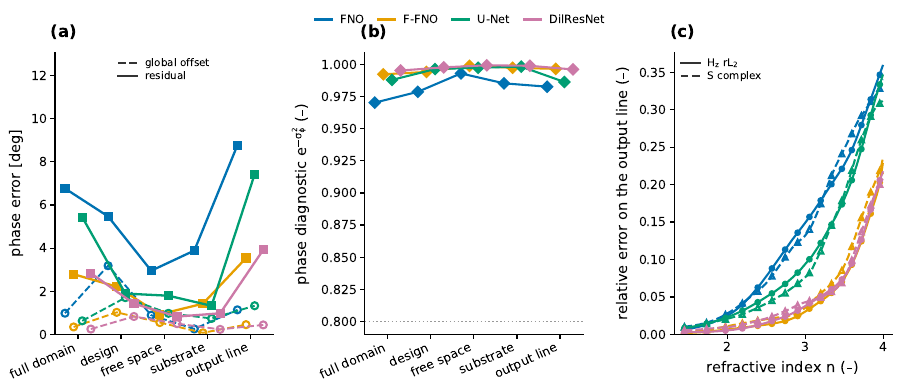}
\caption[]{\textbf{Continued.} Additional test sets using the same quantities and normalization as the preceding panels.}
\end{figure}
\clearpage

\begin{figure}[p]
\ContinuedFloat
\centering
{\small\bfseries Two-material freeform patterns}\par\vspace{1mm}
\includegraphics[width=\textwidth]{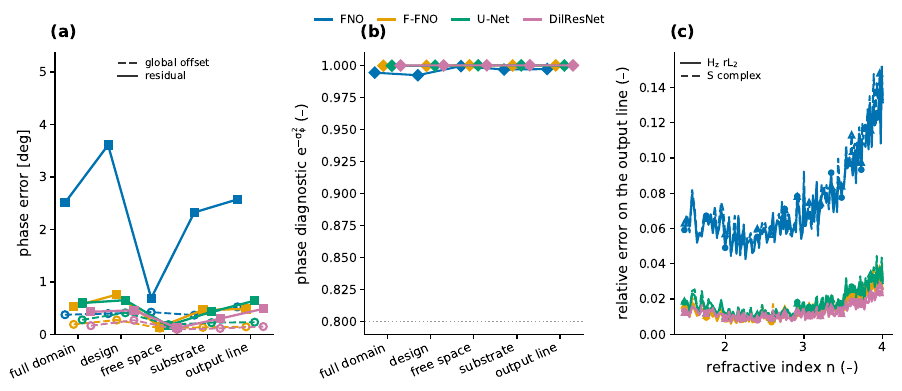}
\par\vspace{4mm}
{\small\bfseries Freeform structures across the structural-scale sweep}\par\vspace{1mm}
\includegraphics[width=\textwidth]{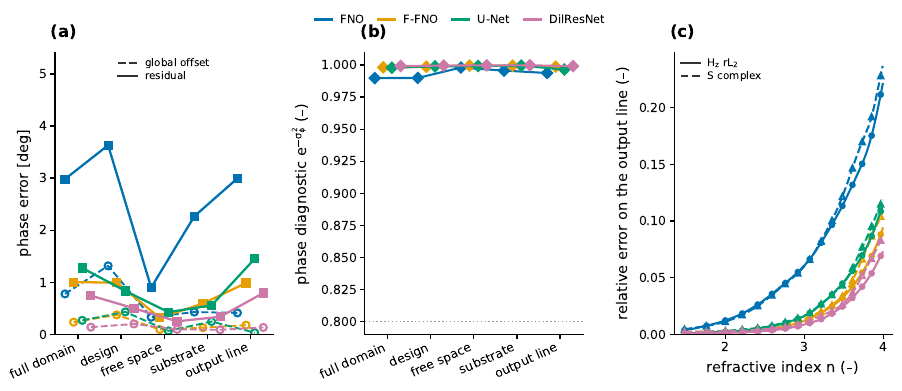}
\caption[]{\textbf{Continued.} Additional test sets using the same quantities and normalization as the preceding panels.}
\end{figure}
\clearpage

\begin{figure}[p]
\centering
{\small\bfseries Held-out geometry--permittivity combinations}\par\vspace{1mm}
\includegraphics[width=\textwidth]{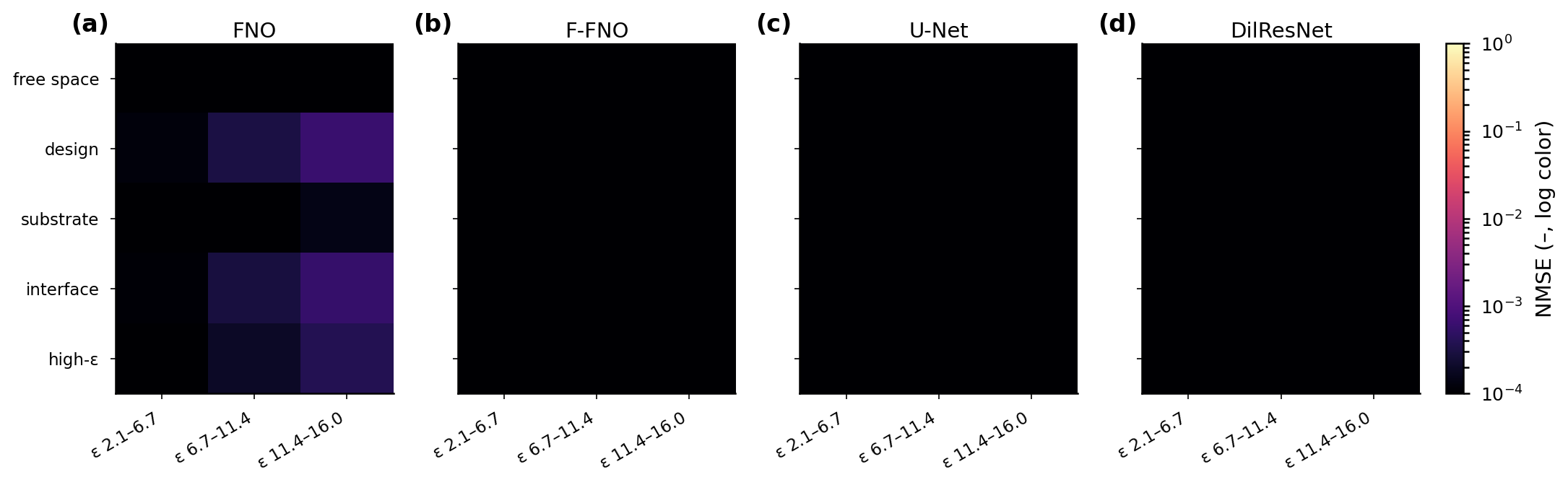}
\caption{\textbf{Field errors by spatial region and permittivity interval.} NMSE values for FNO, F-FNO, U-Net, and DilResNet are resolved over the indicated free-space, design, substrate, interface, and high-permittivity subsets. Columns represent three permittivity intervals bounded by the tertiles of the sampled permittivities, and each cell is the mean per-sample regional NMSE; two-material samples are placed at $\varepsilon_A$. A shared logarithmic scale of $10^{-4}$--1 is used, with lower errors assigned the minimum color value. Region definitions follow Supporting Note~3. The seven sets contain 120,909 samples.}
\label{fig:s7}
\end{figure}
\clearpage

\begin{figure}[p]
\ContinuedFloat
\centering
{\small\bfseries Aperiodic bar arrays}\par\vspace{1mm}
\includegraphics[width=\textwidth]{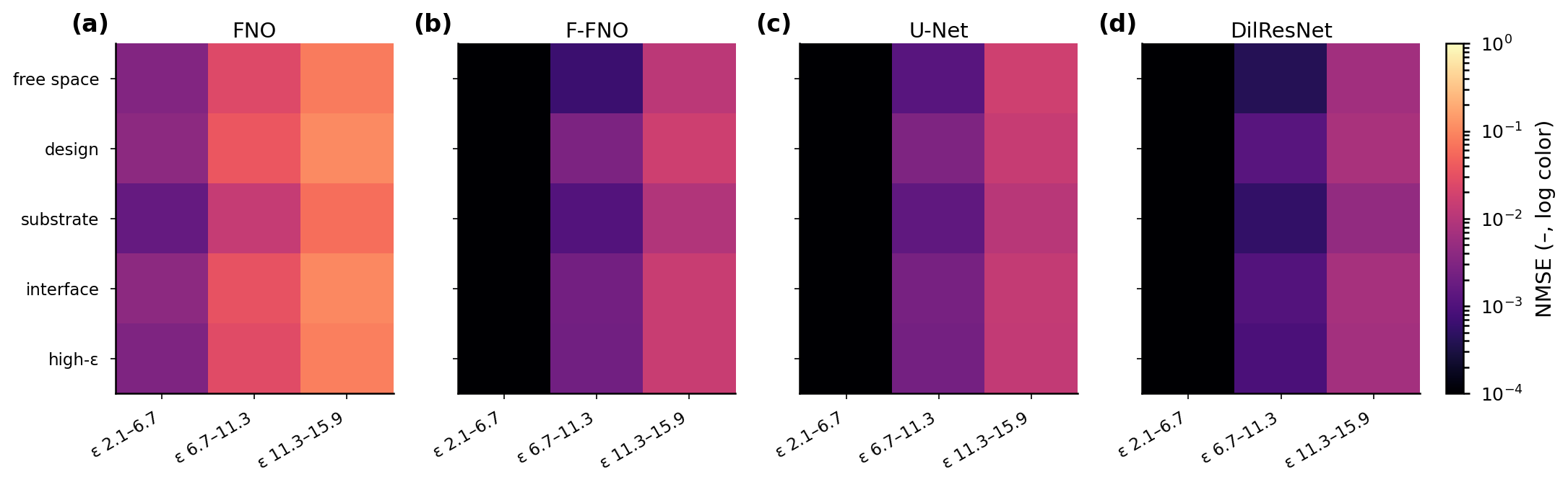}
\par\vspace{4mm}
{\small\bfseries Manually defined patterns}\par\vspace{1mm}
\includegraphics[width=\textwidth]{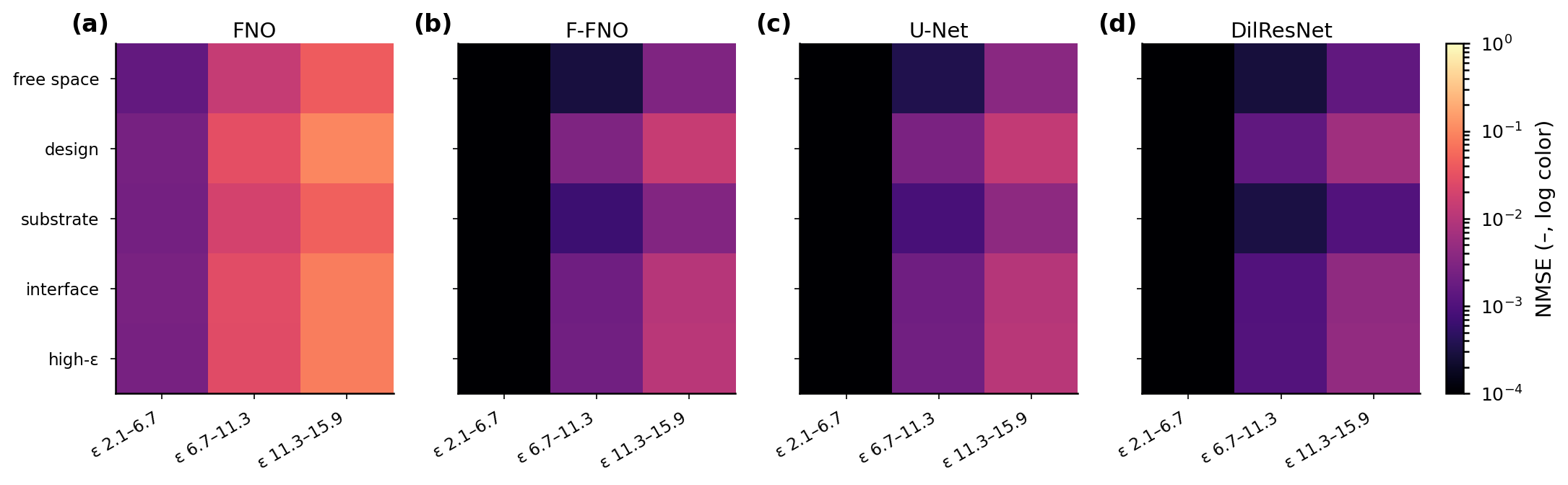}
\caption[]{\textbf{Continued.} Additional test sets using the same quantities and normalization as the preceding panels.}
\end{figure}
\clearpage

\begin{figure}[p]
\ContinuedFloat
\centering
{\small\bfseries Inverse-designed metalenses}\par\vspace{1mm}
\includegraphics[width=\textwidth]{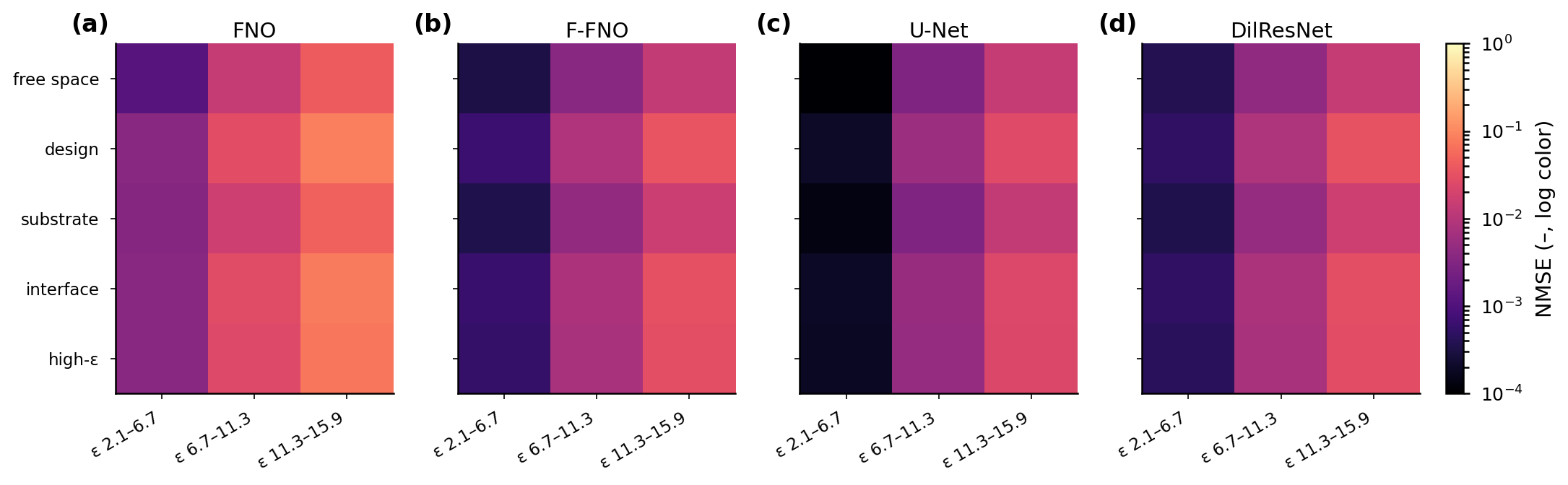}
\par\vspace{4mm}
{\small\bfseries Inverse-designed binary structures}\par\vspace{1mm}
\includegraphics[width=\textwidth]{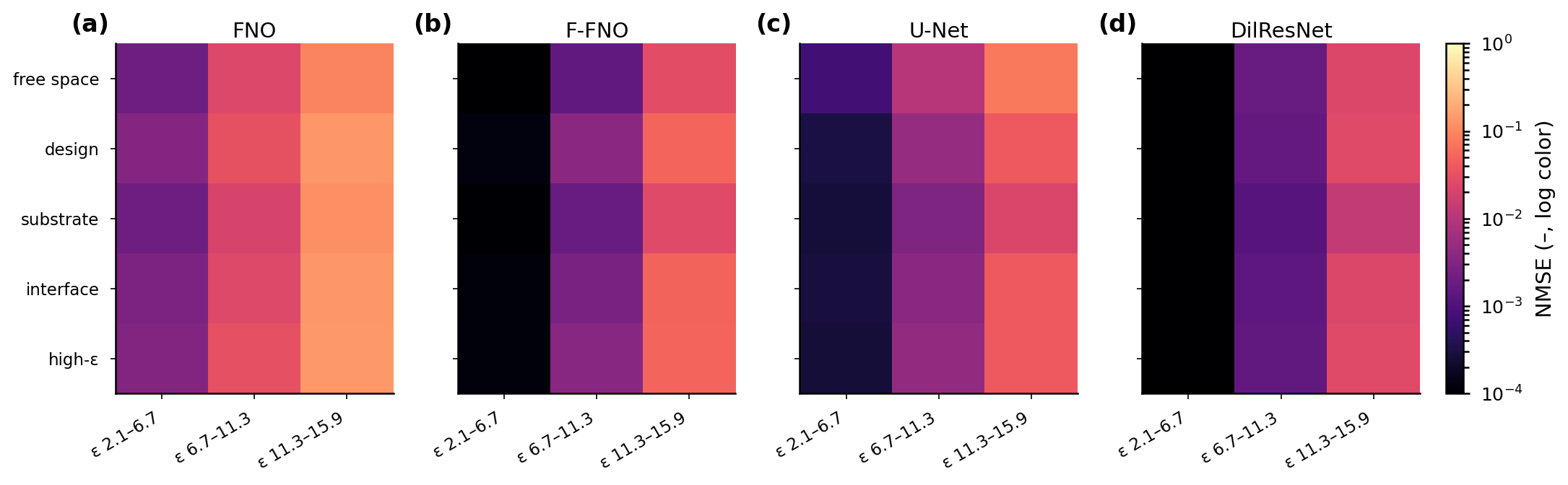}
\caption[]{\textbf{Continued.} Additional test sets using the same quantities and normalization as the preceding panels.}
\end{figure}
\clearpage

\begin{figure}[p]
\ContinuedFloat
\centering
{\small\bfseries Two-material freeform patterns}\par\vspace{1mm}
\includegraphics[width=\textwidth]{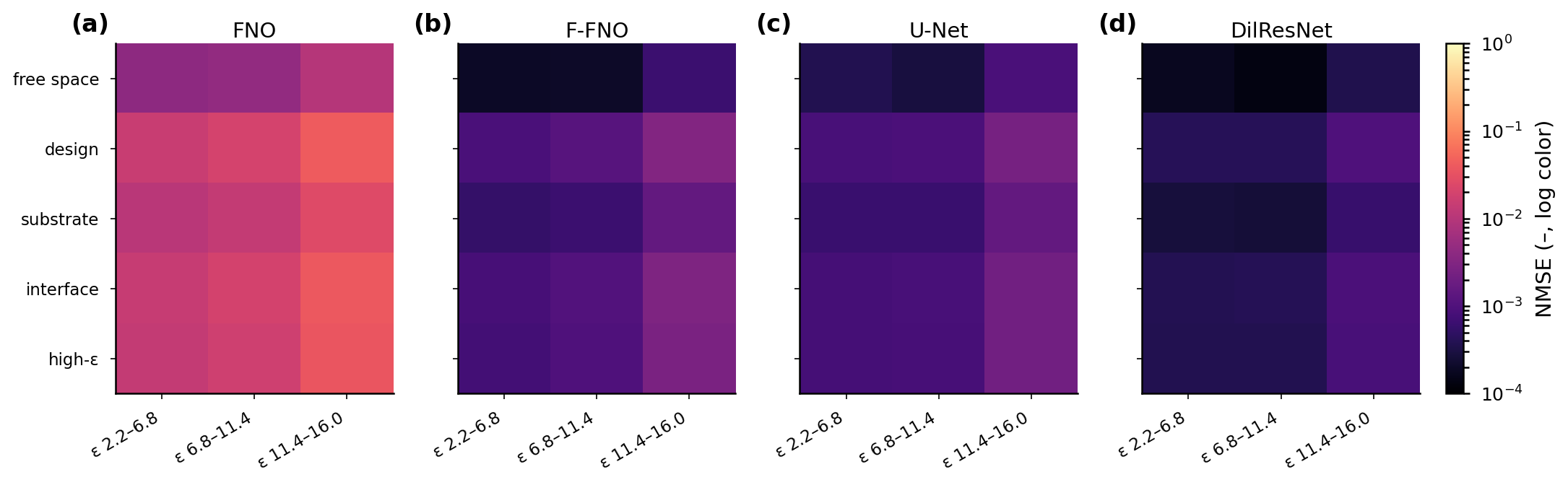}
\par\vspace{4mm}
{\small\bfseries Freeform structures across the structural-scale sweep}\par\vspace{1mm}
\includegraphics[width=\textwidth]{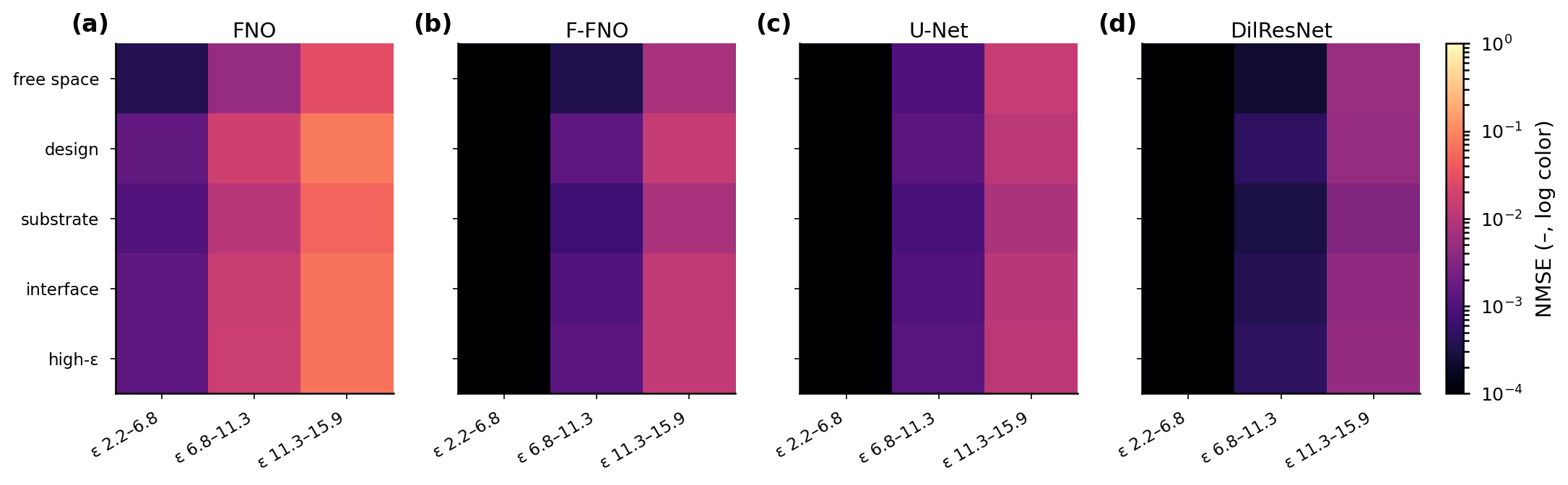}
\caption[]{\textbf{Continued.} Additional test sets using the same quantities and normalization as the preceding panels.}
\end{figure}
\clearpage

\begin{figure}[p]
\centering
{\small\bfseries Held-out geometry--permittivity combinations}\par\vspace{1mm}
\includegraphics[width=\textwidth]{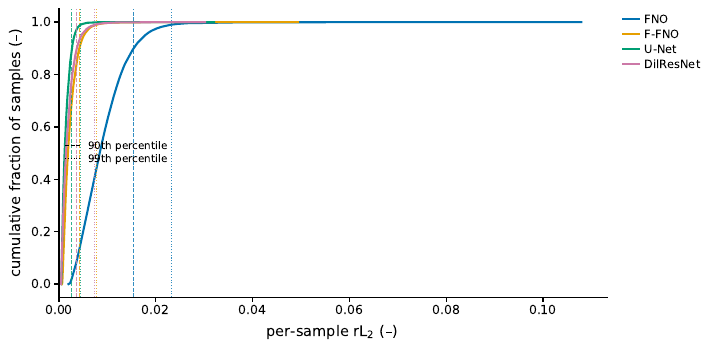}
\caption{\textbf{Distributions of relative complex-field error.} Empirical cumulative distributions for four architectures on seven evaluation sets comprising 120,909 field samples. Horizontal axes show fractional relative $L_2$ error (0.1 = 10\%) with different ranges. Dashed and dotted lines mark the 90th and 99th percentiles. The distributions show sample variation and upper-error tails that are not resolved by the means in Table 1. Each curve represents one trained model and includes repeated geometries evaluated at different permittivities where applicable.}
\label{fig:s8}
\end{figure}
\clearpage

\begin{figure}[p]
\ContinuedFloat
\centering
{\small\bfseries Aperiodic bar arrays}\par\vspace{1mm}
\includegraphics[width=\textwidth]{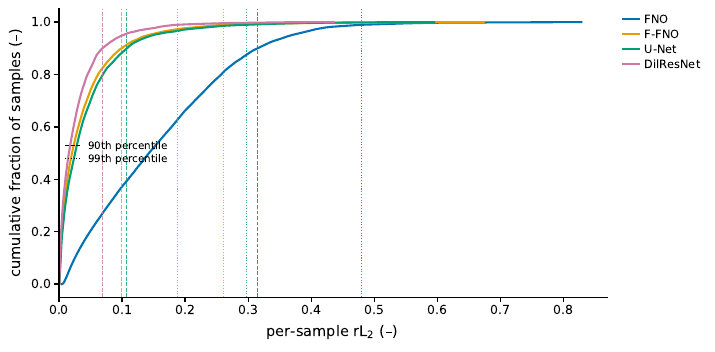}
\par\vspace{4mm}
{\small\bfseries Manually defined patterns}\par\vspace{1mm}
\includegraphics[width=\textwidth]{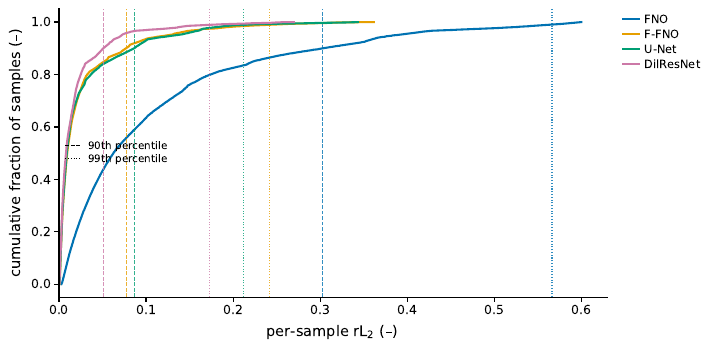}
\caption[]{\textbf{Continued.} Cumulative error distributions for aperiodic bar arrays and manually defined patterns. Axis units and percentile markers follow the preceding page.}
\end{figure}
\clearpage

\begin{figure}[p]
\ContinuedFloat
\centering
{\small\bfseries Inverse-designed metalenses}\par\vspace{1mm}
\includegraphics[width=\textwidth]{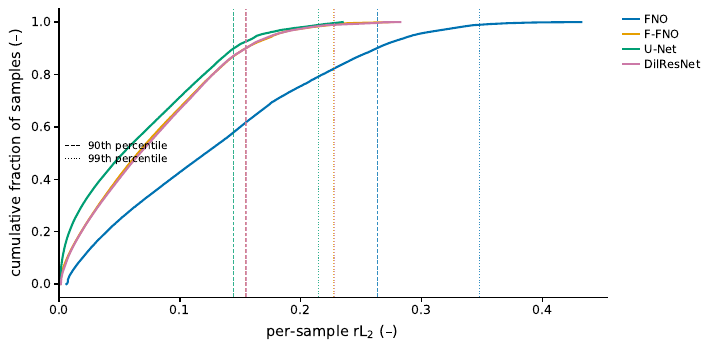}
\par\vspace{4mm}
{\small\bfseries Inverse-designed binary structures}\par\vspace{1mm}
\includegraphics[width=\textwidth]{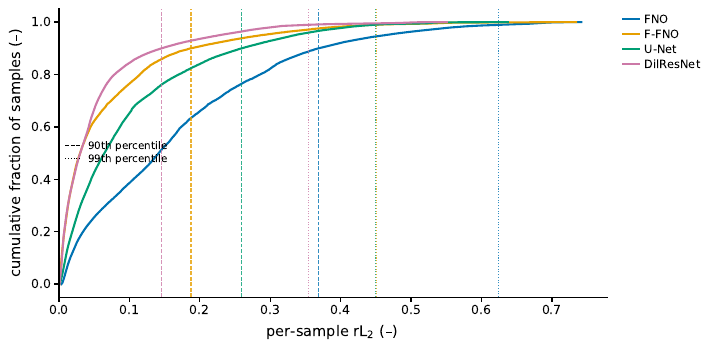}
\caption[]{\textbf{Continued.} Cumulative error distributions for inverse-designed metalenses and inverse-designed binary structures. Axis units and percentile markers follow the preceding pages.}
\end{figure}
\clearpage

\begin{figure}[p]
\ContinuedFloat
\centering
{\small\bfseries Two-material freeform patterns}\par\vspace{1mm}
\includegraphics[width=\textwidth]{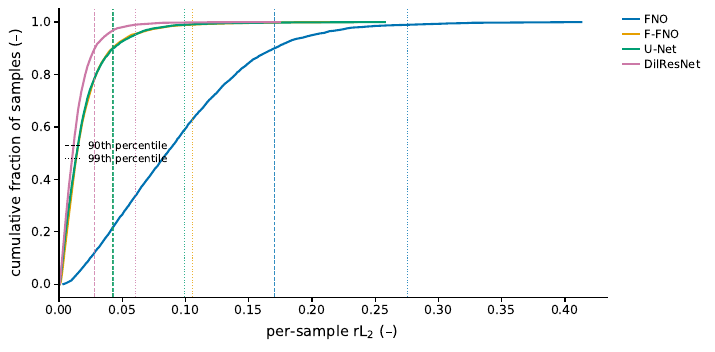}
\par\vspace{4mm}
{\small\bfseries Freeform structures across the structural-scale sweep}\par\vspace{1mm}
\includegraphics[width=\textwidth]{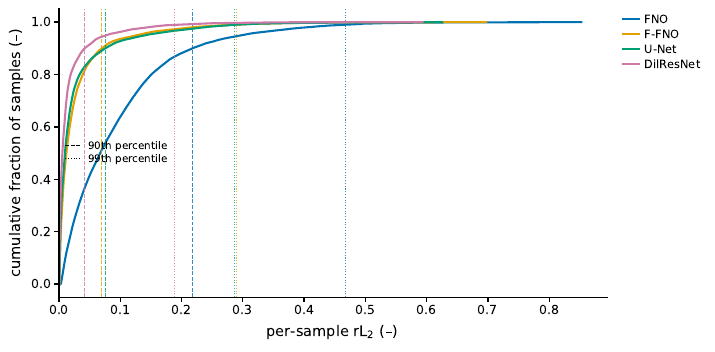}
\caption[]{\textbf{Continued.} Cumulative error distributions for two-material freeform patterns and the structural-scale sweep. These sets test changes in material arrangement and structural scale, respectively. Axis units and percentile markers follow the preceding pages.}
\end{figure}
\clearpage

\begin{figure}[p]
\centering
\includegraphics[width=\textwidth]{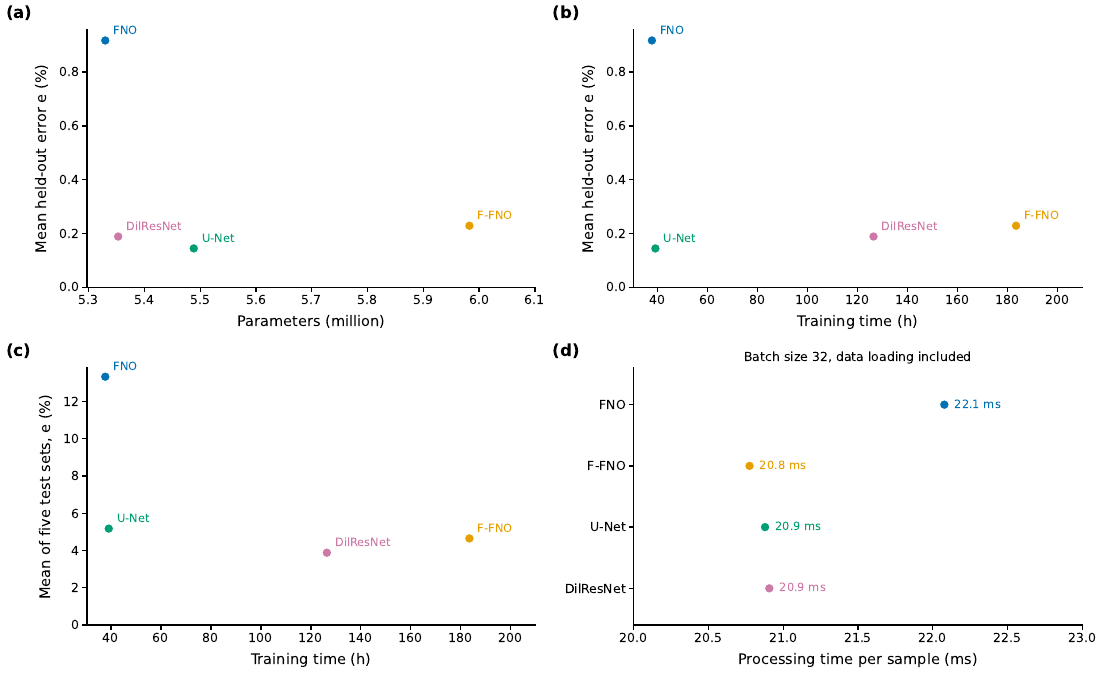}
\caption{\textbf{Accuracy and computational cost of the tested configurations.} Mean per-sample relative field error on 70,012 held-out combinations versus (a) parameter count and (b) training time. (c) Equally weighted mean of the five independent-test means versus training time. The five sets contain 35,441 field samples in total. Error values are the per-sample means of Table 1, with parameters and training times from Table S2. Panels (a,b) and (c) use different vertical scales. (d) Amortized processing time of each configuration, including data loading at batch size 32. All timings use one RTX 6000 Ada. The figure describes surrogate cost and accuracy without a matched FDFD speed comparison.}
\label{fig:s9}
\end{figure}
\clearpage

\begin{figure}[p]
\centering
\includegraphics[width=\textwidth]{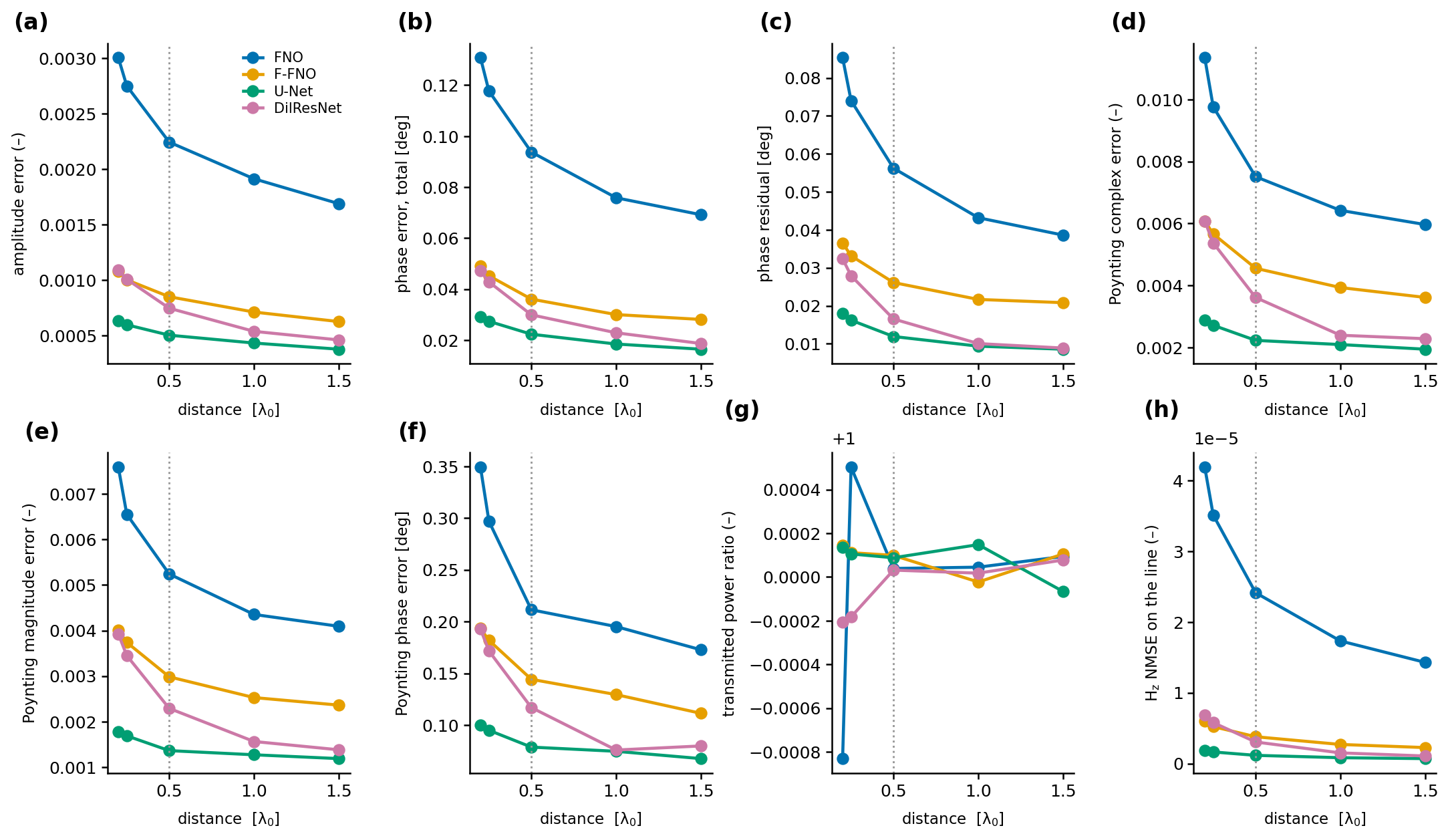}
\caption{\textbf{Optical readout at different distances from the structure.} The output line is varied from 0.2$\lambda_0$ to 1.5$\lambda_0$ for 70,012 held-out combinations. Panels show (a) amplitude error, (b) total phase error, (c) residual phase error, (d--f) complex, magnitude, and phase errors of the Poynting quantity, (g) integrated transmitted-power ratio, and (h) output-line $H_z$ NMSE. The dotted line marks the 0.5$\lambda_0$ readout used in Figure 6. The power ratio varies little over the tested distances.}
\label{fig:s10}
\end{figure}
\clearpage

\begin{figure}[p]
\centering
{\small\bfseries Inverse-designed metalenses}\par\vspace{1mm}
\includegraphics[width=\textwidth]{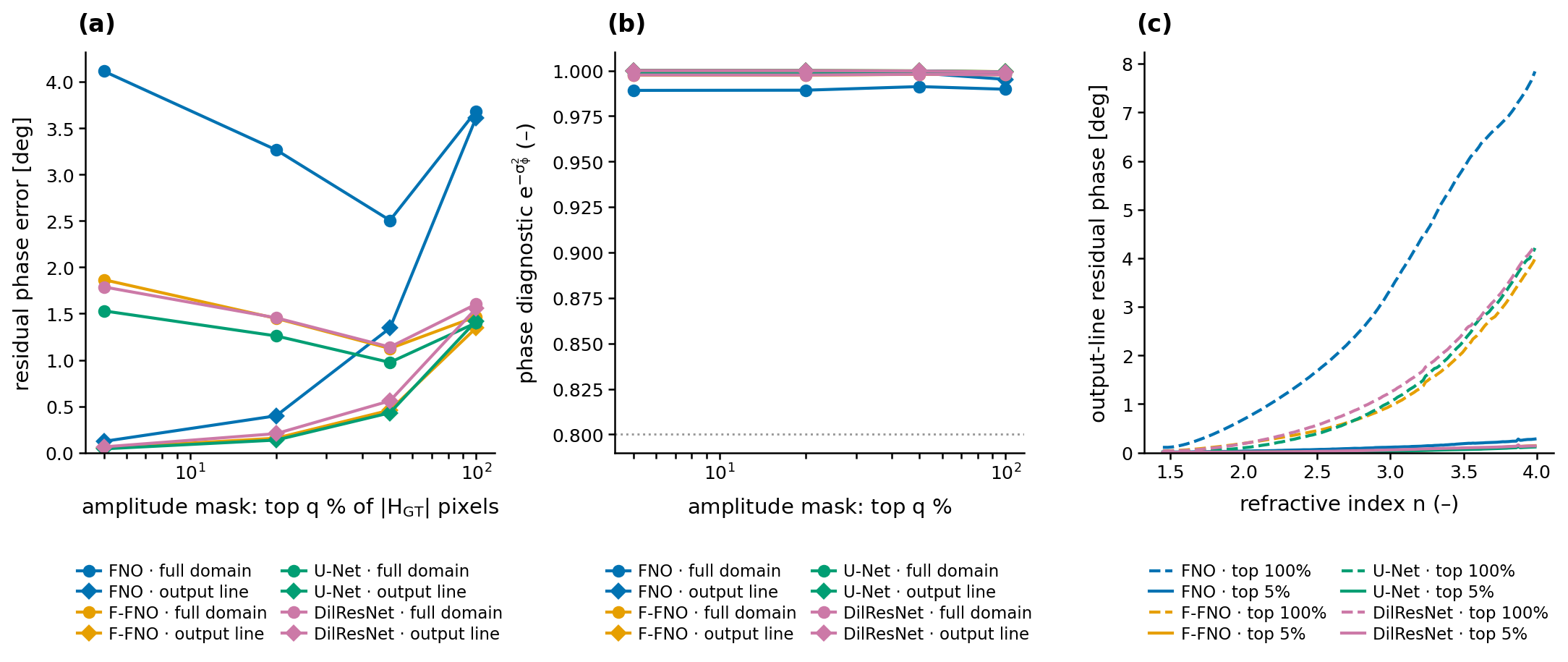}
\par\vspace{4mm}
{\small\bfseries Held-out geometry--permittivity combinations}\par\vspace{1mm}
\includegraphics[width=\textwidth]{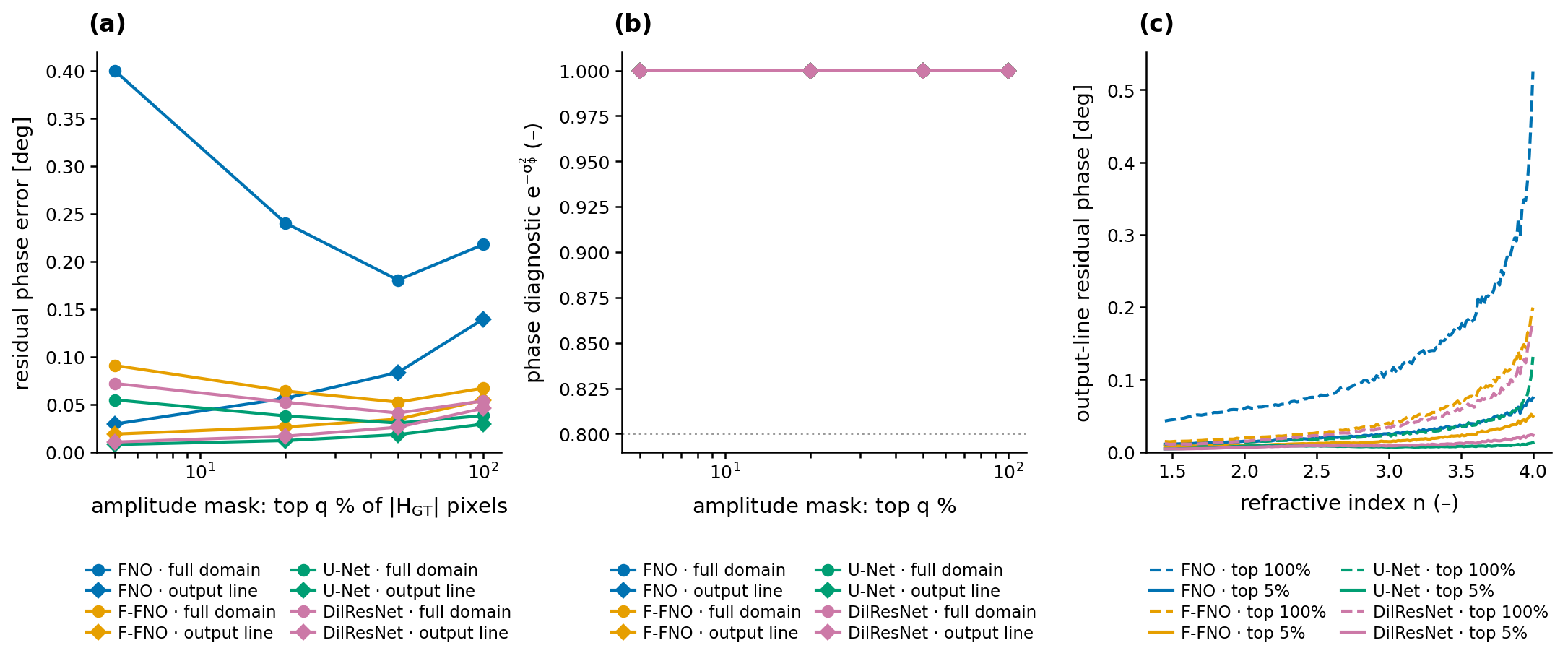}
\caption{\textbf{Phase diagnostics under amplitude masking.} Upper and lower panels show 6,950 metalens samples and 70,012 held-out combinations, respectively. (a) Full-domain and output-line residual phase errors for the highest 5\%, 20\%, 50\%, or 100\% of reference-amplitude pixels. (b) Corresponding phase-derived diagnostics. (c) Output-line residual phase versus refractive index for the 100\% and 5\% masks. The primary analysis uses all pixels. Each mask changes the evaluated pixel set. These diagnostics exclude amplitude error and do not calculate focal intensity.}
\label{fig:s11}
\end{figure}
\clearpage

\begin{figure}[p]
\centering
\includegraphics[width=\textwidth]{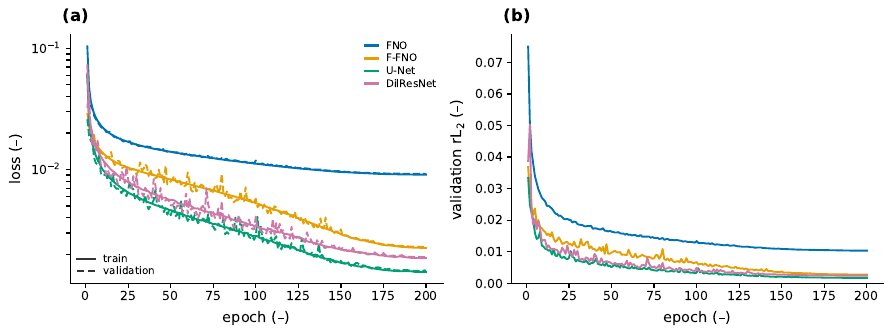}
\caption{\textbf{Training and validation histories.} (a) Training and validation losses and (b) validation relative $L_2$ error for FNO, F-FNO, U-Net, and DilResNet. All models use AdamW, a warmup/cosine schedule, batch size 64, and a maximum of 200 epochs. Each curve represents one training run. The selected checkpoints minimize the validation loss and occur at epochs 196--200. The curves do not quantify variation across training seeds.}
\label{fig:s12}
\end{figure}
\clearpage

\begin{figure}[p]
\centering
\includegraphics[width=\textwidth,height=183mm,keepaspectratio]{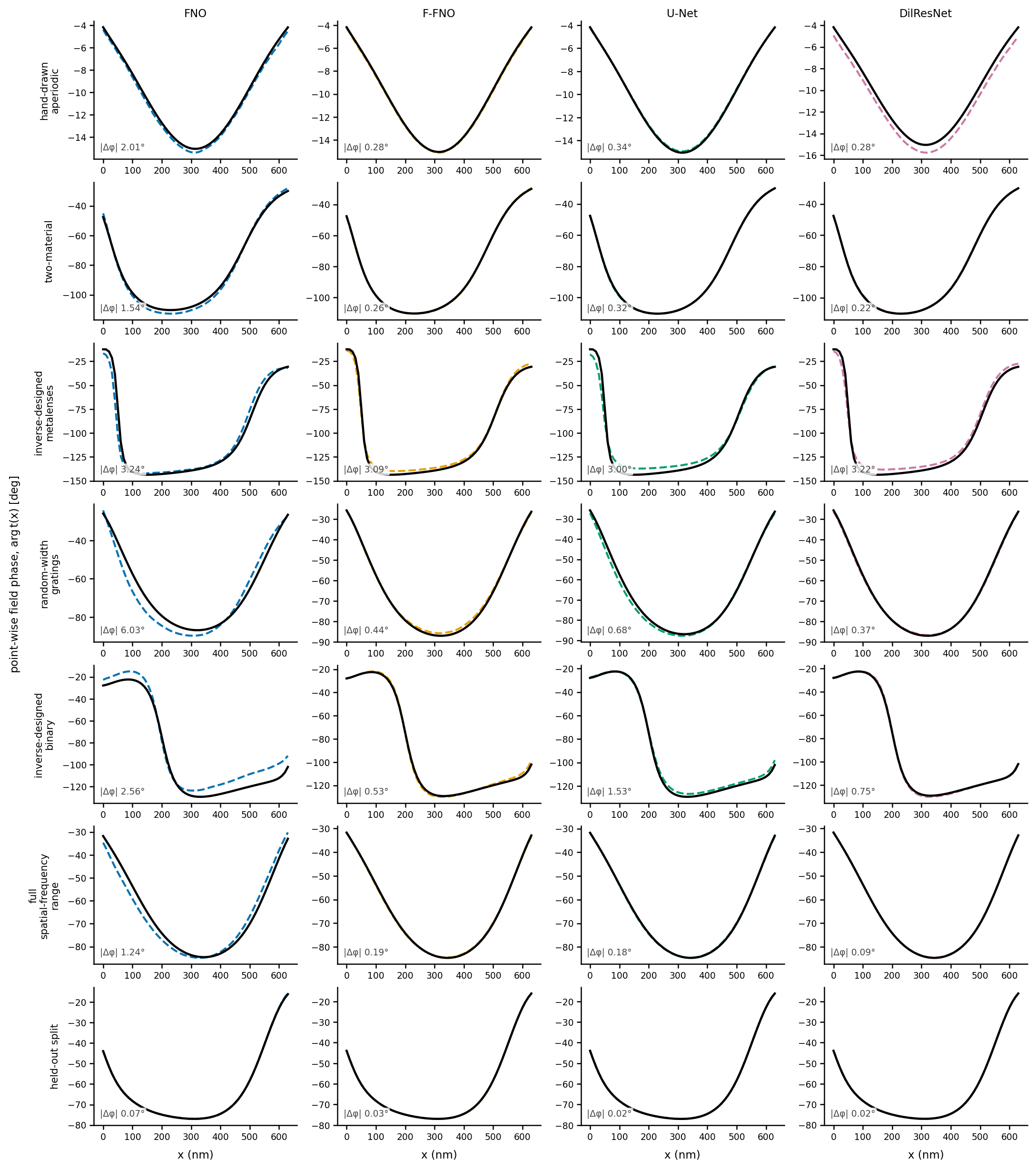}
\caption{\textbf{Transmitted-phase profiles for selected structures.} One sample from each of the seven evaluation sets is shown for all four architectures. Black solid curves denote FDFD references and colored dashed curves denote predictions. Unwrapped phase is given in degrees relative to the incident field propagated through vacuum to the readout line. These profiles supplement Figure 6 and illustrate individual samples.}
\label{fig:s13}
\end{figure}
\clearpage

\begin{figure}[p]
\centering
\includegraphics[width=\textwidth]{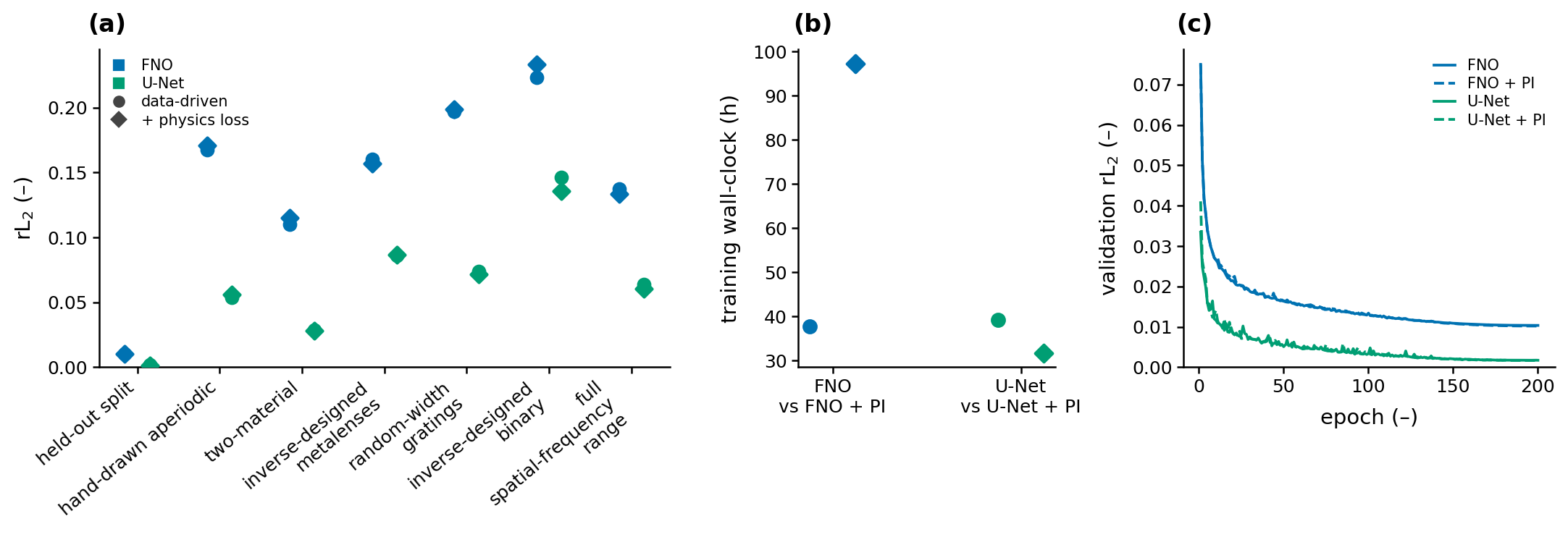}
\caption{\textbf{Comparison with an additional wave-equation residual.} Data-driven FNO and U-Net are compared with parameter-matched variants trained with a residual term. (a) Relative $L_2$ error summaries for seven evaluation sets. (b) Training time. (c) Validation relative $L_2$ histories. Each variant uses one training seed. The statement that errors change by less than 15\% refers to the reported average NMSE (Table~S6), which is a different summary from panel (a). The results characterize the tested loss settings and do not establish a general benefit of residual-based training.}
\label{fig:s14}
\end{figure}

\begin{figure}[p]
\centering
\includegraphics[width=0.75\textwidth]{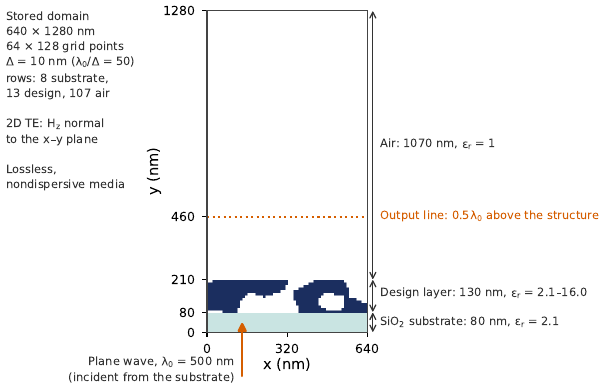}
\caption{\textbf{Simulation domain.} Permittivity map of one training sample on the stored $64\times128$ grid with 10\,nm spacing ($\lambda_0/\Delta=50$), covering $640\times1280$\,nm. The domain consists of an 80\,nm SiO$_2$ substrate ($\varepsilon_r=2.1$, 8 rows), a 130\,nm design layer ($\varepsilon_r=2.1$--16.0, 13 rows), and 1070\,nm of air (107 rows). The reference field is simulated using the Hz-polarized FDFD solver in Ceviche. A TE-polarized plane wave at $\lambda_0=500$\,nm is incident from the substrate side, with $H_z$ normal to the simulation plane, and all media are lossless and nondispersive. The dotted line marks the output line $0.5\lambda_0$ above the structure used for the transmitted-field analysis.}
\label{fig:s15}
\end{figure}
\clearpage

\begin{figure}[p]
\centering
\includegraphics[width=\textwidth]{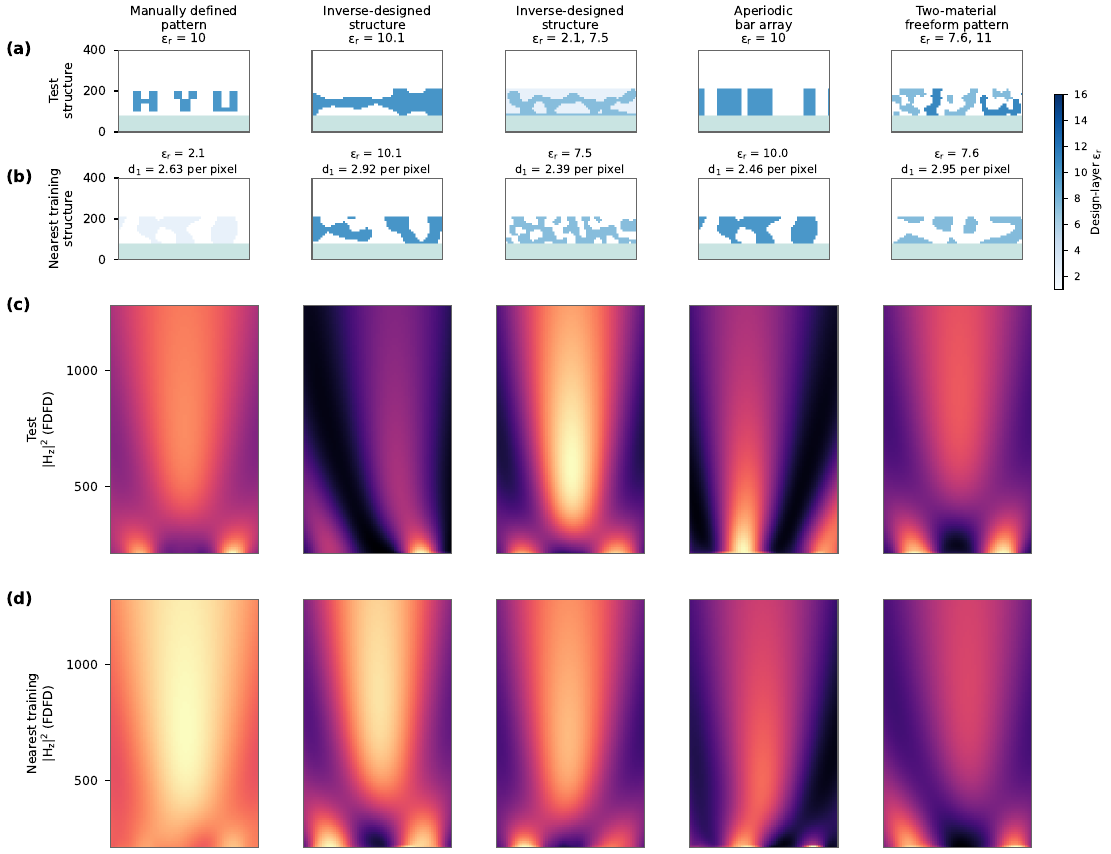}
\caption{\textbf{Nearest training samples for the test structures of main-text Figure~3.} (a) Design-layer permittivity of the five test structures. (b) Training sample with the smallest pixel-wise $L_1$ distance $d_1$ in design-layer permittivity (Equation~\eqref{eq:nearest}), found by exhaustive search over the 560,113 training samples. Titles give the permittivity of the training sample and $d_1$ per design-layer pixel. (c) FDFD reference intensity $|H_z|^2$ of each test structure in the propagation region. (d) FDFD intensity of the corresponding nearest training sample. Each intensity map is normalized by its own maximum. The nearest training fields differ from the test fields by 22\%--77\% in relative complex-field error, compared with 1.3\%--4.2\% for the DilResNet predictions (Supporting Note~10).}
\label{fig:s16}
\end{figure}

{\small
\begin{thebibliography}{54}
\providecommand{\natexlab}[1]{#1}
\providecommand{\url}[1]{\texttt{#1}}
\expandafter\ifx\csname urlstyle\endcsname\relax
  \providecommand{\doi}[1]{doi: #1}\else
  \providecommand{\doi}{doi: \begingroup \urlstyle{rm}\Url}\fi

\bibitem[Molesky et~al.(2018)Molesky, Lin, Piggott, Jin, Vu{\v c}kovi{\'c}, and
  Rodriguez]{molesky2018inverse}
Sean Molesky, Zin Lin, Alexander~Y. Piggott, Weiliang Jin, Jelena Vu{\v
  c}kovi{\'c}, and Alejandro~W. Rodriguez.
\newblock Inverse design in nanophotonics.
\newblock \emph{Nature Photonics}, 12\penalty0 (11):\penalty0 659--670, 2018.
\newblock \doi{10.1038/s41566-018-0246-9}.

\bibitem[Joannopoulos et~al.(2008)Joannopoulos, Johnson, Winn, and
  Meade]{joannopoulos2008photonic}
John~D. Joannopoulos, Steven~G. Johnson, Joshua~N. Winn, and Robert~D. Meade.
\newblock \emph{Photonic Crystals: Molding the Flow of Light}.
\newblock Princeton University Press, Princeton, 2nd edition, 2008.

\bibitem[Smith et~al.(2004)Smith, Pendry, and
  Wiltshire]{smith2004metamaterials}
David~R. Smith, John~B. Pendry, and Mike C.~K. Wiltshire.
\newblock Metamaterials and negative refractive index.
\newblock \emph{Science}, 305\penalty0 (5685):\penalty0 788--792, 2004.
\newblock \doi{10.1126/science.1096796}.

\bibitem[Johnson and Joannopoulos(2001)]{johnson2001block}
Steven~G. Johnson and John~D. Joannopoulos.
\newblock Block-iterative frequency-domain methods for {Maxwell's} equations in
  a planewave basis.
\newblock \emph{Optics Express}, 8\penalty0 (3):\penalty0 173--190, 2001.
\newblock \doi{10.1364/OE.8.000173}.

\bibitem[Yu and Capasso(2014)]{yu2014flat}
Nanfang Yu and Federico Capasso.
\newblock Flat optics with designer metasurfaces.
\newblock \emph{Nature Materials}, 13\penalty0 (2):\penalty0 139--150, 2014.
\newblock \doi{10.1038/nmat3839}.

\bibitem[Khorasaninejad et~al.(2016)Khorasaninejad, Chen, Devlin, Oh, Zhu, and
  Capasso]{khorasaninejad2016metalenses}
Mohammadreza Khorasaninejad, Wei~Ting Chen, Robert~C. Devlin, Jaewon Oh,
  Alexander~Y. Zhu, and Federico Capasso.
\newblock Metalenses at visible wavelengths: Diffraction-limited focusing and
  subwavelength resolution imaging.
\newblock \emph{Science}, 352\penalty0 (6290):\penalty0 1190--1194, 2016.
\newblock \doi{10.1126/science.aaf6644}.

\bibitem[Pestourie et~al.(2018)Pestourie, P{\'e}rez-Arancibia, Lin, Shin,
  Capasso, and Johnson]{pestourie2018inverse}
Rapha{\"e}l Pestourie, Carlos P{\'e}rez-Arancibia, Zin Lin, Wonseok Shin,
  Federico Capasso, and Steven~G. Johnson.
\newblock Inverse design of large-area metasurfaces.
\newblock \emph{Optics Express}, 26\penalty0 (26):\penalty0 33732--33747, 2018.
\newblock \doi{10.1364/OE.26.033732}.

\bibitem[Hughes et~al.(2021)Hughes, Minkov, Liu, Yu, and
  Fan]{hughes2021perspective}
Tyler~W. Hughes, Momchil Minkov, Victor Liu, Zongfu Yu, and Shanhui Fan.
\newblock A perspective on the pathway toward full wave simulation of large
  area metalenses.
\newblock \emph{Applied Physics Letters}, 119\penalty0 (15):\penalty0 150502,
  2021.
\newblock \doi{10.1063/5.0071245}.

\bibitem[Kim et~al.(2026)Kim, Han, Lee, Jang, and Chung]{kim2026suppressing}
Donghyun Kim, Junseo Han, Seunghyun Lee, Min~Seok Jang, and Haejun Chung.
\newblock Suppressing stitching errors in full-sensor-plane color routers via
  optical structural similarity.
\newblock \emph{Nanophotonics}, 15\penalty0 (14):\penalty0 e70220, 2026.
\newblock \doi{10.1002/nap2.70220}.

\bibitem[Huang(1994)]{huang1994coupled}
Wei-Ping Huang.
\newblock Coupled-mode theory for optical waveguides: an overview.
\newblock \emph{Journal of the Optical Society of America A}, 11\penalty0
  (3):\penalty0 963--983, 1994.
\newblock \doi{10.1364/JOSAA.11.000963}.

\bibitem[Jensen and Sigmund(2011)]{jensen2011topology}
Jakob~S. Jensen and Ole Sigmund.
\newblock Topology optimization for nano-photonics.
\newblock \emph{Laser \& Photonics Reviews}, 5\penalty0 (2):\penalty0 308--321,
  2011.
\newblock \doi{10.1002/lpor.201000014}.

\bibitem[Piggott et~al.(2015)Piggott, Lu, Lagoudakis, Petykiewicz, Babinec, and
  Vu{\v c}kovi{\'c}]{piggott2015inverse}
Alexander~Y. Piggott, Jesse Lu, Konstantinos~G. Lagoudakis, Jan Petykiewicz,
  Thomas~M. Babinec, and Jelena Vu{\v c}kovi{\'c}.
\newblock Inverse design and demonstration of a compact and broadband on-chip
  wavelength demultiplexer.
\newblock \emph{Nature Photonics}, 9\penalty0 (6):\penalty0 374--377, 2015.
\newblock \doi{10.1038/nphoton.2015.69}.

\bibitem[Sell et~al.(2017)Sell, Yang, Doshay, Yang, and Fan]{sell2017large}
David Sell, Jianji Yang, Sage Doshay, Rui Yang, and Jonathan~A. Fan.
\newblock Large-angle, multifunctional metagratings based on freeform multimode
  geometries.
\newblock \emph{Nano Letters}, 17\penalty0 (6):\penalty0 3752--3757, 2017.
\newblock \doi{10.1021/acs.nanolett.7b01082}.

\bibitem[Seo et~al.(2026{\natexlab{a}})Seo, Um, Lee, Ye, and
  Chung]{seo2026physics}
Dongjin Seo, Soobin Um, Sangbin Lee, Jong~Chul Ye, and Haejun Chung.
\newblock Physics-guided and fabrication-aware inverse design of photonic
  devices using diffusion models.
\newblock \emph{ACS Photonics}, 13\penalty0 (2):\penalty0 363--372,
  2026{\natexlab{a}}.
\newblock \doi{10.1021/acsphotonics.5c00993}.

\bibitem[Kang et~al.(2024)Kang, Seo, Boriskina, and Chung]{kang2024adjoint}
Chanik Kang, Dongjin Seo, Svetlana~V. Boriskina, and Haejun Chung.
\newblock Adjoint method in machine learning: A pathway to efficient inverse
  design of photonic devices.
\newblock \emph{Materials \& Design}, 239:\penalty0 112737, 2024.
\newblock \doi{10.1016/j.matdes.2024.112737}.

\bibitem[Park et~al.(2026)Park, Miller, and Chung]{park2026nyquist}
Mingyu Park, Owen~D. Miller, and Haejun Chung.
\newblock Nyquist-sampled time-domain adjoint {FDTD} for memory-efficient
  broadband nanophotonic inverse design.
\newblock \emph{arXiv preprint arXiv:2607.08159}, 2026.

\bibitem[Jiang et~al.(2021)Jiang, Chen, and Fan]{jiang2021deep}
Jiaqi Jiang, Mingkun Chen, and Jonathan~A. Fan.
\newblock Deep neural networks for the evaluation and design of photonic
  devices.
\newblock \emph{Nature Reviews Materials}, 6\penalty0 (8):\penalty0 679--700,
  2021.
\newblock \doi{10.1038/s41578-020-00260-1}.

\bibitem[So et~al.(2020)So, Badloe, Noh, Rho, and Bravo-Abad]{so2020deep}
Sunae So, Trevon Badloe, Jaebum Noh, Junsuk Rho, and Jorge Bravo-Abad.
\newblock Deep learning enabled inverse design in nanophotonics.
\newblock \emph{Nanophotonics}, 9\penalty0 (5):\penalty0 1041--1057, 2020.
\newblock \doi{10.1515/nanoph-2019-0474}.

\bibitem[Wiecha and Muskens(2020)]{wiecha2020deep}
Peter~R. Wiecha and Otto~L. Muskens.
\newblock Deep learning meets nanophotonics: a generalized accurate predictor
  for near fields and far fields of arbitrary {3D} nanostructures.
\newblock \emph{Nano Letters}, 20\penalty0 (1):\penalty0 329--338, 2020.
\newblock \doi{10.1021/acs.nanolett.9b03971}.

\bibitem[Chen et~al.(2022)Chen, Lupoiu, Mao, Huang, Jiang, Lalanne, and
  Fan]{chen2022high}
Mingkun Chen, Robert Lupoiu, Chenkai Mao, Der-Han Huang, Jiaqi Jiang, Philippe
  Lalanne, and Jonathan~A. Fan.
\newblock High speed simulation and freeform optimization of nanophotonic
  devices with physics-augmented deep learning.
\newblock \emph{ACS Photonics}, 9\penalty0 (9):\penalty0 3110--3123, 2022.
\newblock \doi{10.1021/acsphotonics.2c00876}.

\bibitem[Kanmaz et~al.(2023)Kanmaz, Ozturk, Demir, and
  Gunduz-Demir]{kanmaz2023deep}
Tevfik~Bulent Kanmaz, Efe Ozturk, Hilmi~Volkan Demir, and Cigdem Gunduz-Demir.
\newblock Deep-learning-enabled electromagnetic near-field prediction and
  inverse design of metasurfaces.
\newblock \emph{Optica}, 10\penalty0 (10):\penalty0 1373--1382, 2023.
\newblock \doi{10.1364/OPTICA.498211}.

\bibitem[Kang et~al.(2026)Kang, Suk, Jang, and Chung]{kang2026neuraladjoint}
Chanik Kang, Hyewon Suk, Ikbeom Jang, and Haejun Chung.
\newblock Neural adjoint inverse design of three-dimensional nanophotonic
  devices.
\newblock \emph{Nanophotonics}, 15\penalty0 (20):\penalty0 e70334, 2026.
\newblock \doi{10.1002/nap2.70334}.

\bibitem[Augenstein et~al.(2023)Augenstein, Rep{\"a}n, and
  Rockstuhl]{augenstein2023neural}
Yannick Augenstein, Taavi Rep{\"a}n, and Carsten Rockstuhl.
\newblock Neural operator-based surrogate solver for free-form electromagnetic
  inverse design.
\newblock \emph{ACS Photonics}, 10\penalty0 (5):\penalty0 1547--1557, 2023.
\newblock \doi{10.1021/acsphotonics.3c00156}.

\bibitem[Kuang et~al.(2026)Kuang, Sun, Fu, Yun, He, Zhao, Wang, Shao, Gao, Fu,
  Liang, Li, Wang, and Lei]{kuang2026rapid}
Shiqi Kuang, Zhi Sun, Boyan Fu, Xue Yun, Minru He, Tianyu Zhao, Shaowei Wang,
  Tianhua Shao, Geze Gao, Yunlai Fu, Yansheng Liang, Tianyue Li, Shuming Wang,
  and Ming Lei.
\newblock Rapid inverse design of large-scale freeform meta-optics with the
  neighborhood-attention transformer.
\newblock \emph{PhotoniX}, 7:\penalty0 69, 2026.
\newblock \doi{10.1186/s43074-026-00291-x}.

\bibitem[Gu et~al.(2022)Gu, Gao, Feng, Zhu, Chen, Boning, and
  Pan]{gu2022neurolight}
Jiaqi Gu, Zhengqi Gao, Chenghao Feng, Hanqing Zhu, Ray~T. Chen, Duane~S.
  Boning, and David~Z. Pan.
\newblock {NeurOLight}: A physics-agnostic neural operator enabling parametric
  photonic device simulation.
\newblock In \emph{Advances in Neural Information Processing Systems
  (NeurIPS)}, volume~35, pages 14623--14636, 2022.
\newblock \doi{10.52202/068431-1063}.

\bibitem[Mao et~al.(2024)Mao, Lupoiu, Dai, Chen, and Fan]{mao2024snapddm}
Chenkai Mao, Robert Lupoiu, Tianxiang Dai, Mingkun Chen, and Jonathan~A. Fan.
\newblock Towards general neural surrogate solvers with specialized neural
  accelerators.
\newblock In \emph{Proceedings of the 41st International Conference on Machine
  Learning (ICML)}, volume 235 of \emph{Proceedings of Machine Learning
  Research}, pages 34693--34711, 2024.

\bibitem[Mao and Fan(2025)]{mao2025multilevel}
Chenkai Mao and Jonathan~A. Fan.
\newblock Accurate and scalable deep {Maxwell} solvers using multilevel
  iterative methods.
\newblock \emph{arXiv preprint arXiv:2509.03622}, 2025.
\newblock \doi{10.48550/arXiv.2509.03622}.
\newblock URL \url{https://arxiv.org/abs/2509.03622}.

\bibitem[Seo et~al.(2026{\natexlab{b}})Seo, Kang, Seo, and
  Chung]{seo2026broadband}
Joonhyuk Seo, Chanik Kang, Dongjin Seo, and Haejun Chung.
\newblock Broadband electromagnetic field prediction at unseen wavelengths via
  physics-guided neural operators.
\newblock \emph{Nanophotonics}, 15\penalty0 (15):\penalty0 e70242,
  2026{\natexlab{b}}.
\newblock \doi{10.1002/nap2.70242}.

\bibitem[Peurifoy et~al.(2018)Peurifoy, Shen, Jing, Yang, Cano-Renteria,
  DeLacy, Joannopoulos, Tegmark, and
  Solja{\v{c}}i{\'c}]{peurifoy2018nanophotonic}
John Peurifoy, Yichen Shen, Li~Jing, Yi~Yang, Fidel Cano-Renteria, Brendan~G.
  DeLacy, John~D. Joannopoulos, Max Tegmark, and Marin Solja{\v{c}}i{\'c}.
\newblock Nanophotonic particle simulation and inverse design using artificial
  neural networks.
\newblock \emph{Science Advances}, 4\penalty0 (6):\penalty0 eaar4206, 2018.
\newblock \doi{10.1126/sciadv.aar4206}.

\bibitem[Lim and Psaltis(2022)]{lim2022maxwellnet}
Joowon Lim and Demetri Psaltis.
\newblock {MaxwellNet}: Physics-driven deep neural network training based on
  {M}axwell's equations.
\newblock \emph{APL Photonics}, 7\penalty0 (1):\penalty0 011301, 2022.
\newblock \doi{10.1063/5.0071616}.

\bibitem[Pestourie et~al.(2020)Pestourie, Mroueh, Nguyen, Das, and
  Johnson]{pestourie2020active}
Rapha{\"e}l Pestourie, Youssef Mroueh, Thanh~V. Nguyen, Payel Das, and
  Steven~G. Johnson.
\newblock Active learning of deep surrogates for {PDEs}: application to
  metasurface design.
\newblock \emph{npj Computational Materials}, 6:\penalty0 164, 2020.
\newblock \doi{10.1038/s41524-020-00431-2}.

\bibitem[Dove and Waller(2026)]{dove2026inductively}
Charles Dove and Laura Waller.
\newblock Inductively scalable, single-step neural surrogates for
  wave-scattering inverse problems.
\newblock \emph{arXiv preprint arXiv:2608.17344}, 2026.
\newblock URL \url{https://arxiv.org/abs/2608.17344}.

\bibitem[Zhang et~al.(2023)Zhang, Yao, Jiang, and Ng]{zhang2023fast}
Huan~Huan Zhang, He~Ming Yao, Lijun Jiang, and Michael Ng.
\newblock Fast full-wave electromagnetic forward solver based on deep
  conditional convolutional autoencoders.
\newblock \emph{IEEE Antennas and Wireless Propagation Letters}, 22\penalty0
  (4):\penalty0 779--783, 2023.
\newblock \doi{10.1109/LAWP.2022.3224983}.

\bibitem[Hughes et~al.(2019)Hughes, Williamson, Minkov, and
  Fan]{hughes2019forward}
Tyler~W. Hughes, Ian A.~D. Williamson, Momchil Minkov, and Shanhui Fan.
\newblock Forward-mode differentiation of {M}axwell's equations.
\newblock \emph{ACS Photonics}, 6\penalty0 (11):\penalty0 3010--3016, 2019.
\newblock \doi{10.1021/acsphotonics.9b01238}.

\bibitem[Rumpf(2012)]{rumpf2012fdfd}
Raymond~C. Rumpf.
\newblock Simple implementation of arbitrarily shaped
  total-field/scattered-field regions in finite-difference frequency-domain.
\newblock \emph{Progress In Electromagnetics Research B}, 36:\penalty0
  221--248, 2012.
\newblock \doi{10.2528/PIERB11092006}.

\bibitem[Shin and Fan(2012)]{shin2012pml}
Wonseok Shin and Shanhui Fan.
\newblock Choice of the perfectly matched layer boundary condition for
  frequency-domain {M}axwell's equations solvers.
\newblock \emph{Journal of Computational Physics}, 231\penalty0 (8):\penalty0
  3406--3431, 2012.
\newblock \doi{10.1016/j.jcp.2012.01.013}.

\bibitem[Albella et~al.(2013)Albella, Poyli, Schmidt, Maier, Moreno, S{\'a}enz,
  and Aizpurua]{albella2013low}
Pablo Albella, M.~Ameen Poyli, Mikolaj~K. Schmidt, Stefan~A. Maier, Fernando
  Moreno, Juan~Jos{\'e} S{\'a}enz, and Javier Aizpurua.
\newblock Low-loss electric and magnetic field-enhanced spectroscopy with
  subwavelength silicon dimers.
\newblock \emph{The Journal of Physical Chemistry C}, 117\penalty0
  (26):\penalty0 13573--13584, 2013.
\newblock \doi{10.1021/jp4027018}.

\bibitem[Perlin(2002)]{perlin2002improving}
Ken Perlin.
\newblock Improving noise.
\newblock \emph{ACM Transactions on Graphics}, 21\penalty0 (3):\penalty0
  681--682, 2002.
\newblock \doi{10.1145/566654.566636}.

\bibitem[Torquato and Lu(1993)]{torquato1993chord}
S.~Torquato and B.~Lu.
\newblock Chord-length distribution function for two-phase random media.
\newblock \emph{Physical Review E}, 47\penalty0 (4):\penalty0 2950--2953, 1993.
\newblock \doi{10.1103/PhysRevE.47.2950}.

\bibitem[Zhu et~al.(2024)Zhu, Cong, Chen, Ning, Chen, Gu, and Pan]{zhu2024pace}
Hanqing Zhu, Wenyan Cong, Guojin Chen, Shupeng Ning, Ray~T. Chen, Jiaqi Gu, and
  David~Z. Pan.
\newblock {PACE}: Pacing operator learning to accurate optical field simulation
  for complicated photonic devices.
\newblock In A.~Globerson, L.~Mackey, D.~Belgrave, A.~Fan, U.~Paquet,
  J.~Tomczak, and C.~Zhang, editors, \emph{Advances in Neural Information
  Processing Systems}, volume~37, pages 67535--67555. Curran Associates, Inc.,
  2024.
\newblock \doi{10.52202/079017-2156}.
\newblock URL
  \url{https://proceedings.neurips.cc/paper_files/paper/2024/hash/7cc16e8635e6f27c295355bd214ef8d8-Abstract-Conference.html}.

\bibitem[Li et~al.(2021)Li, Kovachki, Azizzadenesheli, Liu, Bhattacharya,
  Stuart, and Anandkumar]{li2021fourier}
Zongyi Li, Nikola Kovachki, Kamyar Azizzadenesheli, Burigede Liu, Kaushik
  Bhattacharya, Andrew Stuart, and Anima Anandkumar.
\newblock Fourier neural operator for parametric partial differential
  equations.
\newblock In \emph{International Conference on Learning Representations
  (ICLR)}, 2021.

\bibitem[Tran et~al.(2023)Tran, Mathews, Xie, and Ong]{tran2023factorized}
Alasdair Tran, Alexander Mathews, Lexing Xie, and Cheng~Soon Ong.
\newblock Factorized {F}ourier neural operators.
\newblock In \emph{International Conference on Learning Representations
  (ICLR)}, 2023.

\bibitem[Ronneberger et~al.(2015)Ronneberger, Fischer, and
  Brox]{ronneberger2015unet}
Olaf Ronneberger, Philipp Fischer, and Thomas Brox.
\newblock {U-Net}: Convolutional networks for biomedical image segmentation.
\newblock In \emph{Medical Image Computing and Computer-Assisted Intervention
  (MICCAI)}, pages 234--241, 2015.
\newblock \doi{10.1007/978-3-319-24574-4_28}.

\bibitem[Stachenfeld et~al.(2022)Stachenfeld, Fielding, Kochkov, Cranmer,
  Pfaff, Godwin, Cui, Ho, Battaglia, and
  Sanchez-Gonzalez]{stachenfeld2022learned}
Kimberly Stachenfeld, Drummond~B. Fielding, Dmitrii Kochkov, Miles Cranmer,
  Tobias Pfaff, Jonathan Godwin, Can Cui, Shirley Ho, Peter Battaglia, and
  Alvaro Sanchez-Gonzalez.
\newblock Learned coarse models for efficient turbulence simulation.
\newblock In \emph{International Conference on Learning Representations
  (ICLR)}, 2022.

\bibitem[Loshchilov and Hutter(2019)]{loshchilov2019decoupled}
Ilya Loshchilov and Frank Hutter.
\newblock Decoupled weight decay regularization.
\newblock In \emph{International Conference on Learning Representations
  (ICLR)}, 2019.

\bibitem[Loshchilov and Hutter(2017)]{loshchilov2017sgdr}
Ilya Loshchilov and Frank Hutter.
\newblock {SGDR}: Stochastic gradient descent with warm restarts.
\newblock In \emph{International Conference on Learning Representations
  (ICLR)}, 2017.

\bibitem[Micikevicius et~al.(2018)Micikevicius, Narang, Alben, Diamos, Elsen,
  Garcia, Ginsburg, Houston, Kuchaiev, Venkatesh, and
  Wu]{micikevicius2018mixed}
Paulius Micikevicius, Sharan Narang, Jonah Alben, Gregory Diamos, Erich Elsen,
  David Garcia, Boris Ginsburg, Michael Houston, Oleksii Kuchaiev, Ganesh
  Venkatesh, and Hao Wu.
\newblock Mixed precision training.
\newblock In \emph{International Conference on Learning Representations
  (ICLR)}, 2018.

\bibitem[Zhan et~al.(2016)Zhan, Colburn, Trivedi, Fryett, Dodson, and
  Majumdar]{zhan2016low}
Alan Zhan, Shane Colburn, Rahul Trivedi, Taylor~K. Fryett, Christopher~M.
  Dodson, and Arka Majumdar.
\newblock Low-contrast dielectric metasurface optics.
\newblock \emph{ACS Photonics}, 3\penalty0 (2):\penalty0 209--214, 2016.
\newblock \doi{10.1021/acsphotonics.5b00660}.

\bibitem[Devlin et~al.(2016)Devlin, Khorasaninejad, Chen, Oh, and
  Capasso]{devlin2016broadband}
Robert~C. Devlin, Mohammadreza Khorasaninejad, Wei~Ting Chen, Jaewon Oh, and
  Federico Capasso.
\newblock Broadband high-efficiency dielectric metasurfaces for the visible
  spectrum.
\newblock \emph{Proceedings of the National Academy of Sciences}, 113\penalty0
  (38):\penalty0 10473--10478, 2016.
\newblock \doi{10.1073/pnas.1611740113}.

\bibitem[Malitson(1965)]{malitson1965interspecimen}
I.~H. Malitson.
\newblock Interspecimen comparison of the refractive index of fused silica.
\newblock \emph{Journal of the Optical Society of America}, 55\penalty0
  (10):\penalty0 1205--1209, 1965.
\newblock \doi{10.1364/JOSA.55.001205}.

\bibitem[Green and Keevers(1995)]{green1995optical}
Martin~A. Green and Mark~J. Keevers.
\newblock Optical properties of intrinsic silicon at 300{\,}{K}.
\newblock \emph{Progress in Photovoltaics: Research and Applications},
  3\penalty0 (3):\penalty0 189--192, 1995.
\newblock \doi{10.1002/pip.4670030303}.

\bibitem[Mahajan(1983)]{mahajan1983strehl}
Virendra~N. Mahajan.
\newblock Strehl ratio for primary aberrations in terms of their aberration
  variance.
\newblock \emph{Journal of the Optical Society of America}, 73\penalty0
  (6):\penalty0 860--861, 1983.
\newblock \doi{10.1364/JOSA.73.000860}.

\bibitem[Pestourie et~al.(2023)Pestourie, Mroueh, Rackauckas, Das, and
  Johnson]{pestourie2023physics}
Rapha{\"e}l Pestourie, Youssef Mroueh, Chris Rackauckas, Payel Das, and
  Steven~G. Johnson.
\newblock Physics-enhanced deep surrogates for partial differential equations.
\newblock \emph{Nature Machine Intelligence}, 5\penalty0 (12):\penalty0
  1458--1465, 2023.
\newblock \doi{10.1038/s42256-023-00761-y}.

\bibitem[kim(2026)]{kim2026dataset}
Test datasets and trained models for generalizable electromagnetic field
  prediction.
\newblock GitHub repository, release v1.0.0, 2026.
\newblock URL
  \url{https://github.com/nanophotonics-lab/random-pattern-em-surrogate/releases/tag/v1.0.0}.

\end{thebibliography}
}


\clearpage
{\small
\begin{thebibliography}{11}
\providecommand{\natexlab}[1]{#1}
\providecommand{\url}[1]{\texttt{#1}}
\expandafter\ifx\csname urlstyle\endcsname\relax
  \providecommand{\doi}[1]{doi: #1}\else
  \providecommand{\doi}{doi: \begingroup \urlstyle{rm}\Url}\fi

\bibitem[Gu et~al.(2022)Gu, Gao, Feng, Zhu, Chen, Boning, and
  Pan]{SI:gu2022neurolight}
Jiaqi Gu, Zhengqi Gao, Chenghao Feng, Hanqing Zhu, Ray~T. Chen, Duane~S.
  Boning, and David~Z. Pan.
\newblock {NeurOLight}: A physics-agnostic neural operator enabling parametric
  photonic device simulation.
\newblock In \emph{Advances in Neural Information Processing Systems
  (NeurIPS)}, volume~35, pages 14623--14636, 2022.
\newblock \doi{10.52202/068431-1063}.

\bibitem[Perlin(2002)]{SI:perlin2002improving}
Ken Perlin.
\newblock Improving noise.
\newblock \emph{ACM Transactions on Graphics}, 21\penalty0 (3):\penalty0
  681--682, 2002.
\newblock \doi{10.1145/566654.566636}.

\bibitem[Oskooi et~al.(2010)Oskooi, Roundy, Ibanescu, Bermel, Joannopoulos, and
  Johnson]{SI:oskooi2010meep}
Ardavan~F. Oskooi, David Roundy, Mihai Ibanescu, Peter Bermel, J.~D.
  Joannopoulos, and Steven~G. Johnson.
\newblock Meep: A flexible free-software package for electromagnetic
  simulations by the {FDTD} method.
\newblock \emph{Computer Physics Communications}, 181\penalty0 (3):\penalty0
  687--702, 2010.
\newblock \doi{10.1016/j.cpc.2009.11.008}.

\bibitem[Seo et~al.(2026)Seo, Kang, Seo, and Chung]{SI:seo2026broadband}
Joonhyuk Seo, Chanik Kang, Dongjin Seo, and Haejun Chung.
\newblock Broadband electromagnetic field prediction at unseen wavelengths via
  physics-guided neural operators.
\newblock \emph{Nanophotonics}, 15\penalty0 (15):\penalty0 e70242, 2026.
\newblock \doi{10.1002/nap2.70242}.

\bibitem[Wiecha and Muskens(2020)]{SI:wiecha2020deep}
Peter~R. Wiecha and Otto~L. Muskens.
\newblock Deep learning meets nanophotonics: a generalized accurate predictor
  for near fields and far fields of arbitrary {3D} nanostructures.
\newblock \emph{Nano Letters}, 20\penalty0 (1):\penalty0 329--338, 2020.
\newblock \doi{10.1021/acs.nanolett.9b03971}.

\bibitem[Mahajan(1983)]{SI:mahajan1983strehl}
Virendra~N. Mahajan.
\newblock Strehl ratio for primary aberrations in terms of their aberration
  variance.
\newblock \emph{Journal of the Optical Society of America}, 73\penalty0
  (6):\penalty0 860--861, 1983.
\newblock \doi{10.1364/JOSA.73.000860}.

\bibitem[Born and Wolf(1999)]{SI:born1999principles}
Max Born and Emil Wolf.
\newblock \emph{Principles of Optics}.
\newblock Cambridge University Press, Cambridge, 7 edition, 1999.

\bibitem[Chen et~al.(2022)Chen, Lupoiu, Mao, Huang, Jiang, Lalanne, and
  Fan]{SI:chen2022high}
Mingkun Chen, Robert Lupoiu, Chenkai Mao, Der-Han Huang, Jiaqi Jiang, Philippe
  Lalanne, and Jonathan~A. Fan.
\newblock High speed simulation and freeform optimization of nanophotonic
  devices with physics-augmented deep learning.
\newblock \emph{ACS Photonics}, 9\penalty0 (9):\penalty0 3110--3123, 2022.
\newblock \doi{10.1021/acsphotonics.2c00876}.

\bibitem[Hughes et~al.(2019)Hughes, Williamson, Minkov, and
  Fan]{SI:hughes2019forward}
Tyler~W. Hughes, Ian A.~D. Williamson, Momchil Minkov, and Shanhui Fan.
\newblock Forward-mode differentiation of {M}axwell's equations.
\newblock \emph{ACS Photonics}, 6\penalty0 (11):\penalty0 3010--3016, 2019.
\newblock \doi{10.1021/acsphotonics.9b01238}.

\bibitem[Rumpf(2012)]{SI:rumpf2012fdfd}
Raymond~C. Rumpf.
\newblock Simple implementation of arbitrarily shaped
  total-field/scattered-field regions in finite-difference frequency-domain.
\newblock \emph{Progress In Electromagnetics Research B}, 36:\penalty0
  221--248, 2012.
\newblock \doi{10.2528/PIERB11092006}.

\bibitem[Shin and Fan(2012)]{SI:shin2012pml}
Wonseok Shin and Shanhui Fan.
\newblock Choice of the perfectly matched layer boundary condition for
  frequency-domain {M}axwell's equations solvers.
\newblock \emph{Journal of Computational Physics}, 231\penalty0 (8):\penalty0
  3406--3431, 2012.
\newblock \doi{10.1016/j.jcp.2012.01.013}.

\end{thebibliography}
}
\end{document}